\documentclass[iop,apj,twocolumn,numberedappendix,onecolappendix]{emulateapj}
\usepackage{amsmath}
\usepackage{txfonts}
\usepackage{xspace}
\usepackage{relsize}
\usepackage{enumitem}
\usepackage[usenames,dvipsnames]{color}
\usepackage[normalem]{ulem}
\usepackage{subfigure}
\usepackage[colorlinks = false,
            linkcolor = black,
            urlcolor  = blue,
            citecolor = black,
            anchorcolor = blue]{hyperref}

\newcommand{\kB}{\ensuremath{k_\mathrm{B}}} 
\newcommand{\me}{\ensuremath{m_\mathrm{e}}} 

\newcommand{\code}[1]{\texttt{#1}}
\newcommand{\mesa}{\code{MESA}}
\newcommand{\MESA}{\mesa}
\newcommand{\ADIPLS}{\code{ADIPLS}}
\newcommand{\GYRE}{\code{GYRE}}
\newcommand{\voned}{\code{V1D}}

\newcommand{\mesastar}{\mesa\code{star}}
\newcommand{\mesabinary}{\mesa\code{binary}}
\newcommand{\MESAstar}{\mesastar}

\newcommand{\rates}{\code{rates}}

\newcommand{\dif}{\ensuremath{\mathrm{d}}}

\newcommand{\nuclei}[2]{\ensuremath{\mathrm{^{#1}#2}}}

\newcommand{\helium}[1][4]{\nuclei{#1}{He}}

\newcommand{\nitrogen}[1][14]{\nuclei{#1}{N}}
\newcommand{\oxygen}[1][16]{\nuclei{#1}{O}}
\newcommand{\fluorine}[1][19]{\nuclei{#1}{F}}
\newcommand{\neon}[1][20]{\nuclei{#1}{Ne}}
\newcommand{\sodium}[1][23]{\nuclei{#1}{Na}}
\newcommand{\magnesium}[1][24]{\nuclei{#1}{Mg}}

\newcommand{\silicon}[1][28]{\nuclei{#1}{Si}}

\newcommand{\sulfur}[1][32]{\nuclei{#1}{S}}

\newcommand{\iron}[1][56]{\nuclei{#1}{Fe}}

\newcommand{\nickel}[1][58]{\nuclei{#1}{Ni}}

\newcommand{\zinc}[1][64]{\nuclei{#1}{Zn}}

\newcommand{\unitspace}{\ensuremath{\,}}
\newcommand{\usp}{\unitspace}

\newcommand{\unitstyle}[1]{\ensuremath{\mathrm{#1}}}
\newcommand{\power}[2]{\ensuremath{{#1}^{#2}}}

\newcommand{\centi}{\unitstyle{c}}
\newcommand{\kilo}{\unitstyle{k}}

\newcommand{\meter}{\unitstyle{m}}

\newcommand{\second}{\unitstyle{s}}

\newcommand{\Kelvin}{\unitstyle{K}}
\newcommand{\K}{\Kelvin}  

\newcommand{\cm}{\centi\meter}
\newcommand{\gram}{\unitstyle{g}}

\newcommand{\grampercc}{\gram\usp\power{\cm}{-3}} 

\newcommand{\erg}{\unitstyle{ergs}} 

\newcommand{\ergspersecond}{\erg\unitspace\power{\second}{-1}}

\newcommand{\Msun}{\ensuremath{\unitstyle{M}_\odot}}
\newcommand{\Lsun}{\ensuremath{\unitstyle{L}_{\odot}}}

\newcommand{\hour}{\unitstyle{hr}} 
\newcommand{\kms}{\ensuremath{\mathrm{km}\,\second^{-1}}\xspace}

\input{macros/vectors}
\newcommand{\Jorb}{\ensuremath{J_{\rm orb}}} 
\newcommand{\Jdot}{\ensuremath{\dot{J}}} 
\newcommand{\Jorbdot}{\ensuremath{\dot{J}_{\rm orb}}} 
\newcommand{\Jgrdot}{\ensuremath{\dot{J}_{\rm gr}}} 
\newcommand{\Jmldot}{\ensuremath{\dot{J}_{\rm ml}}} 
\newcommand{\Jmbdot}{\ensuremath{\dot{J}_{\rm mb}}} 
\newcommand{\Jlsdot}{\ensuremath{\dot{J}_{\rm ls}}} 
\newcommand{\porb}{\ensuremath{P_{\rm orb}}} 
\newcommand{\HP}{\ensuremath{H_{\rm P}}} 

\newcommand{\epsnuc}{\ensuremath{\epsilon_{\mathrm{nuc}}}} 
\newcommand{\epsgrav}{\ensuremath{\epsilon_{\mathrm{grav}}}} 
\newcommand{\epsnu}{\ensuremath{\epsilon_{\mathrm{\nu}}}} 

\newcommand{\vhat}{\ensuremath{\hat{v}}}           
\newcommand{\etavisc}{\ensuremath{\eta_{\mathrm{visc}}}} 
\newcommand{\Qvisc}[1]{\ensuremath{Q_{\mathrm{visc}#1}}} 
\newcommand{\taumlt}{\ensuremath{\tau_{\mathrm{MLT}}}}  

\newcommand{\foe}{\ensuremath{10^{51}\,{\rm erg}}}

\newcommand{\Teff}{\ensuremath{T_{\rm eff}}}	

\newcommand{\fov}{\ensuremath{f_{\mathrm{ov}}}} 

\newcommand{\freq}{\ensuremath{\sigma}}           
\newcommand{\freqad}{\ensuremath{\sigma_{{\rm ad}}}}           
\newcommand{\dfreq}{\ensuremath{\omega}}           
\newcommand{\nuosc}{\ensuremath{\nu_{\rm osc}}}
\newcommand{\tosc}{\ensuremath{\tau_{\rm osc}}}
\newcommand{\Posc}{\ensuremath{P_{\rm osc}}}

\newcommand{\repart}{\ensuremath{\operatorname{Re}}}
\newcommand{\impart}{\ensuremath{\operatorname{Im}}}

\newcommand{\BV}{Brunt-V\"{a}is\"{a}l\"{a}}

\newcommand{\tprime}{\ensuremath{t'}} 
\newcommand{\grredshift}{\ensuremath{\rm{z}}} 

\newcommand{\mdotyr}{\ensuremath{\rm \Msun\,yr^{-1}}} 

\newcommand{\alphaMLT}{\ensuremath{\alpha_{\mathrm{MLT}}}}	
\newcommand{\timestep}{\ensuremath{\delta t}} 

\newcommand{\Tc}{\ensuremath{T_{\mathrm{\!c}}}} 
\newcommand{\rhoc}{\ensuremath{\rho_{\mathrm{c}}}} 
\newcommand{\Mdot}{\ensuremath{\dot{M}}} 
\newcommand{\Ye}{\ensuremath{Y_{\mathrm{e}}}} 
\newcommand{\mue}{\ensuremath{\mu_{\mathrm{e}}}} 

\newcommand{\EF}{\ensuremath{E_\mathrm{F}}} 

\newcommand{\mesaone}{Paper~I}  
\newcommand{\mesatwo}{Paper~II} 

\newlength{\apjcolwidth}
\newlength{\figwidth}
\newlength{\doublewide}
\newcommand\add{}

\begin{document}
\title{Modules for Experiments in Stellar Astrophysics (MESA): \\
Binaries, Pulsations, and Explosions}
\author{%
Bill Paxton,\altaffilmark{1}
Pablo Marchant,\altaffilmark{2}
Josiah Schwab,\altaffilmark{3,4}
Evan B. Bauer,\altaffilmark{5}
Lars Bildsten,\altaffilmark{1,5}
Matteo Cantiello,\altaffilmark{1}\\
Luc Dessart,\altaffilmark{6}
R. Farmer,\altaffilmark{7}
H. Hu,\altaffilmark{8}
N. Langer,\altaffilmark{2}
R.H.D.~Townsend,\altaffilmark{9}
Dean M. Townsley,\altaffilmark{10}
and F.X.~Timmes\altaffilmark{7}
}
\altaffiltext{1}{Kavli Institute for Theoretical Physics, University of California, Santa Barbara, CA 93106, USA}
\altaffiltext{2}{Argelander Institut f\"ur Astronomie, Universitat Bonn, Auf dem H\"ugel 71, 53121 Bonn, Germany}
\altaffiltext{3}{Physics Department, University of California, Berkeley, CA 94720, USA}
\altaffiltext{4}{Astronomy Department and Theoretical Astrophysics Center, University of California, Berkeley, CA 94720, USA}
\altaffiltext{5}{Department of Physics, University of California, Santa Barbara, CA 93106, USA}
\altaffiltext{6}{Laboratoire Lagrange, UMR7293, Universit\'e Nice Sophia-Antipolis, CNRS, Observatoire de la C\^{o}te d'Azur, 06300 Nice, France.}
\altaffiltext{7}{School of Earth and Space Exploration, Arizona State University, Tempe, AZ 85287, USA}
\altaffiltext{8}{SRON, Netherlands Institute for Space Research, Sorbonnelaan 2, 3584 CA Utrecht, Netherlands}
\altaffiltext{9}{Department of Astronomy, University of Wisconsin-Madison, Madison, WI 53706, USA}
\altaffiltext{10}{Department of Physics \& Astronomy, University of Alabama, Tuscaloosa, AL 35487, USA}

\email{Corresponding author: pablo@astro.uni-bonn.de}

\begin{abstract}

We substantially update the capabilities of the open-source software
instrument Modules for Experiments in Stellar Astrophysics (\MESA).
\MESA\ can now simultaneously evolve an interacting pair of
differentially rotating stars undergoing transfer and loss of mass and
angular momentum, greatly enhancing the prior ability to model binary
evolution.  New \MESA\ capabilities in fully coupled calculation of
nuclear networks with hundreds of isotopes now allow \MESA\ to
accurately simulate advanced burning stages needed to construct
supernova progenitor models.  Implicit hydrodynamics with shocks can
now be treated with \MESA, enabling modeling of the entire massive
star lifecycle, from pre-main sequence evolution to the onset of core
collapse and nucleosynthesis from the resulting explosion. Coupling of
the \GYRE\ non-adiabatic pulsation instrument with \MESA\ allows for
new explorations of the instability strips for massive stars while
also accelerating the astrophysical use of asteroseismology data.  We
improve treatment of mass accretion, giving more accurate and robust
near-surface profiles.  A new \MESA\ capability to calculate weak
reaction rates ``on-the-fly'' from input nuclear data allows better
simulation of accretion induced collapse of massive white dwarfs and
the fate of some massive stars.  We discuss the ongoing challenge of
chemical diffusion in the strongly coupled plasma regime, and exhibit
improvements in \MESA\ that now allow for the simulation of radiative
levitation of heavy elements in hot stars. We close by noting that the
\MESA\ software infrastructure provides bit-for-bit consistency for
all results across all the supported platforms, a profound enabling
capability for accelerating \MESA's development.
\end{abstract}


\maketitle
\tableofcontents
\section{Introduction}\label{s.introduction}

The development of a relatively complete and quantitative picture of
stellar evolution is one of the great drivers of astrophysics.
On the observational side of this impetus,  the {\it Kepler}
\citep{borucki_2010_aa} and
the CoRoT \citep[][]{baglin_2009_aa}
missions
continuously monitored more than 100,000
stars in a 100 deg$^2$ window with a dynamic range of apparent stellar
brightness of 10$^{6}$. Highlights include the discoveries that nearly all
$\gamma$ Doradus and $\delta$ Scuti stars are hybrid pulsators, 
and the detection of solar-like oscillations in a large sample of red
giants \citep{auvergne_2009_aa,de-ridder_2009_aa,grigahcene_2010_aa,
bedding_2010_aa,
christensen-dalsgaard_2011_aa,chaplin_2013_aa}.
The {\it Dark Energy Survey}
is scanning 5000 deg$^2$ of the southern sky in 5 optical
filters every few days to discover and study thousands of supernovae
\citep[e.g.,][]{papadopoulos_2015_aa,yuan_2015_aa}.
Building upon the legacy of the {\it Palomar Transient Factory}
\citep[][]{law_2009_aa}, the
{\it intermediate Palomar Transient Factory} conducts a fully-automated,
wide-field survey that systematically explores the
transient sky with a 90 second to 5 day cadence
\citep[][]{vreeswijk_2014_aa}.
The forthcoming
{\it Zwicky Transient Facility} will enable a survey more than an order of
magnitude faster at the same depth as its predecessors.  In its
unique orbit the {\it Transiting Exoplanet Survey Satellite} will have an
unobstructed view to scrutinize the light curves of the brightest
100,000 stars with a 1 minute cadence
\citep{ricker_2015_aa}.
The {\it Gaia} mission aims
to provide unprecedented distance and radial velocity measurements
with the accuracies needed to reveal the evolutionary state,
composition, and kinematics of about one billion stars in our Galaxy
\citep[e.g.,][]{creevey_2015_aa,sacco_2015_aa}.
The {\it Large Synoptic Survey Telescope} will image the entire Southern
Hemisphere deeply in multiple optical colors every week with its three
billion pixel digital camera, thus opening a new window on
transient objects such as interacting close binary systems.
\citep[e.g.,][]{oluseyi_2012_aa}.

Interpreting these new observations and predicting new stellar
phenomena propels the theoretical side, in particular the evolution of
the community software instrument Modules for Experiments in Stellar
Astrophysics (\mesa) for research and education.  We introduced
\mesa\ in \citet[][hereafter \mesaone]{paxton_2011_aa} and expanded
its range of capabilities in \citet[][hereafter \mesatwo]{paxton_2013_aa}.  
This paper describes the major new advances for \MESA\ modeling of
binary systems, shock hydrodynamics, explosions of massive stars and
x-ray bursts with large, in situ reaction networks.  Moreover it
details the coupling of \MESA\ with the non-adiabatic pulsation
software instrument \GYRE\ \citep{Townsend:2013aa}.  We also describe
advances made to existing modules since \mesatwo, including 
improved treatments of mass accretion, weak reaction rates,
and particle diffusion.

It has been a little more than 200 years since
\citet{herschel_1802_aa} announced, after 25 years of observation,
that certain pairs of stars displayed evidence of orbital motion
around their common center of mass.  Binary systems allow the masses
of their component stars to be directly determined, which in turn
allows stellar radii to be indirectly estimated.  This allows the
calibration of an empirical mass-luminosity relationship from which
the masses of single stars can be estimated \citep{torres_2010_aa}.
Recent surveys such as \citet{raghavan_2010_aa} suggest 30\% to 50\%
of solar-like systems in the Galactic disk are composed of binaries,
where the binary fraction is higher for more massive stars
\citep{b:sana2012, kobulnicky_2014_aa}.
As argued by  \citet{b:demi2013}, the most rapidly rotating massive stars are  expected
in binary systems as a consequence of accretion-induced spin-up.
Differential rotation has a major impact on the evolution of massive stars
\citep{maeder_2000_aa,Heger:2000,Heger:2005} and
for single stars the corresponding physics has been included into \MESA\ as described in \mesatwo.
On the other hand very few works that include the physics of
differential rotation in binaries have been published \citep{b:petr2005,b:cant2007}.
Our improvements to \mesa\ now allow for the calculation of differentially-rotating binary stars.

The rapid expansion of extra-solar planet research has led to a
revival of interest in the detailed properties of stars probed through
space-based brightness variability studies and radial velocity
measurements.  Stellar properties can be derived from measurements of
the radial and non-radial oscillation modes of a star, but this
requires the accurate and efficient computations of mode frequencies
and their eigenfunctions enabled by the coupling of \GYRE\ with \MESA.

There are many ways $M \gtrsim 8\,\Msun$ stars can end their lives
\citep[e.g.,][]{woosley_2002_aa,smartt_2009_aa,meynet_2010_aa,langer_2012_aa,nomoto_2013_aa,smith_2014_ac}.
Some become electron capture supernovae; others collapse with most of
their extended envelope intact and yield a Type II supernova; others
can lead to pair instability; and some have envelopes thin enough to
allow a jet to break through and appear as a long gamma-ray burst
\citep[][]{macfadyen_1999_aa,woosley_2006_aa,gehrels_2009_aa}.  There
is a pressing need throughout the stellar community to routinely
explore this entire mass range with new supernova progenitor and
explosion models.  The observational facilities discussed above have
found explosions that indicate large amounts of mass are lost within a
few years of explosion \citep{smith_2014_ac}; some show evidence of
optically thick winds present at the moment of explosion
\citep{ofek_2014_aa}, while others have yet to be securely identified
with a specific core collapse scenario.  These mysteries, coupled with
the community's call for new yields from massive stars for galactic
chemical evolution studies motivate the development of implicit shock
hydrodynamics and explosions with large, in situ reactions networks in
\mesa.

The paper is outlined as follows.
Section \ref{s.binaries} describes
the new capability of \mesa\ to evolve binary systems. 
Section \ref{s.pulse} discusses the new non-adiabatic pulsation
capabilities resulting from fully coupling to the \GYRE\ software
instrument.
Section \ref{s.hydro} describes the improvements to
accommodate implicit hydrodynamics with shocks.
New capabilities for advanced burning and X-ray bursts 
with large, in situ
reaction networks are described in Section \ref{s.advburn}.
In Section \ref{s.ccsn} we 
model the pre-supernova evolution of massive stars and 
combine the implicit hydrodynamics module
and the new capabilities for advanced burning to probe the nucleosynthesis
and yields of core-collapse supernovae.
Section \ref{s.accretion}
discusses the improvements for a more robust and efficient treatment of
mass accretion.
Section \ref{s.weak}
presents a new option for an on-the-fly calculation of weak reaction rates
and their application to the Urca process and accretion-induced-collapse
models.
Section \ref{s.mass_diffusion} presents improvements in the physics
implementation of particle diffusion by
including radiative levitation and pushing diffusion methods
into the strongly coupled, electron degenerate
regime.
In Section \ref{s.infrastructure} we discuss improvements to
the \MESA\ software infrastructure, highlighting bit-for-bit consistency
across operating systems and compilers.
We conclude in Section~\ref{s.conclusions} by noting additional
improvements to \mesa\ are likely to occur in the near future.
Important symbols are defined in Table~\ref{t.list-of-symbols}.  We
denote components of \mesa, such as modules and routines, in Courier
font, e.g., \code{evolve\_star}.

{\LongTables
\begin{deluxetable}{clr}
  \tablecolumns{3}
  \tablewidth{\apjcolwidth}
  \tablecaption{Variable Index.\label{t.list-of-symbols}}
  \tablehead{\colhead{Name} & \colhead{Description} & \colhead{First Appears}}
  \startdata
  $a$            & Orbital seperation                                   & \ref{binaries:init} \\ 
$A$            & Atomic mass number                                   & \ref{s.advburn} \\ 
$\alpha$       & Fine structure constant                              & \ref{s.weak-details} \\ 
$c$            & Speed of light                                       & \ref{binaries:gwr} \\ 
$e$            & Specific thermal energy                              & \ref{s.hydro_ener} \\
$\eta$         & Wind mass loss coefficient                           & \ref{s.presn} \\ 
$g$            & Gravitational acceleration                           & \ref{s.xrb} \\ 
$G$            & Gravitational constant                               & \ref{binaries:init} \\ 
$\Gamma$       & Coulomb coupling parameter                           & \ref{s.mass_diffusion} \\ 
$I$            & Moment of inertia                                    & \ref{binaries:tiderot} \\
$\kappa$       & Opacity                                              & \ref{s.introduction} \\
$L$            & Luminosity                                           & \ref{s.hydro_ener}\\ 
$\lambda$      & Reaction rate                                        & \ref{s.advburn} \\ 
$m$            & Lagrangian mass coordinate                           & \ref{s.hydro} \\ 
$M$            & Stellar mass                                         & \ref{binaries:init} \\ 
$M_1$          & Donor mass                                           & \ref{binaries:init} \\ 
$M_2$          & Accretor mass                                        & \ref{binaries:init} \\ 
$\mu$          & Mean molecular weight                                & \ref{binaries:explicit} \\ 
$N$            & Neutron number                                       & \ref{s.presn} \\
$\dfreq$       & Dimensionless eigenfrequency                         & \ref{s.pulse_instab}\\ 
$\Omega$       & Angular frequency                                    & \ref{binaries:tiderot} \\
$Q$            & Nuclear rest mass energy difference                  & \ref{s.weak-details} \\
$P$            & Pressure                                             & \ref{binaries:explicit} \\ 
$q$            & Fractional mass coordinate                           & \ref{s.accretion}\\
$q_1$          & Mass ratio, $M_1 / M_2$                               & \ref{binaries:rlof}\\ 
$q_2$          & Mass ratio, $M_2 / M_1$                               & \ref{binaries:rlof}\\ 
$r$            & Radial coordinate                                    & \ref{binaries:explicit} \\ 
$R$            & Stellar radius                                       & \ref{binaries:rlof} \\ 
$\rho$         & Baryon mass density                                  & \ref{binaries:explicit}\\ 
$s$            & Specific entropy                                     & \ref{s.accretion} \\ 
$\freq$        & Oscillation eigenfrequency                           & \ref{s.pulse_instab} \\ 
$t$            & Time                                                 & \ref{binaries:tiderot}\\
$T$            & Temperature                                          & \ref{binaries:explicit} \\ 
$\tau$         & Timescale                                            & \ref{binaries:tiderot} \\ 
$v$            & Velocity                                             & \ref{s.hydro_cont} \\
$X$            & Baryon mass fraction                                 & \ref{s.advburn} \\ 
$Y$            & Molar abundance                                      & \ref{s.advburn} \\ 
\grredshift    & Gravitational redshift                               & \ref{s.xrb} \\
$Z$            & Atomic number                                        & \ref{s.advburn} \\ 

  \alphaMLT &  Mixing length parameter & \ref{s.pulse_instab} \\
$C_P$ & Mass specific heat at constant pressure & \ref{s.accretion}\\
$\chi_\rho$ & $(\partial\ln P/\partial\ln \rho)_T$ & \ref{s.accretion}\\
$\chi_T$ & $(\partial\ln P/\partial\ln T)_\rho$ & \ref{s.accretion}\\
$dm$ &  Mass of cell & \ref{s.hydro_cont} \\
$\overline{dm}$ &  mass associated with cell face & \ref{s.hydro_mom} \\
$dq$ & Fractional mass of cell & \ref{s.accretion} \\
\timestep &  Numerical timestep & \ref{binaries:ls} \\ 
$\Delta M$ & Change of stellar mass in one step & \ref{s.accretion} \\
$\nabla_{\rm ad}$ & Adiabatic temperature gradient $(\partial \ln T/\partial \ln P)_{\rm ad}$ & \ref{s.accretion} \\
$\nabla_T$ & Stellar temperature gradient $\dif \ln T/\dif \ln P$ & \ref{s.accretion} \\
\EF &  Fermi energy & \ref{s.presn}\\
\epsgrav &  Gravitational heating rate & \ref{s.shocks} \\
\epsnu &  Neutrino energy loss rate & \ref{s.shocks} \\
\epsnuc &  Nuclear energy generation rate & \ref{s.shocks} \\
$\epsilon_{\rm visc}$ & Viscous heating rate & \ref{s.hydro_ener} \\ 
\etavisc &  Artificial dynamic viscosity coefficient & \ref{s.hydro_cont} \\
\fov &  Convective overshoot parameter & \ref{s.pulse_instab} \\
$g_{\rm visc}$ & Viscous acceleration & \ref{s.hydro_mom} \\ 
$g_{\rm rad}$ & Radiative acceleration & \ref{s:Hu} \\ 
$\Gamma_{1}$ & First adiabatic exponent $(\partial \ln P/\partial \ln \rho)_{\rm ad}$ & \ref{binaries:explicit} \\
\HP &  Pressure scale height & \ref{binaries:rlof} \\
\Jdot &  Rate of change of angular momentum & \ref{binaries:evolution} \\
\Jorb &  Orbital angular momentum & \ref{binaries:init} \\ 
\kB &  Boltzmann constant & \ref{binaries:explicit} \\ 
$\lambda_{\rm ion}$ & Mean inter-ion spacing & \ref{s.mass_diffusion} \\
$m_{\rm p}$ & Proton mass & \ref{binaries:explicit} \\ 
$M_{\rm acc}$ & Accreted mass accumulated, $\dot M t$ & \ref{s.accretion_testing}\\
$M_{\rm c}$ & Mass of unmodeled inert core & \ref{s.accretion}\\
$M_{\rm ign}$ & $M_{\rm acc}$ at time of nova runaway & \ref{s.accretion_testing}\\
$\dot{M}_{\rm Edd}$ & Eddington accretion rate & \ref{binaries:init} \\ 
\mue & Electron chemical potential & \ref{s.weak-details} \\ 
$n_{\rm ion}$ & Ion number density & \ref{s.mass_diffusion} \\ 
$\nu_{\rm osc}$ & Linear oscillation frequency & \ref{s.pulse_instab} \\ 
\porb &  Orbital period  & \ref{binaries:init} \\ 
\Qvisc{ } & Artificial viscosity energy & \ref{s.hydro_visc} \\ 
$R_{\rm RL}$ & Roche lobe radius & \ref{binaries:rlof} \\ 
$R_{\rm D}$ & Debye radius & \ref{s:Burgers} \\ 
\rhoc &  Central baryon mass density & \ref{s.ccsn} \\
\tprime & GR corrected time for observer at infinity & \ref{s.xrb} \\
\Tc &  Central temperature & \ref{s.ccsn} \\
\Teff &  Effective temperature & \ref{binaries:explicit} \\ 
$\tau_{\rm acc}$ & Timescale to accrete outer star layer & \ref{s.accretion}\\
$\tau_{\rm MLT}$ & Convective timescale & \ref{s.hydro-limit} \\ 
$\tau_{\rm osc}$ & Oscillation e-folding time & \ref{s.pulse_instab} \\ 
$\tau_{\rm sync}$ & Tidal synchronization timescale & \ref{binaries:tiderot} \\ 
$\tau_{\rm th}$ & Thermal timescale of outer star layer & \ref{s.accretion}\\
\taumlt &  Convection time scale & \ref{s.hydro_conserve} \\
\vhat &  Time centered velocity & \ref{s.hydro_cont} \\
$\Ye$ & Electrons per baryon & \ref{s.advburn}

  \enddata
\end{deluxetable}
}

\section{Binaries}\label{s.binaries}

{\mesabinary} is a {\mesa} module that uses {\mesastar} to evolve
binary systems. It can be used to evolve a full stellar model plus a
point mass companion or to simultaneously evolve the structure of two
stars. It optionally allows the modeling of systems including stellar
rotation, assuming the axis of rotation of each star to be
perpendicular to the orbital plane, accounting for the effects of
tidal interaction and spin-up through accretion.  The implementation
of {\mesabinary} benefits from early contributions by
\citet{b:madh2006} and \citet{b:lin2011} who modeled mass transfer
from a star to a point mass.

Here we provide an overview of the modelled physical processes for
circular binary systems and describe the tests against which we validate
{\mesabinary}.

\subsection{Initialization of a Circular Binary System}\label{binaries:init}

A binary system is initialized by specifying the components and either
the orbital period $\porb$ or separation $a$.  Each component can be a
point mass or a stellar model.  The initial model(s) are provided by a
saved \MESA\ model or a zero-age main-sequence (ZAMS) specification.
For stellar models including rotation, the initial rotational
velocities of each component can be explicitly defined, or set such
that the star is synchronized to the orbit at the beginning of the
evolution.  The orbital angular momentum of the system is
\begin{eqnarray}
\Jorb = M_1 M_2 \sqrt{\frac{G a}{ M_1+M_2 }},\label{binaries:Jorb}
\end{eqnarray}
where $M_1$ and $M_2$ are the stellar masses.  Evolution of $M_1$,
$M_2$, and $J_{\mathrm{orb}}$ is used to update $a$ using Equation
(\ref{binaries:Jorb}).
Masses can be modified both by Roche lobe overflow (RLOF) and winds.
The total time derivatives of the component masses are given by
\begin{eqnarray}
\dot{M}_1 = \dot{M}_{1,\mathrm{w}}+\dot{M}_{\mathrm{RLOF}},\quad
\dot{M}_2 = \dot{M}_{2,\mathrm{w}}-f_{\mathrm{mt}}\dot{M}_{\mathrm{RLOF}},
\end{eqnarray}
where $M_1$ is the donor mass and $M_2$ the accretor mass.  The
stellar wind mass loss rates are $\dot{M}_{1,\mathrm{w}}$ and
$\dot{M}_{2,\mathrm{w}}$ (see {\mesaone} and {\mesatwo}) and
$\dot{M}_{\mathrm{RLOF}}$ is the mass transfer rate from RLOF, all
defined as negative. The factor $f_{\mathrm{mt}}$ represents the
efficiency of accretion and can be used to limit  accretion to
the Eddington rate $\dot{M}_{\mathrm{Edd}}$.

\subsection{Evolution of Orbital Angular Momentum}\label{binaries:evolution}

To compute the rate of change of orbital angular momentum, we consider
the contribution of gravitational waves, mass loss, magnetic braking,
and spin orbit (LS) coupling
\begin{eqnarray}
\Jorbdot = \Jgrdot + \Jmldot + \Jmbdot  + \Jlsdot
\enskip ,
\label{binaries:Jorb_dot}
\end{eqnarray}
from which the change in orbital angular momentum in one step is
calculated as $\Delta\Jorb = \Jorbdot\delta t$, where $\delta t$ is
the timestep.  Unless models with stellar rotation are being used, the
$\Jlsdot$ term is equal to zero, and the contribution of the
individual spins of each star is not directly considered. On the other
hand, the $\Jmbdot$ term implicitly assumes a strong tide that keeps
the orbit synchronized.  The simultaneous usage of $\Jmbdot$ with
stellar rotation is not consistent (see Section \ref{s.jmb}).  We now
describe how these terms are computed.

\subsubsection{Gravitational Wave Radiation  \label{binaries:gwr} }

Very compact binaries can experience significant orbital decay due to
the emission of gravitational waves.  Observations of the Hulse-Taylor
binary pulsar over three decades \citep{b:weis2005} and of the double
pulsar \citep{b:kram2006} have tested the predicted effect from
general relativity to a high precision. The angular momentum loss from
gravitational waves is

\begin{eqnarray}
\Jgrdot = -\frac{32 }{5c^5}\left(\frac{2\pi G}{\porb}\right)^{7/3}\frac{(M_1M_2)^2}{(M_1+M_2)^{2/3}}.
\end{eqnarray}

\subsubsection{Mass Loss}\label{binaries:jdot_ml}

We assume the mass lost in a stellar wind has the specific orbital
angular momentum of its star.  For the case of inefficient mass
transfer, angular momentum loss follows \citet{b:sobe1997}, where
fixed fractions of the transferred mass are lost either as a fast
isotropic wind from each star or a circumbinary toroid with a given radius:
\begin{eqnarray}
\Jmldot & = \left[(\Mdot_{1,\mathrm{w}}+\alpha_{\mathrm{mt}}\dot{M}_{\mathrm{RLOF}})M_1^2+(\Mdot_{2,\mathrm{w}}+\beta_{\mathrm{mt}}\Mdot_{\mathrm{RLOF}})M_2^2\right] \qquad\nonumber \\
&\displaystyle \times \ \frac{a^2}{(M_1+M_2)^2} \frac{2\pi}{\porb} \nonumber\\
& + \ \gamma_{\mathrm{mt}}\delta_{\mathrm{mt}}\Mdot_{\mathrm{RLOF}} \sqrt{G(M_1+M_2)a},\label{binaries:equjdotml}
\end{eqnarray}
where $\alpha_{\mathrm{mt}}$, $\beta_{\mathrm{mt}}$, and
$\delta_{\mathrm{mt}}$ are respectively the fractions of mass
transferred that is lost from the vicinity of the donor, accretor and
circumbinary toroid, and $\gamma_{\mathrm{mt}}^2a$ is the radius of
the toroid. Ignoring winds, the efficiency of mass transfer is then
given by $f_{\mathrm{mt}} = 1 -\alpha_{\mathrm{mt}}-\beta_{\mathrm{mt}}-\delta_{\mathrm{mt}}$.
When accretion is limited to $\dot{M}_{\mathrm{Edd}}$, efficiency of
accretion is given by
\begin{eqnarray}
f_{\mathrm{mt}}=\min(1 -\alpha_{\mathrm{mt}}-\beta_{\mathrm{mt}}-\delta_{\mathrm{mt}},|\dot{M}_{\mathrm{Edd}}/\Mdot_{\mathrm{RLOF}}|),
\end{eqnarray}
and the additional mass being lost is added to the
$\beta_{\mathrm{mt}}\Mdot_{\mathrm{RLOF}}$ term in Equation
(\ref{binaries:equjdotml}), i.e., it is assumed to leave the system
carrying the specific orbital angular momentum of the accretor.

\subsubsection{Spin Orbit Coupling \label{binaries:ls}}

Tidal interaction and mass transfer can significantly modify the spin
angular momentum of the stars in a binary system, acting as both
sources and sinks for orbital angular momentum.  The impact spin-orbit
interactions have on orbital evolution depends on the orbital
separation and the mass ratio, with the effect being greater for
tighter orbits and uneven masses. The corresponding contribution to
$\Jorbdot$ is computed by demanding conservation of the total angular
momentum, accounting for losses due to the other
$\dot{J}_{\mathrm{orb}}$ mechanisms and loss of stellar angular
momentum due to winds.

In a fully conservative system, the change in orbital angular momentum
in one timestep is $\delta \Jorb=-\delta S_1-\delta S_2$, where
$\delta S_1$ and $\delta S_2$ are the changes in spin angular momenta.
This needs to be corrected if mass loss is included, as winds take
away angular momentum from the system. If $S_{1,\mathrm{lost}}$ and
$S_{2,\mathrm{lost}}$ are the amounts of spin angular momentum removed
in a step from each star due to mass loss (including winds and RLOF),
\begin{eqnarray}
\Jlsdot =\frac{-1}{\delta t}\left(\delta S_1-S_{1,\mathrm{lost}}\frac{\Mdot_{1,\mathrm{w}}}{\Mdot_1}+\delta S_2-S_{2,\mathrm{lost}}\right),\label{binaries:equjdotls}
\end{eqnarray}
where the additional factor for the donor accounts for mass lost from
the system, ignoring mass loss due to mass transfer. In the absence of
RLOF this equation becomes symmetric between both stars, as then
$\Mdot_{1,\mathrm{w}}/\Mdot_1=1$.

The form of Equation (\ref {binaries:equjdotls}) is independent of how
tides and angular momentum accretion work, as it is merely a statement
on angular momentum conservation. The details of how we model these
processes and their impact on the spin of each component are described
in Section \ref{binaries:tiderot}.

\subsubsection{Magnetic Braking}\label{s.jmb}

The rotational velocities of low mass stars are strongly correlated
with their ages \citep{b:skum1972}. This spin-down arises from the
coupling of the stellar wind to a magnetic field. If the star is
in a binary system and tidally coupled to the orbit, magnetic braking can provide a
very efficient sink for orbital angular momentum
\citep{b:mest1968,b:verb1981}. We implement this effect
following \citet{b:rapp1983},
who assumed the star being braked is
tidally synchronized:
\begin{eqnarray}
\Jmbdot = -6.82\times 10^{34}\left(\frac{M_1}{\mathrm{M_\odot}}\right)\left(\frac{R_1}{\mathrm{R_\odot}}\right)^{\gamma_{\mathrm{mb}}}\left(\frac{1\;\mathrm{d}}{\porb}\right)^3\;\mathrm{[dyn\;cm]},\label{binaries:Jmbcomp}
\end{eqnarray}
where in the simplest approximation
$\gamma_{\mathrm{mb}}$=\,4 \citep{b:verb1981}. A similar contribution
from the accretor can be included.  As tidal synchronization is assumed,
this formulation is incompatible with the use of LS coupling.

It is normally assumed that once a star becomes fully convective, the
dynamo process that regenerates the field will stop working or at
least behave in a qualitatively different manner. Similarly, magnetic
fields in stars with radiative envelopes are of a significantly
different nature than those seen in stars with convective envelopes,
and there is no simple way to predict even the presence of magnetism
\citep{b:dona2009}.  By default {\mesabinary} only accounts for
magnetic braking as long as the star being braked has a convective
envelope and a radiative core, though the process might still operate
outside of these conditions.

\subsection{Mass Transfer from RLOF}\label{binaries:rlof}

Close binary stars are defined as systems tight enough to interact
through mass transfer, with the most important mechanism being
RLOF. This process is commonly modeled in 1D by considering the
spherical-equivalent Roche lobe radius $R_{\mathrm{RL}}$ of each
object \citep{b:eggl1983}
\begin{equation}
R_{\mathrm{RL},j} = \frac{0.49 q_j^{2/3}}{0.6q_j^{2/3}+\ln\left(1+q_j^{1/3}\right)}a,
\end{equation}
where $j$ is the index identifying each star, $q_1=M_1/M_2$ and
$q_2=M_2/M_1$. This fit is correct up to a few percent for the full
range of mass ratios, $0<q_j<\infty$.  Mass transfer occurs then when
the radius of a star approaches or exceeds $R_{\mathrm{RL}}$.
Depending on several factors, once a star begins RLOF the ensuing mass
transfer phase can proceed on a nuclear, thermal, or dynamical timescale.

The stability of mass transfer is normally understood in terms of
mass-radius relationships \citep[e.g.,][]{b:tout1997,b:sobe1997},
\begin{align}
\zeta_{\mathrm{eq}} & = \left(\frac{d\ln R_1}{d\ln M_1}\right)_{\mathrm{eq}}, \\
\zeta_{\mathrm{ad}} & = \left(\frac{d\ln R_1}{d\ln M_1}\right)_{\mathrm{ad}}, \\
\zeta_{\mathrm{RL}} & = \frac{d\ln R_{\mathrm{RL},1}}{d\ln M_1}.
\end{align}
Here, $\zeta_{\mathrm{eq}}$ gives the radial response of the donor to
mass loss when it happens slowly enough for the star to remain in
thermal equilibrium.  When mass loss proceeds on a timescale much
shorter than the thermal timescale of the star, but still slow enough
for the star to retain hydrostatic equilibrium then the radial
response will be given by $\zeta_{\mathrm{ad}}$.  The dependency of
the Roche lobe radius on mass transfer is encoded in
$\zeta_{\mathrm{RL}}$.  In general $\zeta = d\ln R_{1}/d\ln M_1$ is a
function of $\Mdot_{\mathrm{RLOF}}$, so requiring
$\zeta=\zeta_{\mathrm{RL}}$ will determine the value of
$\Mdot_{\mathrm{RLOF}}$.  If an overflowing star satisfies
$\zeta_{\mathrm{eq}}>\zeta_{\mathrm{RL}}$, then it can remain inside
its Roche lobe by transferring mass while retaining thermal
equilibrium. If on the contrary
$\zeta_{\mathrm{ad}}>\zeta_{\mathrm{RL}}>\zeta_{\mathrm{eq}}$, mass
transfer will proceed on a thermal timescale, while for the extreme
case $\zeta_{\mathrm{RL}}>\zeta_{\mathrm{ad}}$ the star will depart
from hydrostatic equilibrium and the process will be dynamical.
\MESA\ cannot model common envelope or contact binaries.

{\mesabinary} provides both explicit and implicit methods to compute
mass transfer rates. An explicit computation sets the value of
$\Mdot_{\mathrm{RLOF}}$ at the start of a step, while an implicit one
begins with a guess for $\Mdot_{\mathrm{RLOF}}$ and iterates until the
required tolerance is reached.  The composition of accreted material
is set to that of the donor surface, and the specific entropy of
accreted material is the same as the surface of the accretor. In
models with rotation the specific angular momentum of accreted
material is described in Section \ref{binaries:tiderot}.

\subsubsection{Explicit Methods\label{binaries:explicit}}

{\mesabinary} implements two mass transfer schemes: the model of
\citet{b:ritt1988} which we refer to as the Ritter scheme and
\citet{b:kolb1990} which we refer to as the Kolb scheme.  We use the
mass ratio $q_2$ consistent with the Ritter scheme.

\paragraph{Ritter scheme:}
Stars have extended atmospheres therefore RLOF can
take place through the L1 point even when
$R_1<R_{{\mathrm{RL}},1}$.
\citet{b:ritt1988} estimated the mass transfer rate for
this case as

\begin{equation}
\dot{M}_{\mathrm{RLOF}} = -\dot{M}_0\exp\left(\frac{R_1-R_{\mathrm{RL},1}}{H_{P,1}/\gamma(q_2)}\right),
\end{equation}
where $H_{P,1}$ is the pressure scale height at the photosphere of the donor and
\begin{eqnarray}
\dot{M}_0 = \frac{2\pi}{\exp(1/2)}F_1(q_2)
\frac{R_{\mathrm{RL},1}^3}{GM_1}\left(\frac{k_{\rm B}T_{\mathrm{eff}}}{m_{\rm p} \  \mu_{\mathrm{ph}}}\right)^{3/2}\rho_{\mathrm{ph}},
\end{eqnarray}
where
$m_{\rm p}$ is the proton mass,
$T_{\mathrm{eff}}$ is the effective
temperature of the donor, and $\mu_{\mathrm{ph}}$ and $\rho_{\mathrm{ph}}$ are
the mean molecular weight and density at its photosphere. The two fitting functions are
\begin{eqnarray}
F_1(q_2)=1.23+0.5\log q_2,\qquad 0.5 \lesssim q_2 \lesssim 10.\label{binaries:ritterprob}
\end{eqnarray}
and
\begin{eqnarray}
\gamma(q_2)=\qquad\qquad\qquad\qquad\qquad\qquad\qquad\qquad\qquad\nonumber\\
\hspace{-1in}\begin{cases}
0.954+0.025\log q_2-0.038(\log q_2)^2, & 0.04\lesssim q_2 \le 1\\
0.954+0.039\log q_2 + 0.114(\log q_2)^2, & 1\le q_2 \lesssim 20.
\end{cases}
\end{eqnarray}
Outside the ranges of validity $F_1(q_2)$  and $\gamma(q_2)$
are evaluated using the value of $q_2$ at the
edge of their respective ranges.

\paragraph{Kolb scheme:}
\citet{b:kolb1990} extended the Ritter scheme
in order to cover the case $R_1>R_{\mathrm{RL},1}$ according to
\begin{eqnarray}
\dot{M}_{\mathrm{RLOF}}=-\dot{M}_0-2\pi F_1(q_2)
\frac{R_{\mathrm{RL},1}^3}{GM_1}\qquad\qquad\qquad\qquad\nonumber\\
\times\int_{P_{\mathrm{ph}}}^{P_{\mathrm{RL}}} \Gamma_1^{1/2}\left(\frac{2}{\Gamma_1+1}\right)^{(\Gamma_1+1)/(2\Gamma_1-2)}\left(\frac{k_{\rm B}T}{m_{\rm p} \mu}\right)^{1/2} dP
\end{eqnarray}
where $\Gamma_1$ is the first adiabatic exponent, and $P_\mathrm{ph}$
and $P_\mathrm{RL}$ are respectively the pressures at the photosphere
and at the radius for which $r_1=R_\mathrm{RL,1}$.

\subsubsection{Implicit Methods}\label{binaries:implicit}

Explicit schemes exhibit large jumps in $\Mdot_{\mathrm{RLOF}}$ unless
the timestep is severely restricted.  Therefore, if one needs accurate
values of $\Mdot_{\mathrm{RLOF}}$ and stellar radius, this requires
use of an implicit scheme.  Implicit schemes also allow the
calculation these quantities when there is no general closed form
formula for $\Mdot_{\mathrm{RLOF}}$.

These implicit methods use a bisection-based root solve to satisfy
$|f(\Mdot_{\mathrm{RLOF}})| < \xi$ at the end of the step, where $\xi$
is a given tolerance.  The implicit schemes are then defined by the
choice of the function $f(\Mdot_{\mathrm{RLOF}})$. For the Ritter and
the Kolb scheme the function is chosen as
\begin{equation}
f(\Mdot_{\mathrm{RLOF}})=\frac{\Mdot_{\mathrm{end}}-\Mdot_{\mathrm{RLOF}}}{\Mdot_{\mathrm{end}}},
\end{equation}
with $\Mdot_{\mathrm{end}}$ being the mass transfer rate
computed at the end of each iteration.

A different implicit method is also provided. In this case, whenever
the donor star overflows its Roche lobe the implicit solver will
adjust the mass transfer rate until $R_1=R_{\mathrm{RL,1}}$ within
some tolerance \citep[see e.g.,][]{b:whyt1980,b:rapp1982,b:rapp1983}.
In this case
\begin{eqnarray}
f(\Mdot_{\mathrm{RLOF}})=\frac{2(R_1-R_{\mathrm{RL},1})}{R_{\mathrm{RL},1}}+\xi,
\end{eqnarray}
and if $\Mdot_{\mathrm{RLOF}}$ is below a certain threshold and
$f(\Mdot_{\mathrm{RLOF}})<-\xi$ then the system is assumed to detach
and $\Mdot_{\mathrm{RLOF}}$ is set to zero.

\subsection{Effect of Tides and Accretion on Stellar Spin}\label{binaries:tiderot}

To model tidal interaction we adjusted the model of \citet{b:hut1981}
to include the case of differentially rotating stars.  The time
evolution of the angular frequency for each component
is
\begin{eqnarray}
\frac{d\Omega_{i,j}}{d t} =  \frac{\Omega_{\mathrm{orb}}-\Omega_{i,j}}{\tau_{\mathrm{sync},j}},\quad \frac{1}{\tau_{\mathrm{sync},j}}=\frac{3}{(q_j r_{\mathrm{g},j})^2} \left(\frac{k}{T}\right)_{\mathrm{c},j}\left(\frac{R_j}{a}\right)^6,\label{binaries:equtide}
\end{eqnarray}
where $j=1,2$ is the index of each star, $\Omega_{i,j}$ is the angular
frequency at the face of cell $i$ towards the surface,
$r_{\mathrm{g},j}^2=I_j/(M_jR_j^2)$ is the radius of gyration (with
$I_j$ being the moment of inertia of each star), and the ratio of the
apsidal motion constant to the viscous dissipation timescale,
$(k/T)_{\mathrm{c},j}$, is computed as in \citet{b:hurl2002}.
Similarly to \citet{b:detm2008}, we assume constant
$\tau_{\mathrm{sync},j}$ and $\Omega_\mathrm{orb}$ through a step and
therefore
$\Delta \Omega_{i,j}=[1 - \exp(-\delta
t/\tau_{\mathrm{sync},j})](\Omega_{\mathrm{orb}}-\Omega_{i,j})$.
This extension of Hut's work to differentially rotating stars is not
formally derived but merely applies his result for solid body rotators
independently to each shell. The formulation of \citet{b:hut1981} can
be recovered from Equation~\eqref{binaries:equtide}, by forcing solid
body rotation with a large diffusion coefficient for angular momentum
throughout the star. In reality tides would act mostly on the outer
layers, and whether the core synchronizes or not depends on the
coupling between the core and the envelope.

To compute the specific angular momentum carried by accreted material,
we consider the possibility of both ballistic and Keplerian disk mass
transfer \citep[e.g.,][]{Marsh04,b:demi2013}. To distinguish which
occurs, we compare the minimum distance of approach of the accretion
stream \citep{b:lubo1975, b:ulri1976}\footnote{Note that there is a
  small typo in the fit given by \citet{b:ulri1976}. The corrected fit
  given here fits the results of \citet{b:lubo1975} to the 4\%
  accuracy claimed by \citet{b:ulri1976}.}
\begin{equation}
R_{\mathrm{min}} = 0.0425 a \left(q_2+q_2^2\right)^{1/4},\qquad 0.0667 \le q_2 \le 15
\end{equation}
to the radius of the accreting star.  When outside the range of
validity, $R_{\mathrm{min}}$ is computed using the value of $q_2$ at
the respective edge.  Accretion is assumed to be ballistic whenever
$R_2>R_{\mathrm{min}}$ and the specific angular momentum is
$(1.7GM_2R_{\mathrm{min}})^{1/2}$.  When $R_2<R_{\mathrm{min}}$ the
specific angular momentum is taken as that of a Keplerian orbit at the
surface $(GM_2R_2)^{1/2}$.

\subsection{Treatment of Thermohaline Mixing in Accreting Models}\label{binaries:thermohaline}

In stars with radiative envelopes accreted material with a high mean
molecular weight is expected to mix inwards due to thermohaline
mixing, a process that is very sensitive to the $\mu$-gradient
\citep[see e.g.,][]{b:kipp1970,Cantiello:2010}.  Thermohaline mixing
is included in \MESA\ (see \mesaone).  However, as mass with
homogeneous composition is added during the accretion process, a jump
is produced at the boundary between new and old material.  {\MESAstar}
computes mixing coefficients explicitly at the start of each step, so
this results in thermohaline mixing only operating near this boundary,
leading to unphysical compositional staircases.  To avoid this issue,
we artificially soften the composition gradient in the outer
$(\Delta q)_{\mathrm{large}}$ fraction of the star by mass.  We do
this starting at the surface and homogeneously mixing inwards a region
of size $(\Delta q)_\mathrm{small}$. Then, moving towards the center,
the process is repeated at each cell while linearly (with respect to
mass) reducing the size of the small mixed region such that it is zero
after going $(\Delta q)_{\mathrm{large}}$ inwards.  All the binary
models where the accretor is not a point mass are calculated using
$(\Delta q)_{\mathrm{large}}=0.05$ and
$(\Delta q)_{\mathrm{small}}=0.03$.

\subsection{Numerical Tests}\label{binaries:tests}
Here we describe tests designed to validate the implementation of the
physics described in Section \ref{binaries:evolution}.  We check
orbital evolution in the presence of gravitational waves and mass loss
by comparing to analytical solutions. We also verify total angular
momentum conservation in calculations that include the physics of
tides and spinup by accretion.  To test for the thermal response of
stellar models undergoing mass transfer, we compare {\mesabinary}
results to those from the \code{STARS} code
\citep{b:eggl1971,b:pols1995,b:stan2009}.



\subsubsection{Gravitational Wave Radiation}\label{binary:check_jdot_gr}

If gravitational waves are the only source of angular momentum loss
and the masses of each component remain constant, Equation
(\ref{binaries:Jorb_dot}) can be integrated to obtain the time
evolution of orbital separation \citep{b:pete1964}.  We model a system
consisting of a $0.5\,\mathrm{M_\odot}$ star and a
$0.8\,\mathrm{M_\odot}$ point mass with $a = 2\,\mathrm{R_\odot}$. We
ignore all effects on the evolution of orbital angular momentum except
its loss due to gravitational waves.  In $3.5$ Gyr the orbital
separation of this system reduces to $a = 1.3\,\mathrm{R_\odot}$, at
which point the $0.5\,\mathrm{M_\odot}$ star begins mass transfer. We
terminate the run at the onset of RLOF. The maximum error in $a$ is
$0.35\%$ relative to the analytical result.

\subsubsection{Inefficient Mass Transfer}\label{binary:ineff_mt}

\begin{figure}[]
	\begin{center}
		\includegraphics[width=\columnwidth]{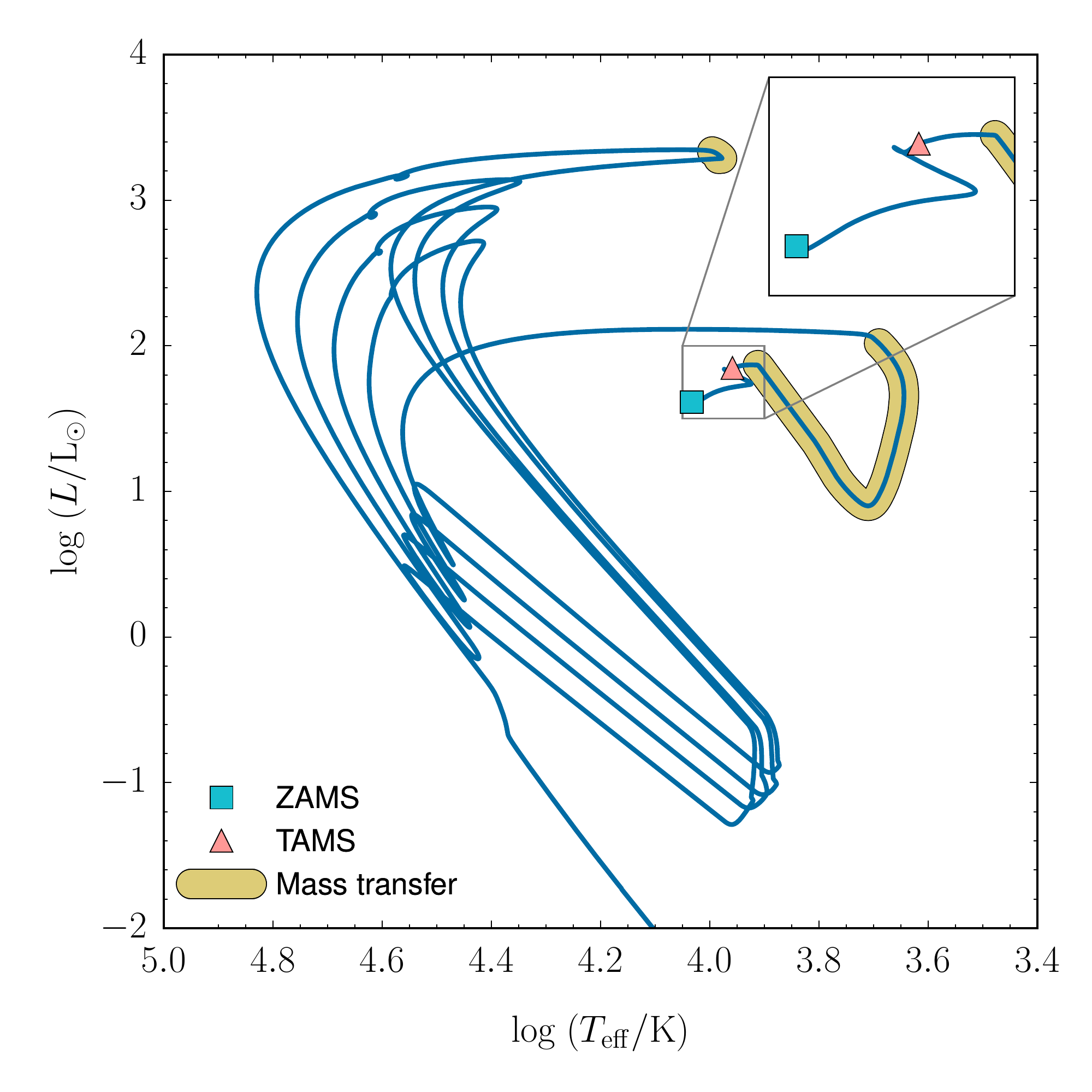}
		\caption{ Evolution in the HR diagram for a
          $2.5\,\mathrm{M_\odot}$ star transferring mass to a
          $1.4\,\mathrm{M_\odot}$ point mass, assuming a mass transfer efficiency of $1\%$. Symbols are shown at zero-age
          main sequence (ZAMS) and terminal-age main sequence (TAMS),
          together with parts of the track where RLOF is
          occurring. The inset shows evolution from ZAMS up to the
          beginning of the first phase of mass transfer.
          \label{binaries:hr_imxb}}
	\end{center}
\end{figure}

An analytical expression for the evolution of orbital separation can
be derived if inefficient mass transfer is the only contribution to
the angular momentum evolution \citep{b:taur2006}.  We model a
$2.5\,\mathrm{M_\odot}$ main sequence star together with a
$1.4\,\mathrm{M_\odot}$ point mass with an initial orbital separation
of $10\,\mathrm{R_\odot}$.  We choose $\alpha_{\mathrm{mt}}=0.03$,
$\beta_{\mathrm{mt}}=0.95$, $\delta_{\mathrm{mt}}=0.01$ and
$\gamma_{\mathrm{mt}}^2=2$, which give a low mass transfer efficiency
of $f_{\mathrm{mt}}=0.01$.  Such a system is representative of the
evolution of an intermediate mass X-ray binary (IMXB). The model
initiates mass transfer just after the end of the main sequence,
interrupting the evolution of the star through the Hertzsprung gap and
producing a low mass white dwarf (WD) ($M_{\mathrm{He}}=0.289\,\Msun$)
with a small amount of hydrogen on its surface. As the WD evolves to
the cooling track, it experiences several hydrogen flashes, one of
them strong enough to produce an additional phase of RLOF (see Figure
\ref{binaries:hr_imxb}).

\begin{figure}[]
\begin{center}
\includegraphics[width=\columnwidth]{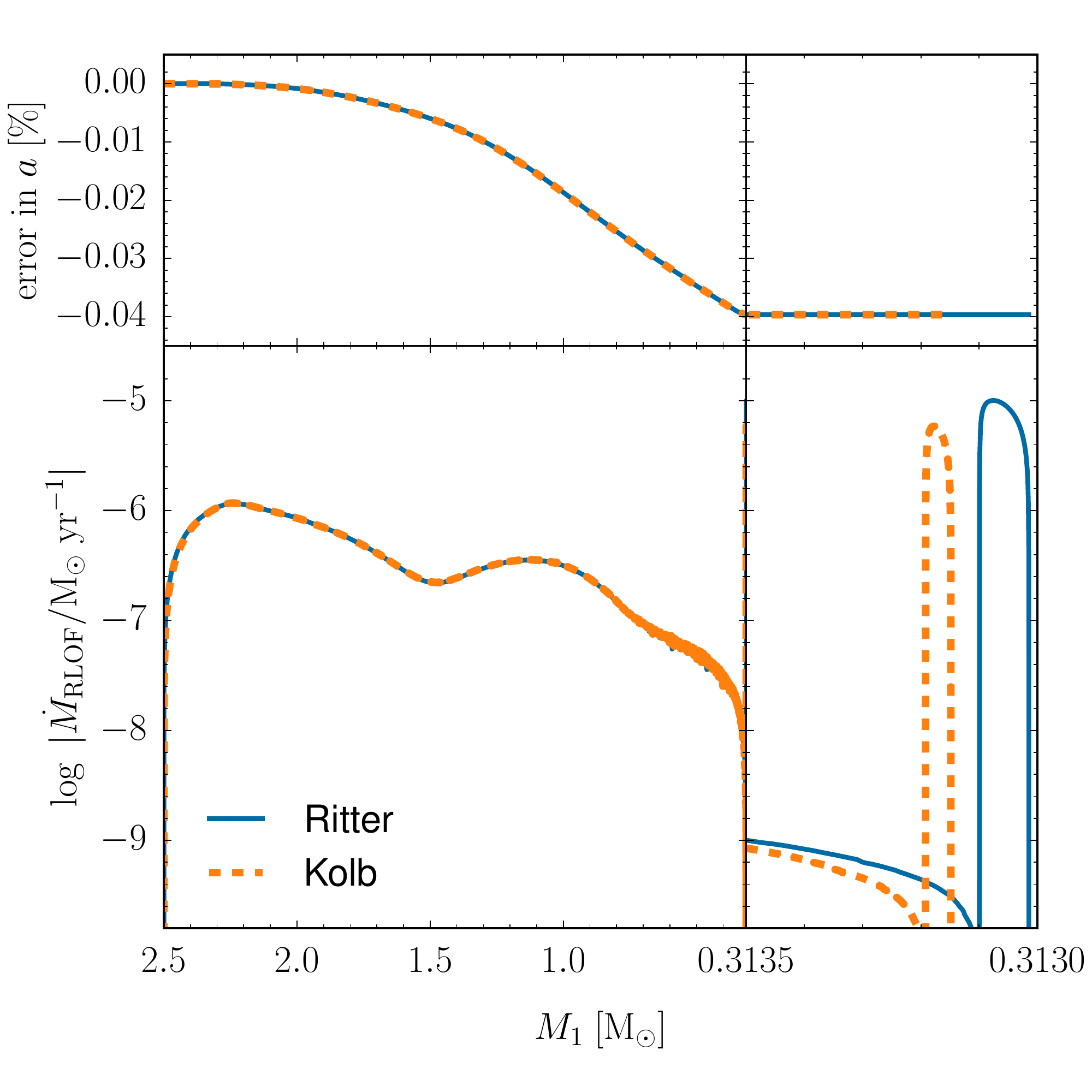}
\caption{
Evolution of mass transfer rate from a $2.5\,\mathrm{M_\odot}$ to a
$1.4\,\mathrm{M_\odot}$ point mass, assuming a mass transfer efficiency of $1\%$. The
upper panel shows the difference between the computed orbital
separation and the analytical solution while the bottom one displays
the evolution of the mass transfer rate, using two different schemes.
\label{binaries:mdot_imxb}}
\end{center}
\end{figure}

Figure \ref{binaries:mdot_imxb} shows that {\mesabinary} computes the
orbital evolution to a precision of a few parts in $10^{4}$. We run
this system using both the Ritter and the Kolb implicit schemes to
display that under some circumstances the precise choice of mass
transfer scheme does not play a big role in the evolution.

\subsubsection{Spin Orbit Coupling}
We now test angular momentum conservation by ignoring all the
mechanisms that remove angular momentum from the binary system.  For
this purpose we model an $8\,\mathrm{M_\odot}+6\,\mathrm{M_\odot}$
binary with rotating components and an initial orbital period of $1.5$
days. Due to the short orbital separation we assume the initial spin
periods of the two stars are equal to the orbital period.  The primary
undergoes RLOF during the main sequence, initiating a phase of mass
transfer on a thermal timescale. After transferring just
$0.3\,\mathrm{M_\odot}$ the accretor also fills its Roche lobe,
producing a contact system.  At this point we terminate the evolution.

\begin{figure}[]
	\begin{center}
		\includegraphics[width=\columnwidth]{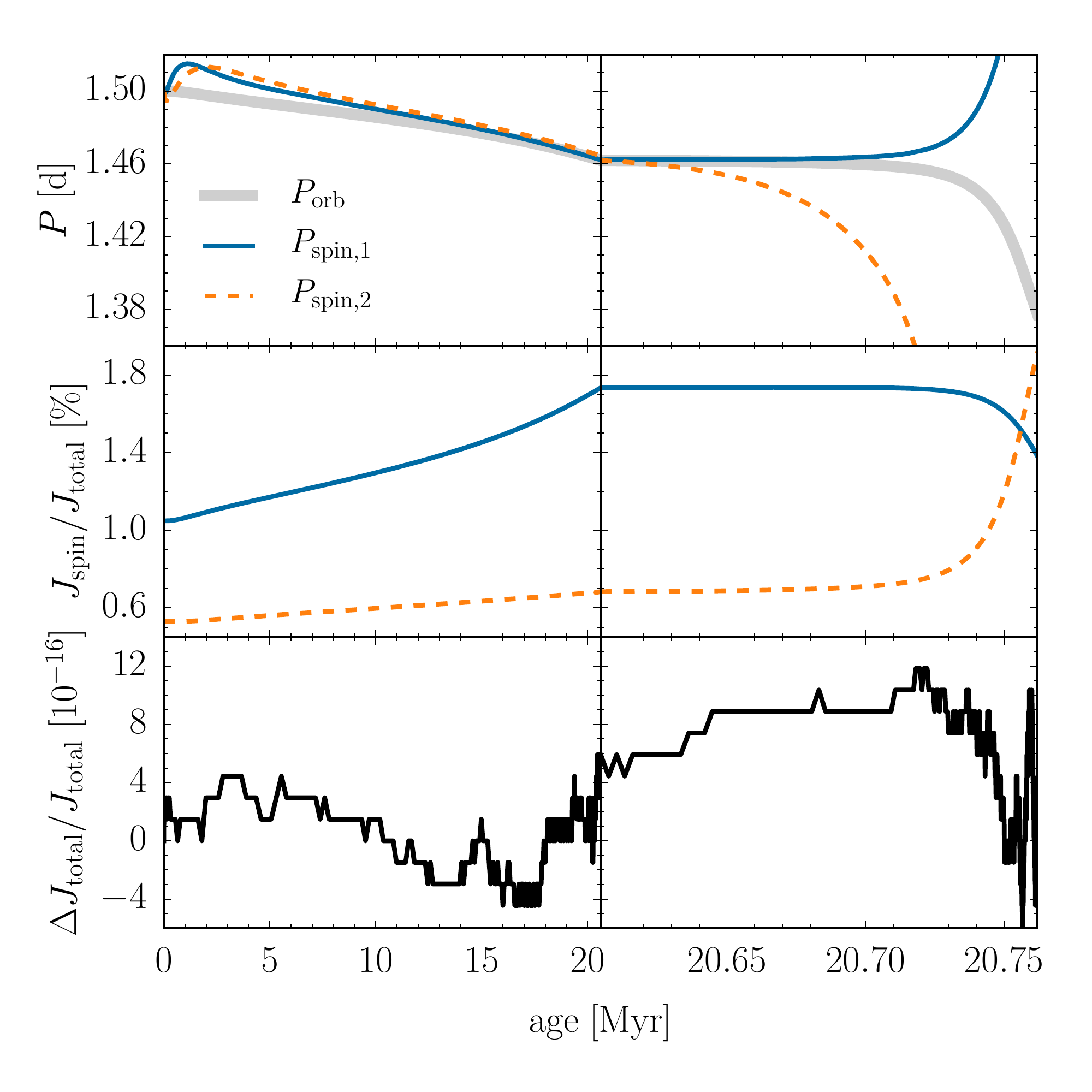}
		\caption{
			Angular momentum evolution in an $8\,\mathrm{M_\odot}+6\,\mathrm{M_\odot}$
			binary with an initial orbital period of $1.5$ days. Left panels show the evolution
			before the onset of RLOF, while right panels display
			evolution from the beginning of RLOF until contact, when both components fill their Roche lobe.
            The fractional error in the total angular momentum is plotted in the bottom panel and is of order machine-precision.
			\label{binary:jdot_ls_plot}}
	\end{center}
\end{figure}

Figure \ref{binary:jdot_ls_plot} shows that spin angular momentum in
both components increases during the pre-interaction phase, which is
due to both stars expanding on the main sequence while remaining
tidally locked. During Roche lobe overflow, the secondary is rapidly
spun-up, reaching nearly $80\%$ of critical rotation before contact.
The calculation of total angular momentum requires the summation of
different contributions (orbital angular momentum and spin of both
components).  Therefore the maximum accuracy to which we can conserve
angular momentum is limited by rounding errors.  Figure
\ref{binary:jdot_ls_plot} shows that conservation of angular momentum
in the run is very close to machine precision.

\subsubsection{Thermal Response to Mass Loss}

The fate of binary systems depends largely on the precise value of
$\dot{M}$ during an interaction phase, which depends on the thermal
response of the donor star to mass loss.  For WDs there is a limited
range of accretion rates for stable hydrogen burning
(\citealt{b:nomo2007,b:shen2007}).  In main sequence binaries the
evolution of the accretor radius depends on the mass transfer rate,
and expansion during the interaction phase can lead to contact or even
a merger (\citealt{b:well2001}).

\begin{figure}[]
\begin{center}
\includegraphics[width=\columnwidth]{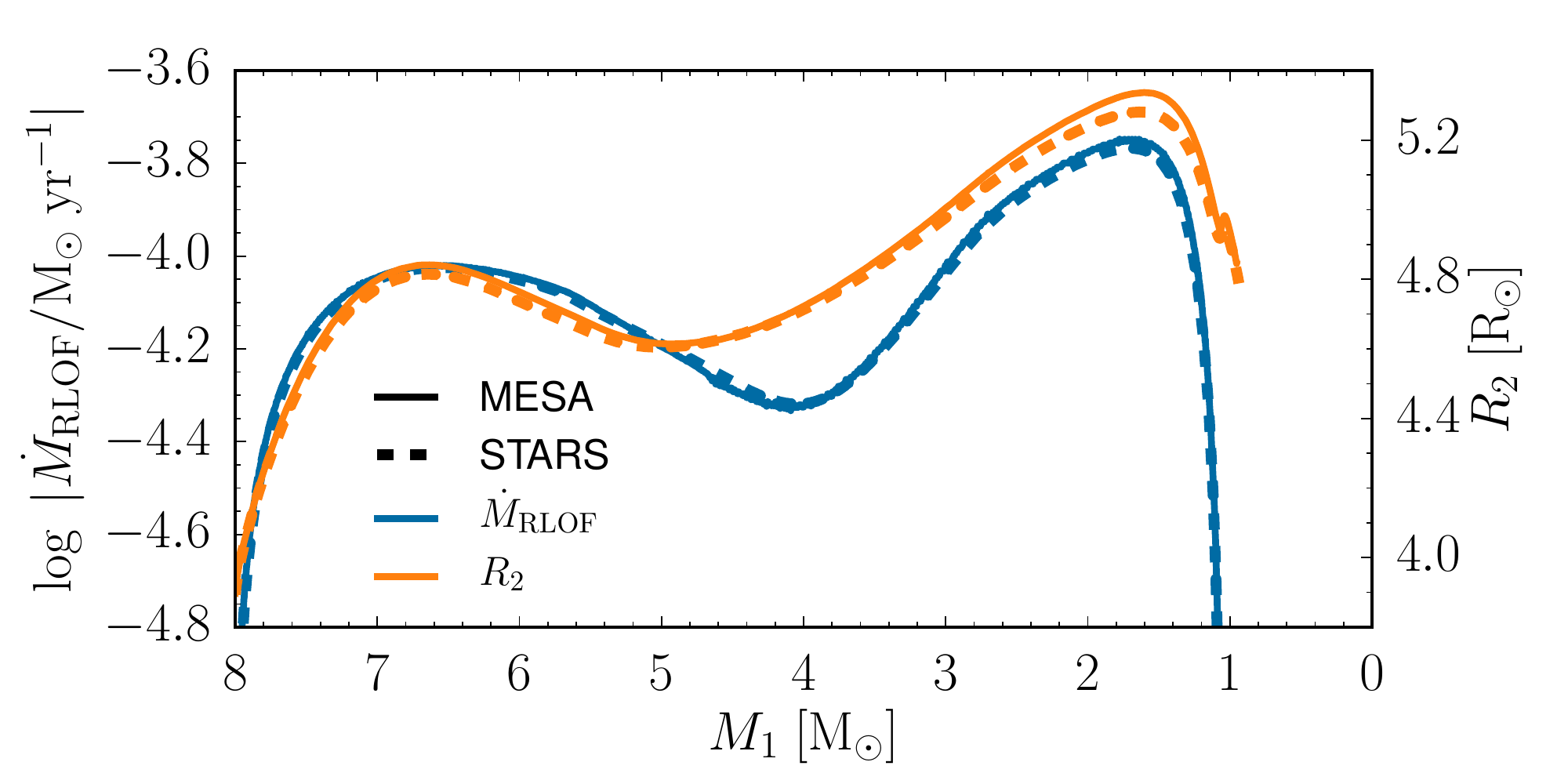}
\caption{
Mass transfer rate and accretor radius as computed by \code{MESA} and
\code{STARS} for an $8\,\mathrm{M_\odot}+6.5\,\mathrm{M_\odot}$ binary with
an initial orbital period of $3$ days. All internal mixing processes (including convective mixing) are turned off in the calculations.
\label{binary:mdot_comp}}
\end{center}
\end{figure}

We calculated an $8\,\mathrm{M_\odot}+6.5\,\mathrm{M_\odot}$ binary
system with an initial orbital period of $1.5$ days using both
{\mesabinary} and \code{STARS}.  To minimize the modeling differences
and focus on the thermal response of both components, we use an
extremely simplified model that ignores internal mixing (including
convective mixing). Under these conditions, the more massive star
quickly depletes its central hydrogen and begins shell hydrogen
burning, reaching RLOF and undergoing a phase of mass transfer on the
thermal timescale. The resulting mass transfer rates are shown in
Figure \ref{binary:mdot_comp}. The agreement is very good, despite
mass transfer rates being computed in slightly different ways.  Masses
at detachment show a small difference, with the {\mesabinary} model
terminating mass transfer when $M_1=0.952\,\mathrm{M_\odot}$ while the
\code{STARS} calculation when $M_1=0.935\,\mathrm{M_\odot}$.  The
figure also shows the change in radius of the accreting star, with two
prominent peaks at $R_2/\mathrm{R_\odot} = 4.84,5.34$ for
{\mesabinary} and $4.82,5.28$ for \code{STARS}. The larger radius of
the {\mesabinary} model is likely associated to the slightly higher
mass transfer rates.

\subsection{Period Gap of Cataclysmic Variables}\label{binaries:cvs}

Although cataclysmic variables (CVs) span a wide range of periods,
observations show a lack of systems in the range
$2\,\mathrm{h} <\porb < 3\,\mathrm{h}$ \citep[see, for
instance,][]{b:gans2009}.  Such a feature is commonly explained by
having an angular momentum loss mechanism ``turn off'' or become
inefficient at some point. The most popular model for such a mechanism
is magnetic braking \citep{b:rapp1983}, as the magnetic field of the
donor is assumed to change quickly when the star loses enough mass to
become fully convective.

\begin{figure}[]
\begin{center}
\includegraphics[width=\columnwidth]{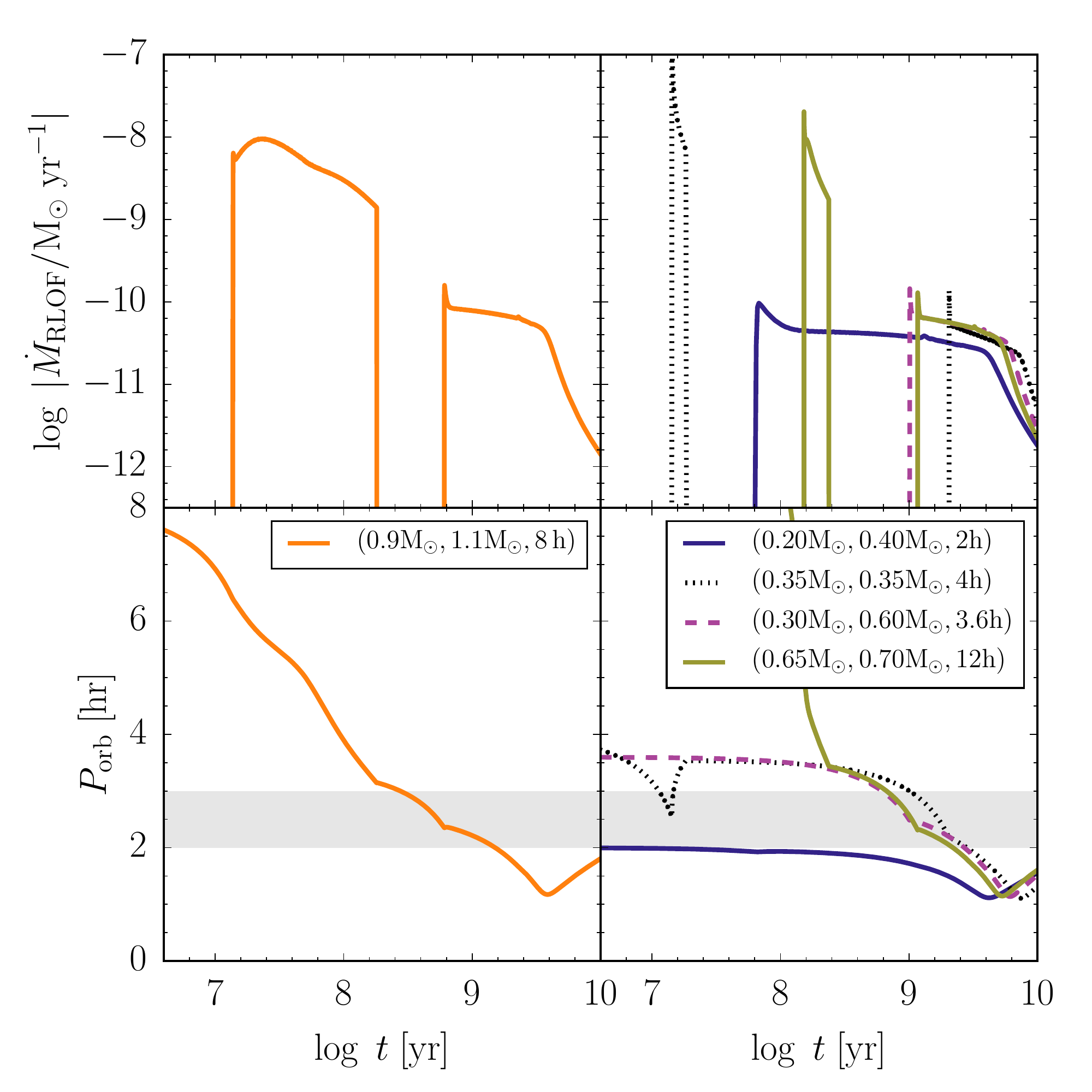}
\caption{
Evolution of CV models under the effect of magnetic braking and
gravitational wave radiation. For each track the label gives
the donor mass, the WD mass, and the initial orbital period respectively.
The grey band shows the observed period gap for CVs.
These results reproduce figure 1 in \citet{b:howe2001}. 
\label{binaries:period_gap_plot}}
\end{center}
\end{figure}

We now compare to the results of \citet{b:howe2001}, who performed a
population synthesis study to explore in detail the standard scenario
involving magnetic braking.  In Figure \ref{binaries:period_gap_plot}
we show the evolution of mass transfer rates and orbital periods for a
set of CV models with different component masses and orbital periods.
We run all models using $\beta_{\mathrm{mt}}=1$ and
$\gamma_\mathrm{mb}=3$ and magnetic braking is turned off when the
donor star becomes fully convective. As an example the system with a
$0.9\,\mathrm{M_\odot}$ donor (left panel in Figure
\ref{binaries:period_gap_plot}) experiences a first phase of mass
transfer induced by magnetic braking between $10^{7.1}$ and
$10^{8.3}\,\mathrm{yr}$, a non-interacting phase (the gap) between
$10^{8.3}$ and $10^{8.8}\,\mathrm{yr}$, and a subsequent phase of mass
transfer dominated by gravitational wave radiation, reaching a minimum
orbital period of about 1 hour at $10^{9.6}\,\mathrm{yr}$. As a
comparison, for the same model \citet{b:howe2001} obtain a first phase
of mass transfer between $10^{7.3}$ and $10^{8.4}\,\mathrm{yr}$, the
gap occurs between $10^{8.4}$ and $10^{8.8}\,\mathrm{yr}$ and a period
minimum is reached at $10^{9.4}\,\mathrm{yr}$.  Figure
\ref{binaries:period_gap_plot} shows that our CV models spend most
time away from the observed period gap.

\subsection{Evolution of Massive Binaries}\label{binaries:massbin}

In massive stars, binary interactions have dramatic effects on the
evolution of both components.  \citet{b:kipp1967} introduced the term
``case A'' to refer to a mass transfer phase occurring in systems
tight enough such that RLOF starts during the main sequence.  This
results in a large amount of mass being transferred on a thermal
timescale, followed by a phase of mass transfer that proceeds on the
nuclear timescale until the end of core H-burning. An additional phase
of thermal-timescale mass transfer then follows (the so-called ``case
AB''), which strips the donor and produces an almost-naked helium
star.

\begin{figure}[]
\begin{center}
\includegraphics[width=\columnwidth]{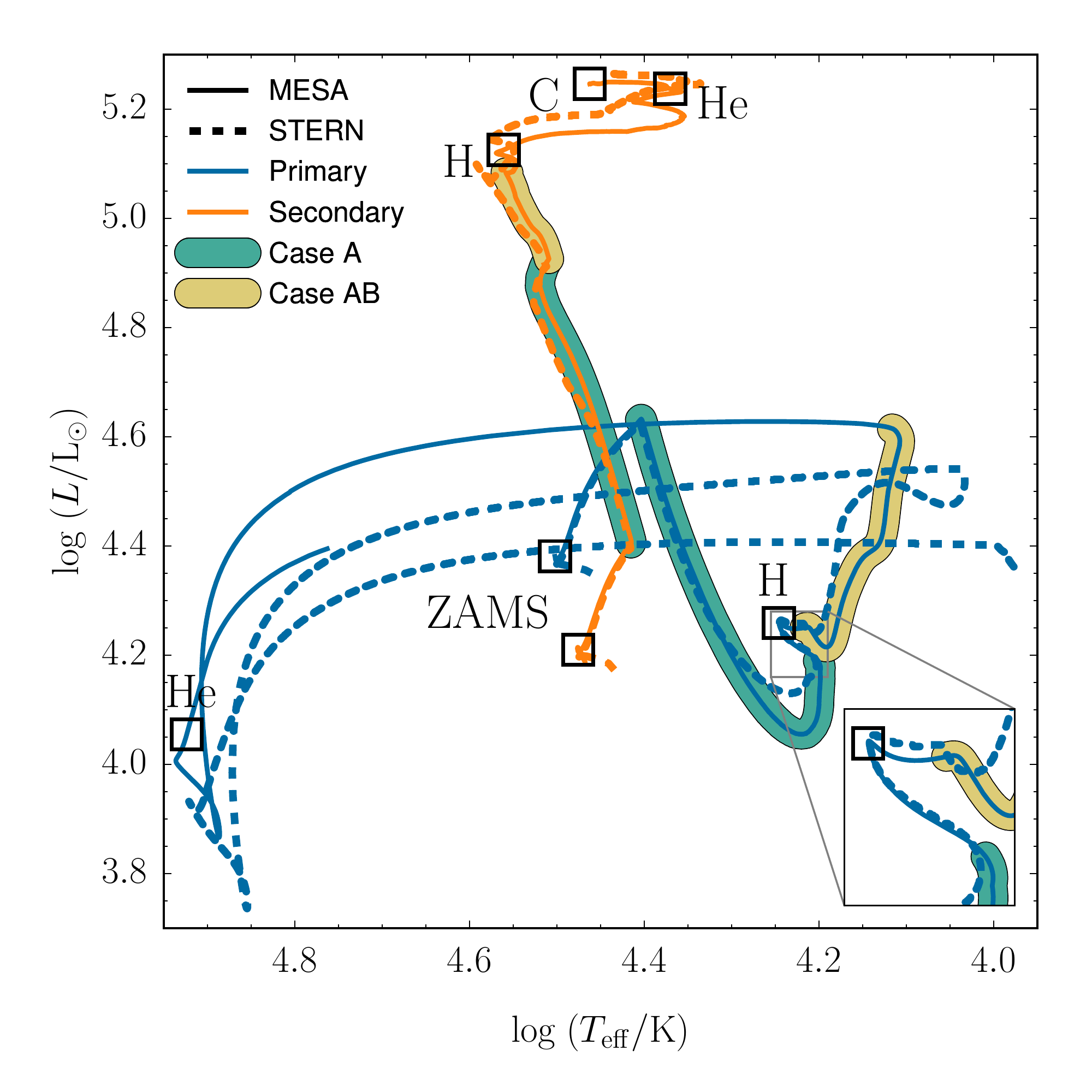}
\caption{
Evolution of a $16\,\mathrm{M_\odot}+14\,\mathrm{M_\odot}$ system with a
$3$ day initial orbital period. {\mesabinary} models are compared to the results
of \citet{b:well2001}, which were calculated using the \code{STERN} code. The terms primary
and secondary are used throughout the evolution to describe
 the initially more massive and the less massive
components, respectively. For each component in the {\mesabinary} model, squares mark
the ZAMS and the depletion of the indicated nuclear fuel in the core.
\label{binary:massive_HR}}
\end{center}
\end{figure}

\begin{figure}[]
\begin{center}
\includegraphics[width=\columnwidth]{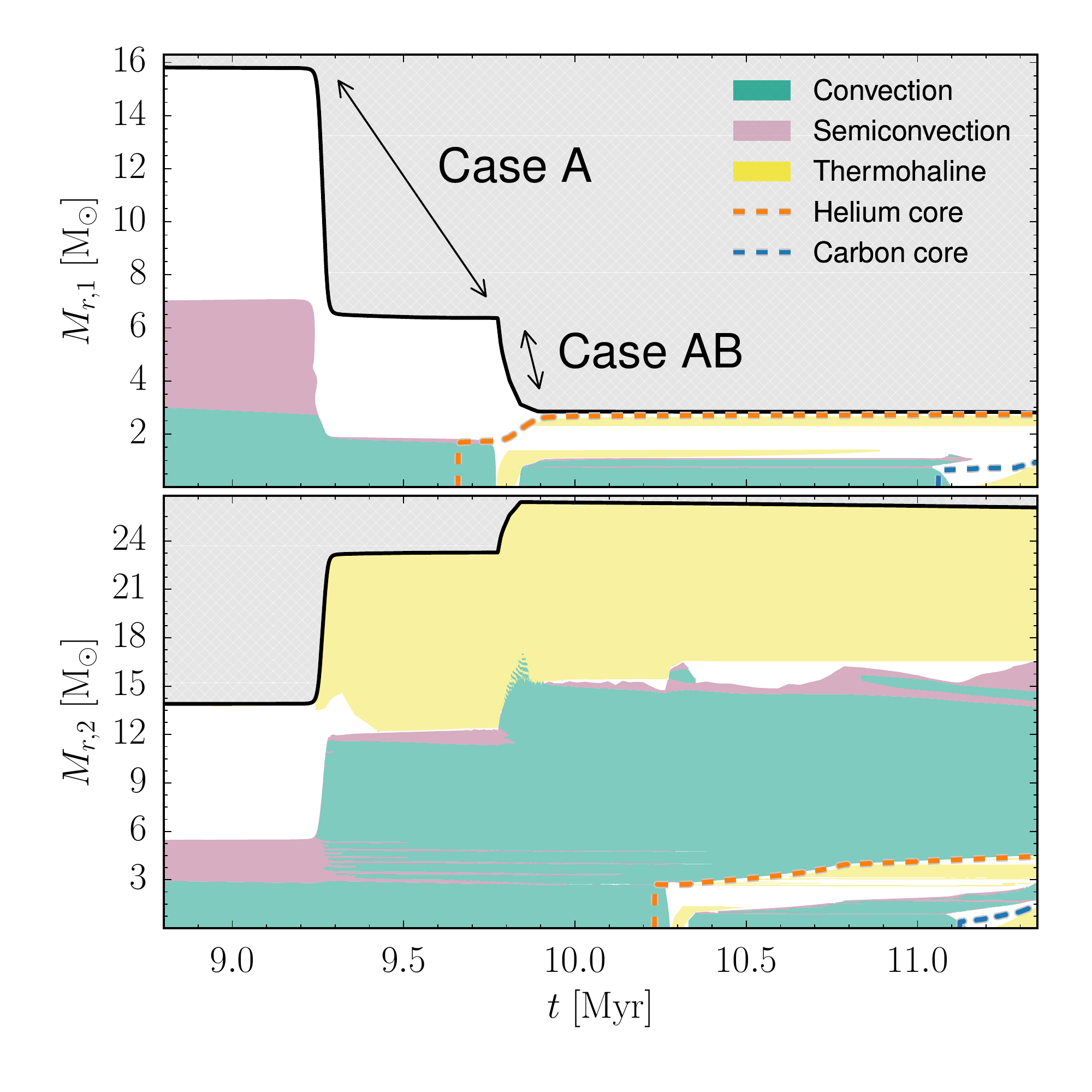}
\caption{
Kippenhahn diagram for the evolution of a
$16\,\mathrm{M_\odot}+14\,\mathrm{M_\odot}$ system with a $3$ day initial
orbital period. Most of the pre-interaction phase is not shown in this
figure. The upper plot shows the evolution of the donor, while the
lower plot displays that of the accretor.
\label{binary:massive_kipp}}
\end{center}
\end{figure}

Here we show that \mesabinary\ can calculate the evolution of massive
interacting binaries.  We reproduce one of the models from
\citet{b:well2001}, a $16\,\mathrm{M_\odot}+14\,\mathrm{M_\odot}$
system with an initial period of $3$ days, using the same
semiconvection efficiency of $\alpha_{\mathrm{sc}}=0.01$.  As shown in
Figures \ref{binary:massive_HR} and \ref{binary:massive_kipp} this
system experiences case A and AB mass transfer, and the accretor
becomes a blue supergiant after core hydrogen depletion.  The accretor
depletes carbon before its donor.

Figure \ref{binary:massive_kipp} illustrates the prevalence of both
thermohaline mixing and semiconvection in the accreting star. Newly
accreted material is efficiently mixed inwards by thermohaline mixing.
On the other hand the $\mu$-gradient formed before interaction
prevents the convective core from growing, with the efficiency of
semiconvection controlling whether or not the star rejuvenates. Due to
the choice of inefficient semiconvection, the core remains small,
preventing the star from becoming a red supergiant.  The star accretes
a large amount of CNO-processed and helium-rich material. After being
mixed through the envelope this material results in the surface being
nitrogen rich and carbon depleted, with a slight enhancement in
helium.

\subsection{Rotating Binaries and the Efficiency of Mass Transfer}\label{binaries:binrot}

The efficiency of mass transfer plays a key role in close binary
systems, but the processes by which material is lost from the system
are not well-understood.  In particular, whenever an accreting star
approaches $\Omega/\Omega_{\mathrm{crit}} = 1$, it is uncertain
whether accretion can continue, one option being the development of a
strong wind that prevents accretion
\citep[e.g.,][]{b:petr2005,b:cant2007}.  Whenever
$\Omega/\Omega_{\mathrm{crit}}$ approaches one, we use an implicit
method to iteratively reduce $\dot{M}_2$ until this ratio falls below
a threshold.

Tides counteract the effect of spin-up from accretion. Whether or not
an accreting object reaches critical rotation depends on the
efficiency of tidal coupling. Here we model a
$16\,\mathrm{M_\odot}+15\,\mathrm{M_\odot}$ binary system including
differential stellar rotation, with an initial orbital period of 3 days and
assuming initial orbital synchronization.  \citet{b:lang2003} argue
that turbulent processes in the radiative envelope can significantly
enhance tidal strength.  They  model the same system using the simple
estimate for the synchronization timescale for a star with a
convective envelope given by \citet{b:zahn1977},
$\tau_{\mathrm{sync},j}=1\,\mathrm{yr}\times q_j^2(a/R)^6$. For our
implicit modeling of stellar winds we use a threshold of
$(\Omega/\Omega_{\mathrm{crit}})=0.99$.

\begin{figure}[]
\begin{center}
\includegraphics[width=\columnwidth]{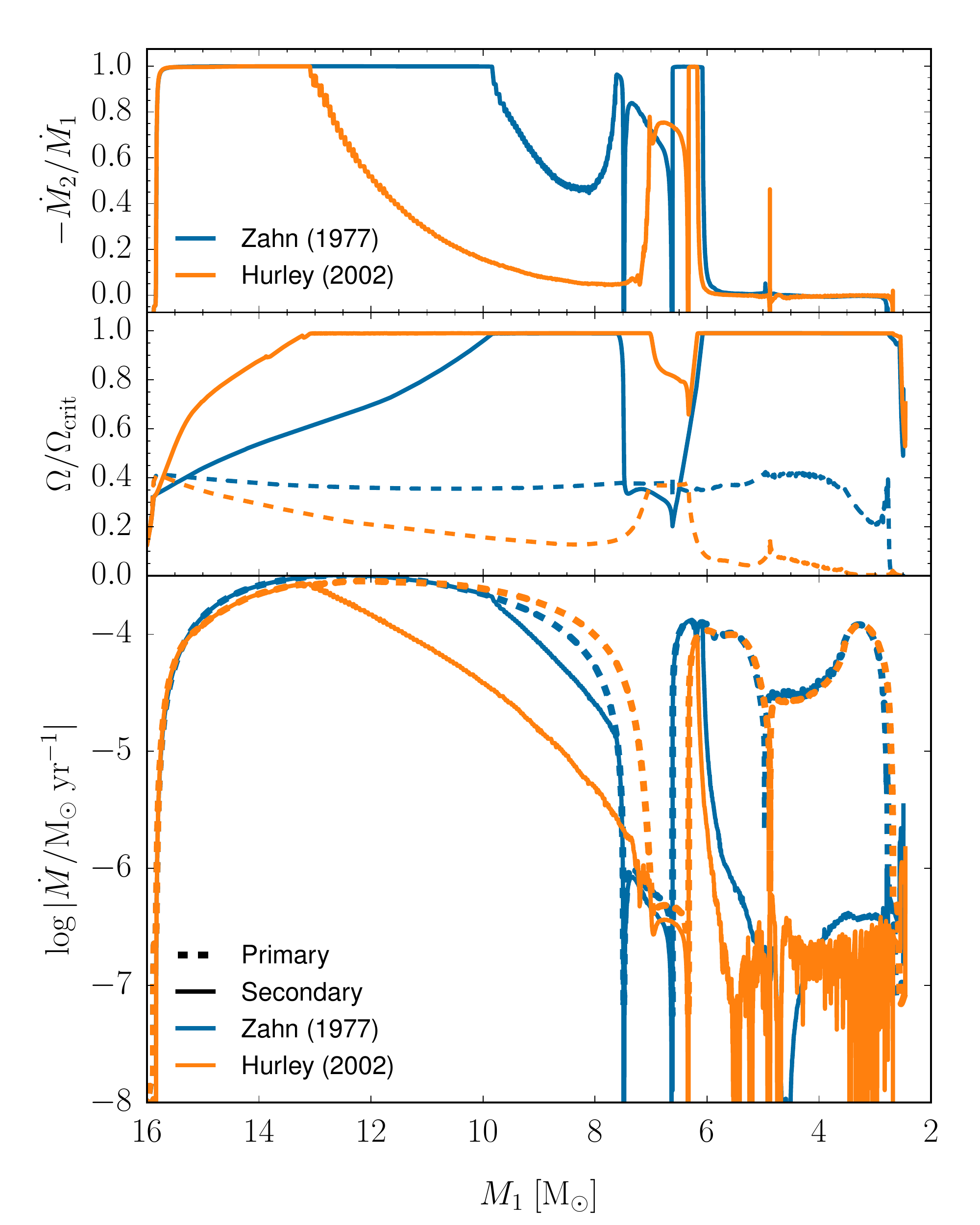}
\caption{ Efficiency of mass transfer in a
  $16\,\mathrm{M_\odot}+15\,\mathrm{M_\odot}$ binary system including
  differential rotation. The system is modeled with tides as described
  by \citet{b:hurl2002} for radiative envelopes, and also with the
  simple tidal timescale given by \citet{b:zahn1977}. The upper panel
  shows the efficiency of mass transfer, the middle panel the angular
  frequency of each star in terms of its critical value, while the
  lower panel shows the evolution of $\Mdot$ for both components.
\label{binary:massive_rot}}
\end{center}
\end{figure}

Figure \ref{binary:massive_rot} shows that \mesabinary\ models using
both the \citet{b:zahn1977} and \citet{b:hurl2002} timescales for
tidal coupling.  These models experience highly non-conservative
phases of mass transfer, corresponding to the accreting star evolving
very close to critical rotation. In particular during case AB mass
transfer the accretor needs to switch from mass accretion to mass loss
in order to remain sub-critical. As expected, the system with the
tidal timescale from \citet{b:zahn1977} has a significantly higher
mass transfer efficiency, and during the first phase of RLOF it only
experiences a brief period in which the accretor reaches critical
rotation. This is in broad agreement with the model by
\citet{b:lang2003}.

\subsection{Description of a Binary Run}\label{binaries:onestep}

{\mesabinary} performs each evolution step by independently solving
the structure of each component and the orbital parameters, using the
same timestep $\delta t$ for each. This approach differs from
\code{STARS}, which simultaneously solves for the structure of both
stars and the orbit in a single Newton-Raphson solver. Our choice to
solve for each star separately gives a significant amount of
flexibility and simplicity, as the examples in this paper demonstrate.

The top-level algorithm for evolving a star is described in appendix
B1 of \mesatwo.  We modified this algorithm to support the new
implementation of binary interactions, which is described in detail in
the \MESA\ documentation. Additional timestep limits are imposed in
{\mesabinary} that consider relative changes between the radius and
Roche lobe radius of both components, the total orbital angular
momentum, the orbital separation, and the envelope mass in the donor.


\section{Pulsations} \label{s.pulse}

The study of stellar pulsations (also termed oscillations) offers
unique insights into the interiors of stars \citep{Aerts:2010aa}. In
some classes of star (e.g., solar-type, red giant), the stochastic
excitation of hundreds of oscillation modes, typically by convective
motions, allows remarkably detailed measurements to be made of the
interior, including nuclear burning state \citep{Bedding:2011aa} and
internal rotation \citep{Beck:2012aa}.  In other classes (e.g.,
classical Cepheid, $\beta$ Cephei, $\delta$ Scuti and $\gamma$ Doradus
pulsators), modes are instead excited by linear instabilities, most
often linked to opacity variations in the envelope (the $\kappa$
mechanism). In these latter objects, typically too few modes are
excited for detailed asteroseismic analysis to be feasible;
nevertheless, mapping out the regions of the theoretical HR diagram
where the instabilities are expected to operate, and then comparing
these instability strips against observational surveys, can often lead
to new science.

\mesatwo\ introduced the \code{astero} extension to \MESAstar, which
permits on-the-fly refinement of stellar model parameters in order to
fit a set of observed oscillation frequencies and spectroscopic
constraints.  Subsequent improvements to the \code{astero}
capabilities include frequency correction recipes from
\citet{Ball:2014aa}; implementation of the downhill simplex
\citep{Nelder:1965aa} and NEWYUO \citep{Powell:2007aa} algorithms for
$\chi^{2}$ minimization; parameter optimization using only
spectroscopic constraints (e.g., \Teff\ and surface gravity); and
coupling to the \GYRE\ oscillation code, as an alternative to the
\code{ADIPLS} code \citep{Christensen-Dalsgaard:2008aa} used in the
original implementation.

\GYRE\ calculates the normal-mode eigenfrequencies \freq\ of a stellar
model by solving the system of linearized equations and boundary
conditions governing small periodic perturbations ($\propto \exp[{\rm i}\freq t]$)
to the equilibrium state. It is based on a novel Magnus Multiple
Shooting (MMS) numerical scheme which is robust and accurate, and
makes full use of all available processors on multicore computer
architectures. The MMS scheme and the initial release of the code,
which focuses on adiabatic pulsations, is described in
\citet{Townsend:2013aa}; extensions to the code to support
non-adiabatic pulsations are described in \citet{Goldstein:2015aa}.

\MESAstar\ couples to \GYRE\ via two mechanisms. Loose coupling is
achieved simply by \MESAstar\ writing models out to disk, and
\GYRE\ subsequently reading these models in; we use this process below
to map out massive-star instability
strips. Tight coupling removes the intermediate disk usage, by
handling all communication between \MESAstar\ and \GYRE\ in-memory;
this permits fully closed-loop calculations, where the changes in the
oscillation eigenfrequencies of an evolving stellar model are used to
guide the further evolution of the model. Tight coupling allows
\GYRE\ to function as an alternative to \code{ADIPLS} in the
\code{astero} extension, and opens up the possibility of
other kinds of novel calculations, such as the automated location of
instability-strip boundaries.

\subsection{Massive-Star Instability Strips} \label{s.pulse_instab}

\begin{figure*}[!htb]
\includegraphics[width=7in]{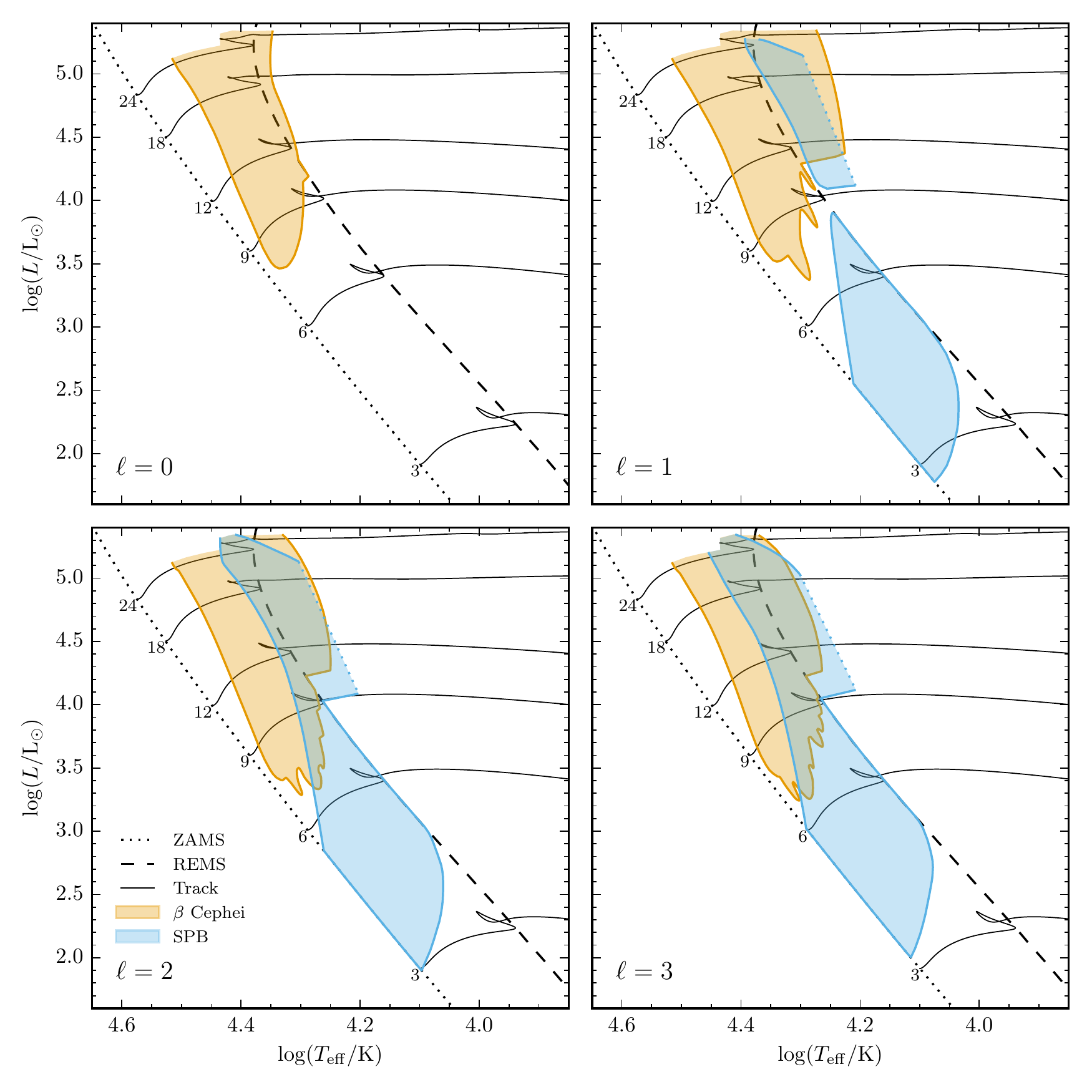}
\caption{Instability strips for $\ell=0\text{--}3$ oscillation modes in
  the upper part of the Hertzsprung-Russell diagram. Separate strips
  are shown for the $\beta$ Cephei ($\dfreq >$\,1) and slowly pulsating B-type (SPB; $\dfreq <$ 1)
  classes of pulsating stars. The ZAMS and red edge of the main
  sequence (REMS) are shown for reference, as are evolutionary tracks
  for models with selected masses (labeled in solar units along the
  ZAMS). The red edges of the post-MS SPB strips are drawn with a
  dotted line, indicating that the positioning of these edges is an
  artifact of our numerical procedure.}
\label{f.instab}
\end{figure*}

\begin{figure}[!htb]
\includegraphics[width=\apjcolwidth]{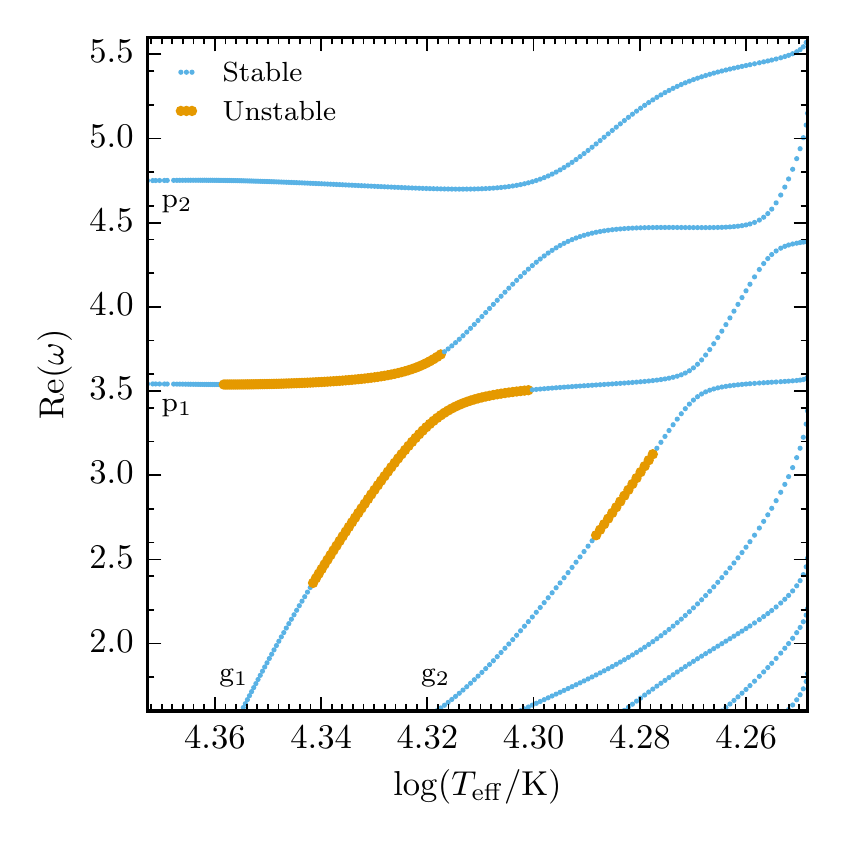}
\caption{The $\ell$ = 1 dimensionless frequency spectrum of an
  8.5\,\Msun\ stellar model as it evolves from the ZAMS to the REMS.
  Blue (orange) dots indicate which modes are stable (unstable);
  selected modes are labeled along the
  left/bottom edge using their classification.}
\label{f.fingers}
\end{figure}

\begin{figure}[!htb]
\includegraphics[width=\apjcolwidth]{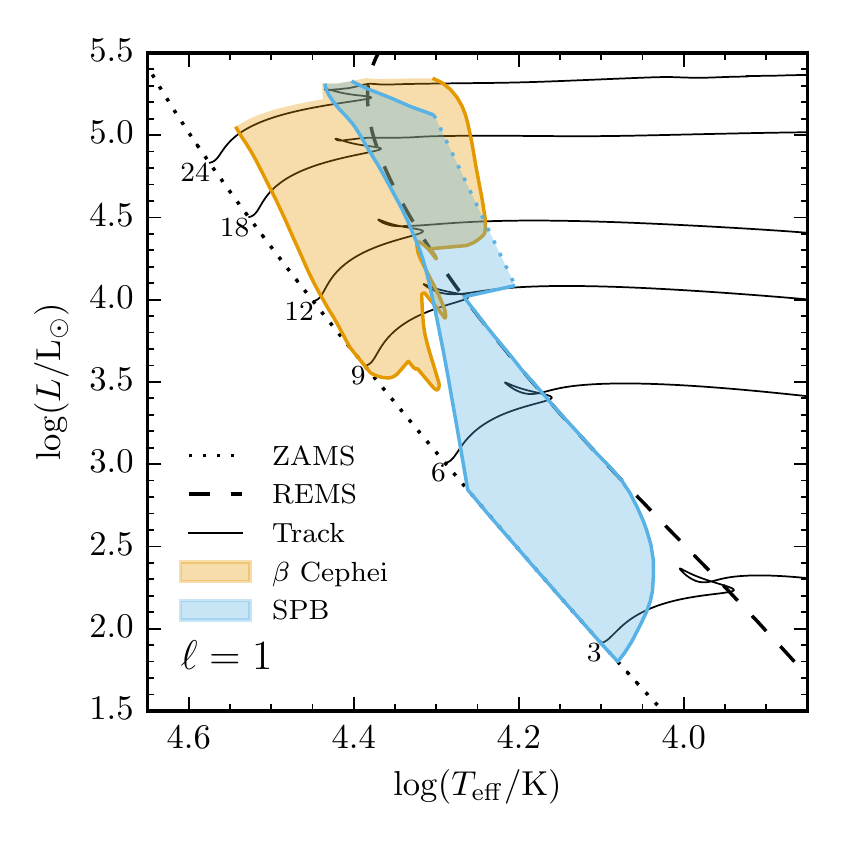}
\caption{Instability strips for dipole ($\ell=1$) oscillation modes in
  the upper part of the HR diagram, but calculated
  using OP rather than OPAL opacities (cf. Figure~\ref{f.instab}).}
\label{f.instab-op}
\end{figure}

As an illustration of a large-scale calculation using \MESAstar\ and
\GYRE\ loosely coupled, Figure~\ref{f.instab} plots the instability
strips for massive stars on and near the upper main sequence, for
oscillation modes with harmonic degrees $\ell=0\text{--}3$. These
strips are based on a set of 182 evolutionary tracks, each extending
from the ZAMS across to a red limit at $\log (\Teff/\K) = 3.75$, with 101
tracks spanning the initial mass range $2.5\,\Msun \le M \le 25\,\Msun$
in uniform logarithmic increments, and the remaining 81 tracks
spanning the mass range $6\,\Msun \le M \le 10\,\Msun$ in uniform
linear increments (the latter set is designed to adequately resolve
the ``fingers'' discussed below). OPAL opacity tables are used with the
proto-solar abundances from \cite{Asplund:2009aa}, and for simplicity
we neglect any rotation or mass loss. Convection is modeled with a
mixing-length parameter $\alphaMLT$ = 1.5 and an exponential overshoot
parameter $\fov = 0.024$, and the Schwarzschild stability criterion is
assumed.

We select points $i=i_{1},i_{2},i_{3},\ldots$ along each of the 182
tracks (where $i$ is the timestep index; see section 6.4 of \mesaone),
chosen so that $i_{1}$ corresponds to the ZAMS,
\begin{equation}
\sum_{i=i_{1}}^{i_{2}-1} \left\{
\left[ \frac{\log (T_{{\rm eff},i+1} / T_{{\rm eff},i})}{\Delta_{T}} \right]^2 +
\left[ \frac{\log (L_{i+1}/L_{i})}{\Delta_{L}} \right]^{2} \right\}^{1/2}
\approx 1
\end{equation}
across the $(i_{1},i_{2})$ pair, and similarly for subsequent
pairs. Here, $\Delta_{T}$ and $\Delta_{L}$ are dimensionless weights
which control the spacing of points in effective temperature and
luminosity; we adopt the values 0.004 and 0.011, respectively, for
these weights. At the selected points, \GYRE\ searches for unstable
oscillation modes with the harmonic degrees considered. First,
\GYRE\ solves the adiabatic oscillation equations to find
eigenfrequencies \freqad\ falling in the range extending from the
asymptotic frequency of the gravity (g) mode with radial order
$n=400$, up to the asymptotic frequency of the pressure (p) mode with
radial order $n=10$. Each \freqad\ is then used as an initial guess in
finding a corresponding eigenfrequency \freq\ of the full
non-adiabatic oscillation equations. The real and imaginary parts of
\freq\ give the linear frequency \nuosc\ and the growth $e$-folding
time \tosc\ of a mode:
\begin{equation}
\nuosc = \frac{\repart(\freq)}{2\pi}, \qquad
\tosc = -\frac{1}{\impart(\freq)}.
\end{equation}
If \tosc \ is negative, the mode is damped.

Separate strips are shown in Figure~\ref{f.instab} for regions exhibiting unstable
modes with $\repart(\dfreq) > 1$ and $\repart(\dfreq) < 1$,
where
\begin{equation}
\dfreq = \freq \sqrt{\frac{R^{3}}{GM}}
\end{equation}
is the dimensionless eigenfrequency; these correspond, respectively,
to the $\beta$ Cephei and slowly pulsating B-type (SPB) classes of
pulsating stars. In $\beta$ Cephei stars during the MS phase, p and g
modes with periods of a few hours and radial orders $n \approx
1\text{--}3$ are excited by a $\kappa$ mechanism operating on the iron
opacity bump situated in the outer envelope at $\log (T/\K) \approx
5.3$ \citep{Cox:1992aa,Dziembowski:1993aa}. In SPB stars during the MS
phase, g modes with periods of a few days and radial orders $n \approx
20\text{--}50$ are excited by the same mechanism
\citep{Dziembowski:1993ab}. For masses $M \gtrsim 9\,\Msun$ the strips
for both classes of stars extend into the post-MS domain.  During this
phase, unstable modes couple with g modes trapped near the boundary of
the inert helium core. In the case of the SPB stars this leads to very
high overall radial orders, $n \gtrsim 100$, and ultimately limits our
ability to follow the instability strips all the way to the red edge
(our calculations are restricted to $n \lesssim 400$ for computational
efficiency reasons). Hence, in Figure~\ref{f.instab} we plot the red
edges of the post-MS SPB strips with dotted lines, to highlight that
these are not the true red edges.

Allowing for differences in adopted abundances and other modeling
parameters, the instability strips plotted in Figure~\ref{f.instab}
are in general agreement with those published in the literature
\citep[e.g.,][]{Pamyatnykh:1999aa,Zdravkov:2008aa,Saio:2011aa}.  The
notable difference is the presence of fingers in the lower boundaries
of our $\beta$ Cephei strips for $\ell \geq 1$.  Their appearance here
is due to the unprecedented resolution in HR-diagram space of our
stability calculations.  To elucidate their origin,
Figure~\ref{f.fingers} plots part of the $\ell=1$ frequency spectrum
of an $8.5\,\Msun$ stellar model as it evolves from the ZAMS to the
red edge of the main sequence (REMS), showing which modes are stable
and which are unstable. The p$_{1}$ mode is unstable over the
effective temperature range $4.358 \geq \log (\Teff/\K) \geq 4.317$,
and the g$_{1}$ mode over the cooler but overlapping range $4.341 \geq
\log (\Teff/\K) \geq 4.301$. The star then passes through a phase with
no unstable modes, before the instability reappears in the range
$4.288 \geq \log (\Teff/\K) \geq 4.278$ for the g$_{2}$ mode.

This alternation between instability and stability, seen as fingers in
Figure~\ref{f.instab}, stems from the fact that the $\kappa$ mechanism
only excites modes whose eigenfrequencies fall in a narrow range
$[\freq_{\rm lo},\freq_{\rm hi}]$. At frequencies $\repart(\freq) >
\freq_{\rm hi}$, the pulsation period becomes comparable to the local
thermal timescale in the envelope region above the iron opacity peak,
and this region behaves as a damping zone, stabilizing the
modes. Conversely, at frequencies $\repart(\freq) < \freq_{\rm lo}$,
modes couple with gravity waves trapped in the $\mu$-gradient zone
developing at the core boundary, and are likewise damped.  The
intermediate stable phase in Figure~\ref{f.fingers}, between $\log
(\Teff/\K) = 4.301$ and $\log (\Teff/\K)=4.288$ occurs when there are
no modes in the $[\freq_{\rm lo},\freq_{\rm hi}]$ range.  As the star
evolves, the unstable range narrows: $\freq_{\rm hi}$ decreases due to
lower \Teff, while $\freq_{\rm lo}$ increases due to the growth of the
$\mu$-gradient zone.

Figure~\ref{f.instab-op} shows a version of the $\ell=1$ panel
calculated using OP opacity tables rather than OPAL tables. There is
an overall shift of the instability strips toward higher luminosities,
an effect already noted by \citet{Pamyatnykh:1999aa}. The fingers
persist with much the same structure, supporting the fact that they
are physical effects rather than numerical artifacts.

Returning now to Figure~\ref{f.instab}, the post-MS extension of the
SPB strips has been attributed in the literature to features in the
\BV\ frequency which reflect gravity waves at the boundary of the
helium core, preventing them from penetrating into the core and being
dissipated by strong radiative damping. \citet{Saio:2006aa} and
\citet{Godart:2009aa} argue that the necessary feature is an
intermediate convection zone (ICZ) associated with the
hydrogen-burning shell, but more recently \citet{Daszynska-Daszkiewicz:2013aa}
have shown that even a local minimum in the \BV\ frequency is
sufficient to reflect modes. In the present case, the empirical mass
threshold $M \gtrsim 9\,\Msun$ required for formation of an ICZ
coincides with the lower boundaries of the SPB strip extensions. In
the lowest-mass models above this threshold, the ICZ vanishes shortly
after its appearance, but it leaves behind a narrow region with a
steep molecular weight gradient. This gradient causes a spike in the
\BV\ frequency, which serves in a similar manner to prevent gravity
waves from entering into the core and being dissipated.

The corresponding post-MS extension of the $\beta$ Cephei strips was
first noted by \citet{Dziembowski:1993aa}, but has not received much
attention in the literature. Figure~\ref{f.instab} shows that this
extension has a well defined lower boundary, much like the SPB stars
although situated at slightly higher masses, $M \gtrsim 10.5\,\Msun$.
We have determined that the extension is also a consequence of ICZ
formation; the shift to higher masses arises because it appears that
multiple convection zones, rather than a single one, are necessary to
reflect waves at the core boundary in the case of $\beta$ Cephei pulsators.

\subsection{Asteroseismic Optimization} \label{s.pulse_astero}

\begin{figure}[!htb]
\includegraphics[width=\apjcolwidth]{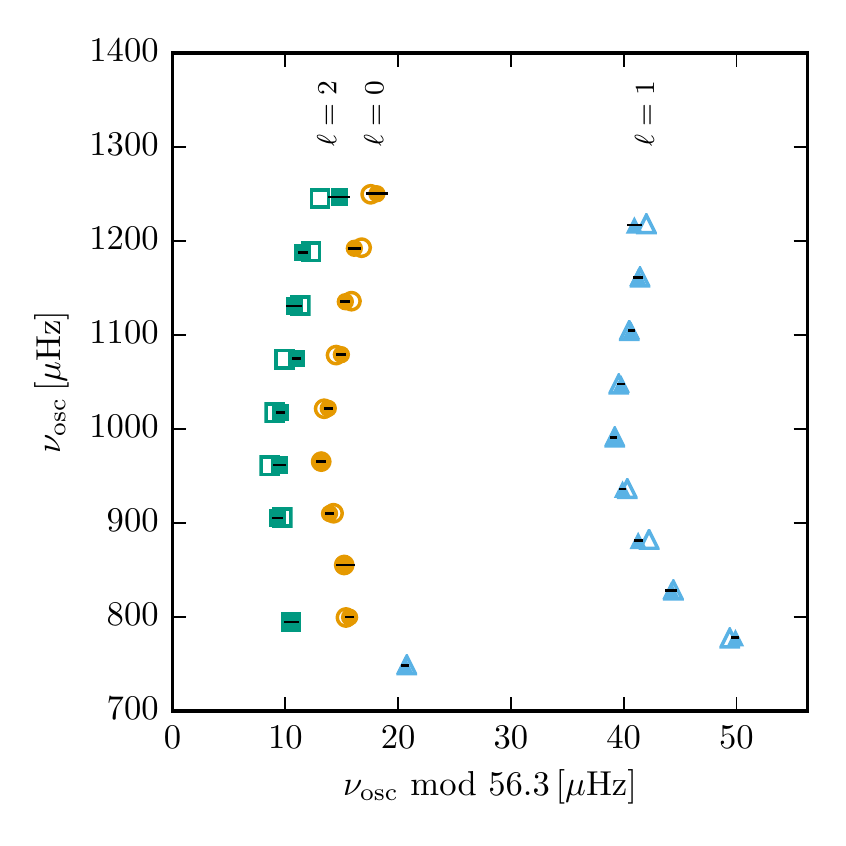}
\caption{Echelle diagram for the subgiant star HD\,49385. Observed
  frequencies are shown as filled circles ($\ell=0)$, triangles ($\ell$\,=\,1) and squares ($\ell$\,=\,2);
  black horizontal lines indicate the
  1$\sigma$ error bars. Calculated frequencies of the best-fit model
  are overplotted as the corresponding open symbols.}
\label{f.astero}
\end{figure}

To illustrate the updated asteroseismic capabilities of \MESA,
Figure~\ref{f.astero} plots the echelle diagram for the subgiant star
HD\,49385, showing both the frequencies of $\ell=0-2$ modes measured by
\citet{Deheuvels:2010aa}, and the corresponding frequencies of the
best-fit model determined using the \code{astero} extension. The
calculations follow the same procedure detailed in section 3.2
of \mesatwo; the only significant differences are that the initial
mass, helium abundance, metal abundance and mixing length parameter
are refined using the downhill simplex algorithm rather than the
Hooke-Jeeves algorithm; oscillation frequencies are calculated using
\GYRE\ rather than \ADIPLS; and the surface corrections to frequencies
are evaluated using equation 4 of \citet{Ball:2014aa} rather than with
the \citet{kjeldsen_2008} scheme.

Comparing Figure~\ref{f.astero} against figure 8 of \mesatwo\ reveals
only small differences between the two. The $\chi^{2}$ of the best-fit
models reported by \code{astero} is 2.3 in the
former case, compared to 2.4 in the latter (cf. table 2 of \mesatwo).

\subsection{Automated Strip Location} \label{s.pulse_auto}

\begin{figure}[!htb]
\includegraphics[width=\apjcolwidth]{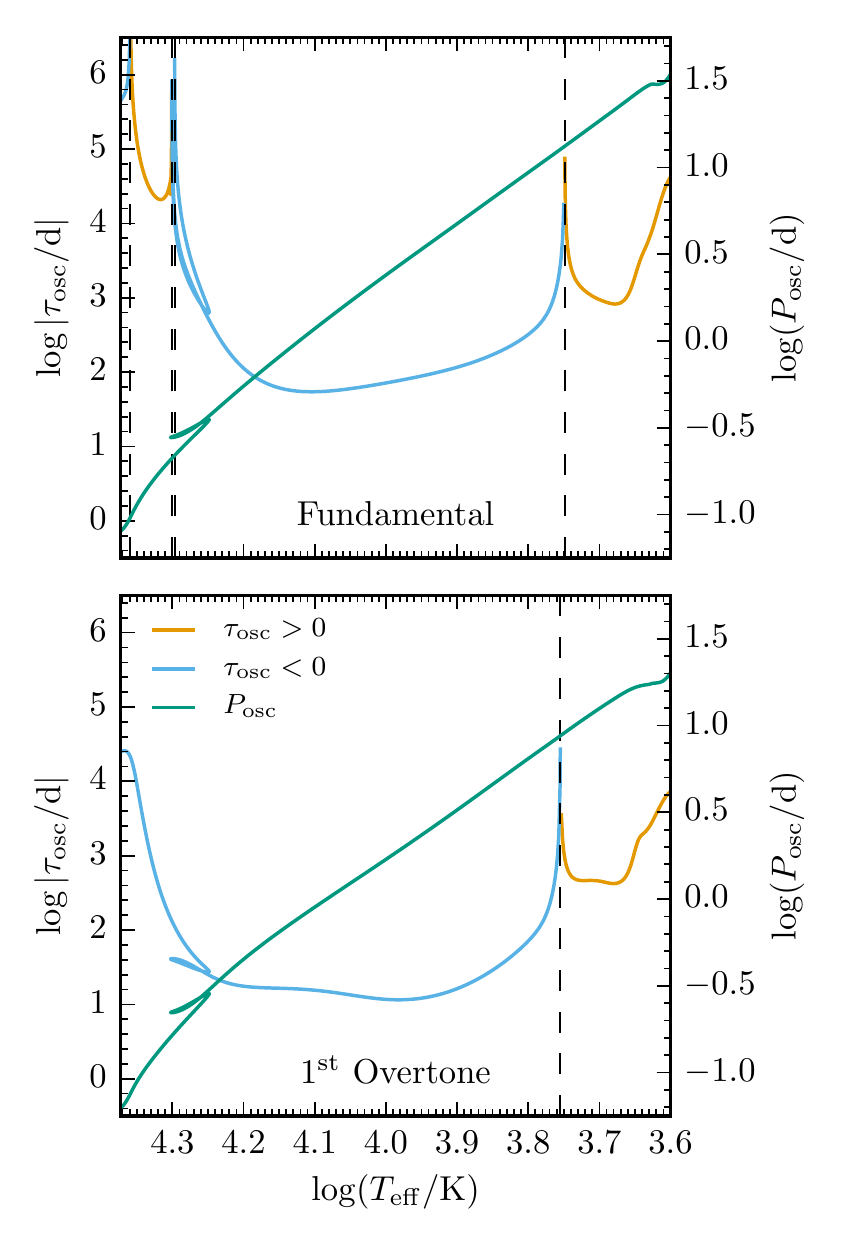}
\caption{The growth timescale \tosc\ (left axis) and oscillation
  period \Posc\ (right axis) of the fundamental and first-overtone
  radial modes of a 8.5\,\Msun\ model, plotted as a function of
  \Teff\ as the star evolves away from the ZAMS.
  The vertical dashed lines, determined automatically,
  show the points where the modes switch from
  stable ($\tosc < 0$) to unstable ($\tosc > 0$), and vice versa.}
\label{f.growth}
\end{figure}

\begin{figure}[!htb]
\includegraphics[width=\apjcolwidth]{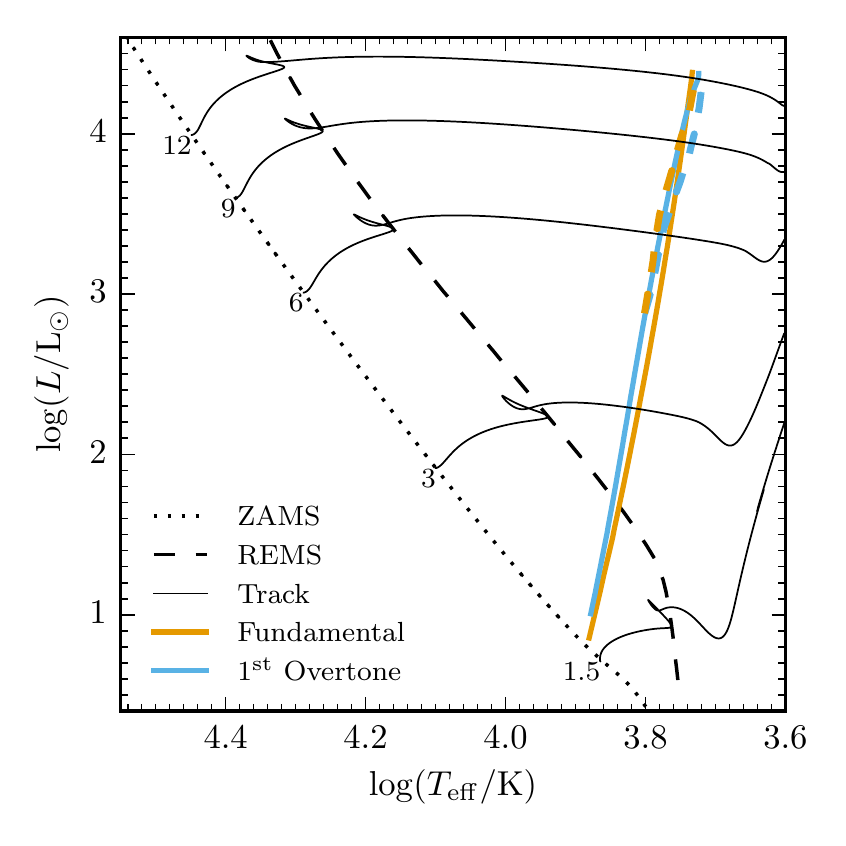}
\caption{The calculated blue edge of the classical instability strip,
  for fundamental and first-overtone radial modes.
  The corresponding dashed lines show the predictions from set B
  of \citet[][their figure 1]{Smolec:2008aa}.}
\label{f.ceph-blue}
\end{figure}

The instability strips presented above involved the examination of
$\sim 11$ million modes of $\sim$ 40,000 stellar models.  To partially
automate the process we can leverage tight coupling between \GYRE\ and
\MESAstar.  This is achieved by making small modifications to the {\tt
  extras\_check\_model} callback routine in \MESAstar\ (see Appendix
B.1 of \mesatwo), so that \GYRE\ is run after each time step to
determine the set of eigenfrequencies $\{\freq\}$ of a user-specified
group of modes.  When $\impart(\freq)$ changes sign from one time step
to the next for any of these modes, indicating that an
instability-strip boundary has been crossed, a search is
performed to find \impart(\freq)\,$\approx$\,0.

Figure~\ref{f.growth} presents an application of the tight coupling
to the fundamental and first-overtone radial
modes of the 8.5\,\Msun\ model considered in
Section~\ref{s.pulse_instab}, showing how the growth timescales
\tosc\ and oscillation periods \Posc\~=\~ 1/\nuosc \ of the modes
change as the star evolves from the ZAMS into the post-MS. The
second-overtone radial mode remains stable, $\tosc<0$, over the range
plotted. The vertical lines show where \tosc\ changes sign.
The blue and red edges of the
($\ell=0$) $\beta$ Cephei instability strip can be seen in the upper
panel of Figure~\ref{f.growth} at $\log (\Teff/\K) = 4.36$ and $\log
(\Teff/\K) = 4.30$, respectively. The blue edge\footnote{For classical
  ($\delta$) Cepheid pulsators, the observational blue edge of the
  classical instability strip is in fact established by stars evolving
  to higher \Teff\ on their first blue loop, rather than stars on
  their first crossing of the Hertzsprung gap. However, the purpose of
  the present section is to demonstrate the capability of tightly coupling, and in this
  context the distinction between the blue edges from multiple
  crossings is unimportant.} of the classical instability strip can
likewise be seen in both panels at $\log (\Teff/\K) = 3.75$. The corresponding
red edge is not found because \GYRE\ does not include a
treatment of the pulsation-convection interaction --- a necessary
ingredient for modeling the classical red edge \citep[see, e.g,
  section 3.7.3 of][and references therein]{Aerts:2010aa}.

As a further demonstration of automated instability strip location,
Figure~\ref{f.ceph-blue} plots the blue edges of the classical
instabi\-li\-ty strip in the HR diagram, for fundamental and
first-overtone radial modes. The edges are calculated
for 51 evolutionary tracks spanning the initial mass range
$1.25\,\Msun \le M \le 12.5\,\Msun$ in uniform logarithmic
increments. At luminosities $\log (L/\Lsun) \gtrsim 2.5$ corresponding
to classical Cepheid pulsators, these edges show good agreement with
the set B results published by \citet[][their figure 1]{Smolec:2008aa}.
 At luminosities $\log (L/\Lsun) \lesssim 1.6$
corresponding to $\delta$ Scuti stars, the edges are somewhat cooler
than results published in the literature; however, this is because we
consider only fundamental and first-overtone modes, whereas the blue
edge is typically set by higher overtones which are displaced toward
hotter \Teff\ \citep[e.g.,][their figure 1]{Dupret:2004aa}.

Ideally, the same automated approach could be used to locate the
boundaries of the non-radial ($\ell > 0$) instability strips plotted
in Figure~\ref{f.instab}. In practice it is very challenging
to devise a robust algorithm that can unambiguously interpret the
eigenfrequencies produced by \GYRE. Sometimes, acoustic glitches in a
model can trap modes in surface layers, where they are strongly
excited; however, these modes are very sensitive to model parameters,
and it is unclear whether they are physically meaningful or not.

\def\msun{M$_{\odot}$}
\def\rsun{R$_{\odot}$}
\def\lsun{L$_{\odot}$}

\section{Implicit Hydrodynamics}\label{s.hydro}

Shocks happen in stars, such as after a massive star collapses, or
cyclically in the outer envelopes of stars pulsating at sufficiently
large amplitude. Previous versions of $\MESAstar$ allowed large
velocities such as those encountered in the last few seconds leading
to a core collapse ($\approx 1000\ {\rm km \ s^{-1}}$), but there was
no provision for large jumps in velocities leading to shocks.  In this
section we describe the changes that have been made to support an
implicit treatment of hydrodynamic shocks that includes careful
attention to conservation of energy.  We demonstrate that the revised
equations are intrinsically conservative in the sense that deviations
from exact energy balance can only arise from residual numerical
errors in the approximate solutions rather than from the form of the
equations themselves.  Following the description of the changes, we
show a series of envelope shocks as a test of the implementation.  The
form of the equations and the demonstration of intrinsic conservation
closely follow \citet{fraley_1968_aa} and \citet{grott_2005_aa}.  The
treatment of artificial viscosity is based on \citet{weaver_1978_aa}.

\begin{figure}[!htb]
\begin{center}
\includegraphics[width=\apjcolwidth]{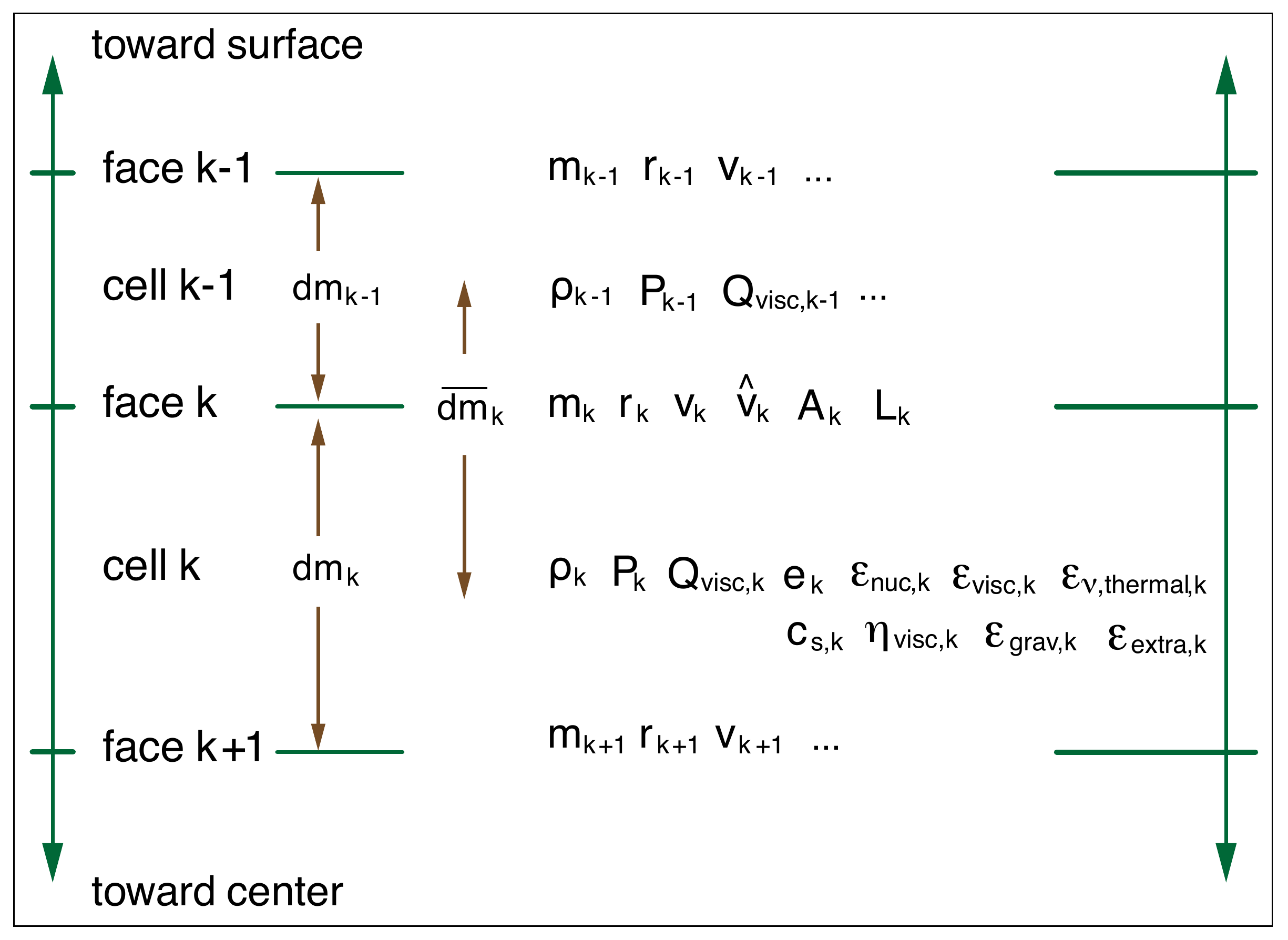}
\caption{Schematic of relevant cell and face variables relevant for hydrodynamics in $\MESAstar$.\label{f.structure}}
\end{center}
\end{figure}

\subsection{Mass Continuity}\label{s.hydro_cont}

The specific volume of cell $k$ is
\begin{equation}
\frac{1}{\rho_k} = \frac{4 \pi}{3} \left( \frac{r_k^3 - r_{k+1}^3}{dm_k} \right )
\enskip,
\label{e.specvol}
\end{equation}
where $r_k$ is the outer face radius, $r_{k+1}$ is the inner face
radius, $dm_k$ is the cell mass, and $\rho_k$ is the cell average
density (see Figure~\ref{f.structure} for the layout of cells in
\MESAstar).  
We create an initial algebraic form of the continuity
differential equation
by dividing the change in the specific volume in a step by
the length of time $\delta t$, using step start and end values for
$r_k$, $r_{k+1}$, and $\rho_k$ and the value $dm_k$ which is constant
during the step:
\begin{equation}
\frac{1/\rho_k - 1/\rho_{{\rm start},k}}{\delta t} =
\frac{4 \pi}{3 dm_k}
\frac{\left[ \left(r_k^3 - r_{{\rm start},k}^3\right) - \left(r_{k+1}^3 - r_{{\rm start},k+1}^3\right) \right ]} {\delta t}
\enskip.
\label{e.foo}
\end{equation}
Next, we rewrite the right hand side introducing variables for the time centered velocity $\vhat$ and the effective area $A$
to get the final form of the mass continuity equation as used in $\MESAstar$:
\begin{equation}
\frac{1/\rho_k - 1/\rho_{{\rm start},k}}{\delta t} =
\frac{1}{dm_k} \left( A_k {\vhat}_{k} - A_{k+1} {\vhat}_{k+1} \right)
\enskip,
\label{e.cont}
\end{equation}
where
\begin{equation}
\vhat_{k} = ({v_k + v_{{\rm start},k}})/2
\end{equation}
and $r_k$ is evaluated as
\begin{equation}
r_k = r_{{\rm start},k} +  \vhat_{k} \delta t
\enskip.
\label{e.tcvel}
\end{equation}
Algebraic simplification then shows that
\begin{equation}
A_k = \frac{4 \pi}{3} \left ( r_k^2 + r_k r_{{\rm start},k} + r_{{\rm start},k}^2 \right )
\enskip .
\label{e.Aeff}
\end{equation}
To be consistent with the mass continuity equation, we use these
expressions for effective area and time centered velocity in the
following momentum and energy equations.  It will be shown below that
to get intrinsic energy conservation, we must time center the velocity
and use special combinations of starting and ending radius in a couple
of places, but all other terms in the equations can remain fully
implicit to avoid degrading the numerical stability as would happen in
a uniformly time-centered scheme.

\subsection{Artificial Viscosity}
\label{s.hydro_visc}

In $\MESAstar$, the artificial dynamic viscosity coefficient \etavisc\ (which has the dimensions $\rm g\ cm^{-1} s^{-1}$)
\begin{equation}
\eta_{{\rm visc}, k} = \eta_{{\rm {\rm visc},linear}, k} + \eta_{{\rm {\rm visc},quad}, k}
\enskip,
\label{e.etavisc}
\end{equation}
where the linear term is
\begin{equation}
\eta_{{\rm {\rm visc},linear}, k} =
\frac{3}{4} l_1 \rho_k r_{{\rm mid},k} c_{{\rm s},k}
\label{e.etavisclinear}
\end{equation}
and the quadratic term is
\begin{equation}
\eta_{{\rm {\rm visc},quad}, k} =
\frac{3}{4}\,l_2^2 \, \frac{\rho_k^2 4 \pi   r_{{\rm mid},k}^2 }{dm_k}
 \ \max\left(0, r_{k+1}^2 \vhat_{k+1} - r_k^2 \vhat_k\right)~,
\label{e.etaviscquad}
\end{equation}
with $r_{{\rm mid},k} = (r_{k+1} + r_k)/2$, $c_{{\rm s},k}$ the sound
speed in cell $k$, and $l_1$ ($l_2$) is a dimensionless coefficient
for the linear (quadratic) term. The linear term is rarely used; it
provides for a general damping of pulsations.  The quadratic term is
only nonzero in regions of compression and is the primary control for
the strength of artificial viscosity.  Assuming the usual case of
$l_1$=0, the shock front is spread over a distance $\sim$$l_2
r_k$.  
We follow \citet{dorfi_98} in opting for a shock spread
proportional to the local radius $r$
rather than the local cell width.  This choice is dictated by
the fact that step-by-step adjustments to the mesh resolution
lead to dynamically changing, non-monotonic variations in cell widths
of up to a factor of 2 or more between neighboring cells.  
Making the shock spread directly dependent
on the local cell widths would produce numerically intolerable 
dynamically changing, non-monotonic variations in cell-to-cell
values for the shock spread.  Use of a local running average cell
width is also ruled out by the need to keep algebraic equations
dependent only on nearest neighbors to allow a block tridiagonal matrix solution.
The use of a small fraction of the local radius gives a smoothly varying
shock spread that avoids the numerical problems associated with using the cell width.

We define the quantity $Q_{{\rm visc},k}$ (having dimensions of energy), in cell $k$ as
\begin{equation}
Q_{{\rm visc},k} =
 \eta_{{\rm visc}, k}  \frac {\rho_k 4 \pi r_{{\rm mid},k}^6 } {dm_k} \left(\frac{\vhat_k}{r_k} - \frac{\vhat_{k+1}}{r_{k+1}}\right)
\enskip .
\label{e.qvisc}
\end{equation}
The momentum equation uses $Q_{\rm visc}$ in an expression that defines an artificial acceleration
analogous to the pressure gradient term, and the energy equation uses
it to define an artificial viscous heating analogous to the mechanical
work term.

\subsection{Specific Linear Momentum Equation}\label{s.hydro_mom}

The local linear momentum conservation equation
at face $k$ between inner cell $k$ and outer cell $k-1$ is
\begin{equation}
\frac{v_k - v_{{\rm start},k}}{\delta t} =
 - \frac{G m_k}{r_k r_{{\rm start},k}}
 - A_k \left ( \frac{P_{k-1} - P_k}{\overline{dm}_k} \right )  + g_{{\rm visc},k}
\enskip .
\label{e.momentum}
\end{equation}
where ${\overline{dm}_k}=(dm_k+dm_{k-1})/2$ is the mass associated with face $k$, and the
 viscous acceleration term at face $k$ is
\begin{equation}
g_{{\rm visc},k} =
\frac{4 \pi}{r_k} \left ( \frac{Q_{{\rm visc},k-1} - Q_{{\rm visc},k}}{\overline{dm}_k} \right )
\enskip.
\label{e.gvisc}
\end{equation}
The use of the product $r_k r_{{\rm start},k}$ in the denominator of
the gravitation term is necessary for intrinsic energy conservation as
will be shown below.

\subsection{Specific Energy Equation}\label{s.hydro_ener}

The local energy conservation equation
for cell $k$ between outer face $k$ and inner face  $k+1$ is
\begin{equation}
\begin{split}
\frac{e_k - e_{{\rm start},k}}{\delta t} =&
 - \frac{L_k - L_{k+1}}{dm_k}
 - P_k \left ( \frac{A_k \vhat_k - A_{k+1} \vhat_{k+1}}{dm_k} \right )\\
 &+ \epsilon_{{\rm visc},k}
+ \epsilon_{{\rm nuc},k}
- \epsilon_{\nu,k}
+ \epsilon_{{\rm extra},k}
\enskip ,
\end{split}
\label{e.energy}
\end{equation}
where $e_k$ is the specific thermal energy for cell $k$. The viscous
heating rate for cell $k$ is
\begin{equation}
\epsilon_{{\rm visc},k} =
\frac{4 \pi Q_{{\rm visc},k}}{dm_k} \left( \frac{\vhat_k}{r_k} - \frac{\vhat_{k+1}}{r_{k+1}} \right)
\enskip.
\label{e.eviscous}
\end{equation}
Energy loss from weak reaction neutrinos is already subtracted from
the nuclear burning term, $\epsilon_{{\rm nuc},k}$, so only the
neutrino energy loss rate from thermal processes, $\epsilon_{\nu,k}$, is
explicitly accounted for in Equation~\eqref{e.energy}.  An example of
$ \epsilon_{{\rm extra},k}$ is artificial injection of energy to
trigger a shock.

An alternative form of the energy equation equates the model $dL/dm$
to the expected value
\begin{equation}
\frac{L_k- L_{k+1}}{dm_k} = \left ( \frac{dL}{dm} \right ) _{{\rm expected},k}
\enskip,
\label{e.dldm}
\end{equation}
where
\begin{equation}
\left ( \frac{dL}{dm} \right ) _{{\rm expected},k} =
\epsilon_{{\rm grav},k}
+ \epsilon_{{\rm visc},k}
+ \epsilon_{{\rm nuc},k}
- \epsilon_{\nu,k}
+ \epsilon_{{\rm extra},k}
\enskip,
\label{e.dldm_expect}
\end{equation}
and
\begin{equation}
\epsilon_{{\rm grav},k} =
 - \frac{e_k - e_{{\rm start},k}}{\delta t}
 - P_k \left ( \frac{A_k \vhat_k - A_{k+1} \vhat_{k+1}}{dm_k} \right )
\enskip.
\label{e.epsgrav}
\end{equation}
Using Equation~\eqref{e.cont}, the expression for $\epsilon_{\rm grav}$
can be rewritten
\begin{equation}
\epsilon_{{\rm grav},k} =
 - \frac{e_k - e_{{\rm start},k}}{\delta t}
 - P_k \frac{(1/\rho_k - 1/\rho_{{\rm start},k})}{\delta t}
\label{e.altepsgrav}
\end{equation}
thereby avoiding the use of velocities and thus be appropriate for
hydrostatic cases.

\subsection{Intrinsic Energy Conservation}\label{s.hydro_conserve}

The summed kinetic, potential, internal energies are
\begin{align}
\mathrm{KE} &= \sum_k \frac{1}{2} \ \overline{dm}_k \ v_k^2,\\
\mathrm{PE} &= \sum_k -\frac{G m_k \overline{dm}_k}{r_k}, \\
\mathrm{IE} &= \sum_k e_k \ dm_k ~,
\label{e.energydefs}
\end{align}
and thus the total energy of the star is
${\rm E} = {\rm KE} + {\rm PE} + {\rm IE}$.  We now explicitly
demonstrate that the equations we solve are formulated in such a way
that the rate of change of total energy exactly equals the combined
energy sources and sinks.

Multiplying Equation~\eqref{e.momentum} by $\vhat_k \overline{dm}_k$
gives an equation with units of luminosity:
\begin{equation}
\begin{split}
\frac{1}{2}  \ \overline{dm}_k \ \left ( \frac{v_k^2 - v_{{\rm start},k}^2}{\delta t} \right ) =&
 - G m_k \overline{dm}_k \left ( \frac{\vhat_k }{r_k r_{{\rm start},k}} \right ) \\
 &-  A_k \vhat_k ( P_{k-1} - P_k ) \\
 &+ 4 \pi \left( \frac{\vhat_k}{r_k} \right) ( Q_{{\rm visc},k-1} - Q_{{\rm visc},k}).
\label{e.momentum_power}
\end{split}
\end{equation}
Using Equation~\eqref{e.tcvel} to eliminate $\vhat_k$ in the first
term on the right,
\begin{equation}
- G m_k \overline{dm}_k \left ( \frac{\vhat_k }{r_k r_{{\rm start},k}} \right ) =
\frac{G m_k \overline{dm}_k}{\delta t} \left ( \frac{1}{r_k} - \frac{1}{r_{{\rm start},k}}  \right )~,
\label{e.firstterm}
\end{equation}
shows that this term is the negative of the rate of change of
potential energy, a result that is made possible by the use of
the $G m_k / r_k r_{{\rm start},k}$  in Equation~\eqref{e.momentum}
instead of an alternative such as $G m_k / r_k^2$.
Thus, Equation~\eqref{e.momentum_power} can be written as
\begin{equation}
\begin{split}
&\frac{1}{2}  \ \overline{dm}_k \ \frac{v_k^2 - v_{{\rm start},k}^2}{\delta t}
-\frac{G m_k \overline{dm}_k}{\delta t} \left ( \frac{1}{r_k} - \frac{1}{r_{{\rm start},k}}  \right ) \\
=& - A_k \vhat_k ( P_{k-1} - P_k )
+ 4 \pi \left ( \frac{\vhat_k}{r_k} \right ) ( Q_{{\rm visc},k-1} - Q_{{\rm visc},k})
~.
\label{e.momentum_power_final}
\end{split}
\end{equation}
Similarly, multiplying equation~\eqref{e.energy} by $dm_k$ also yields
an equation with units of luminosity:
\begin{equation}
\begin{split}
\frac{( e_k - e_{{\rm start},k} )}{\delta t} dm_k =&
 - (L_k - L_{k+1})\\
&- P_k \left ( A_k \vhat_k - A_{k+1} \vhat_{k+1} \right ) \\
&+
4 \pi Q_{{\rm visc},k} \left( \frac{\vhat_k}{r_k} - \frac{\vhat_{k+1}}{r_{k+1}} \right ) \\
&+ ( \epsilon_{{\rm nuc},k} - \epsilon_{\nu,k}  + \epsilon_{{\rm extra},k} ) \ dm_k~,
\label{e.power_final}
\end{split}
\end{equation}
Adding Equations \eqref{e.momentum_power_final} and \eqref{e.power_final} and summing
over the grid index $k$  gives
\begin{equation}
\begin{split}
\mathlarger{\sum}_k
-& \frac{G m_k \overline{dm}_k}{\delta t} \left ( \frac{1}{r_k} - \frac{1}{r_{{\rm start},k}} \right ) \\
+& \frac{1}{2}  \ \overline{dm}_k \ \frac{v_k^2 - v_{{\rm start},k}^2}{\delta t} \\
+& \frac{( e_k - e_{{\rm start},k})} {\delta t} dm_k \\
= \mathlarger{\sum}_k
-& (L_k - L_{k+1})  \\
+& P_k A_{k+1} \vhat_{k+1} - P_{k-1} A_k \vhat_k\\
+& 4 \pi Q_{{\rm visc},k-1} \left ( \frac{\vhat_k}{r_k} \right )
- 4 \pi Q_{{\rm visc},k} \left(\frac{\vhat_{k+1}}{r_{k+1}} \right) \\
+& ( \epsilon_{{\rm nuc},k} - \epsilon_{\nu,k} + \epsilon_{{\rm extra},k}) \ dm_k~.
\label{e.sumeqs}
\end{split}
\end{equation}
The sum over the pressure terms is
\begin{equation}
\begin{split}
&\mathlarger{\sum}_k \left(P_k A_{k+1} \vhat_{k+1} - P_{k-1} A_k \vhat_k\right)\\
=& -\left[\left(P A \vhat\right)_\mathrm{surface} -  \left(P A \vhat\right)_\mathrm{center}\right] \\
=&  -(L_{\rm acoustic, surface} - L_{\rm acoustic, center})~,
\label{e.acoustic}
\end{split}
\end{equation}
which cancels term by term except at the boundaries.  We define
$L_{\rm acoustic, surface}$ as the work done by the model on the
atmosphere at the surface and $L_{\rm acoustic, center}$ as the work
done on the model at the center, for example, by an artificial piston.
The sum over the artificial viscosity terms leads to a similar
expression, but because $Q_\mathrm{visc}$ vanishes at the surface and
the center, the sum equals zero. That is, the energy added by
artificial viscous heating in the energy equation exactly balances the
loss of kinetic energy by artificial viscous acceleration in the
momentum equation.

The terms on the left hand side of Equation~\eqref{e.sumeqs} are the
difference in the total energy between the start and end of a step
divided by the length of the step, in other words, the average rate of
change of the total energy of the model.  Therefore
Equation~\eqref{e.sumeqs} can be written as
\begin{equation}
\begin{split}
(E_{\rm final} - E_{\rm initial})/\delta t =
&- (L_{\rm surface} - L_{\rm center}) \\
&- (L_{\rm acoustic, surface} - L_{\rm acoustic, center}) \\
&+ \sum_k \left(\epsilon_{{\rm nuc},k} - \epsilon_{\nu,k} + \epsilon_{{\rm extra},k}\right) \ dm_k
\enskip.
\label{e.econserv}
\end{split}
\end{equation}
This equation embodies conservation of energy in $\MESAstar$: the rate
of change of total energy equals the combined energy sources and sinks.
This demonstrates that in the given form, the algebraic equations
intrinsically conserve energy
in the sense that failure to get energy
balance can only arise from the residual numerical errors that are
inherent in using approximate solutions to the equations.  This in
turn means that to control energy balance errors, we can focus on
reducing residuals either by changes in the Newton solver or by
timestep reductions.

\subsection{Controlling the Accuracy of Energy Conservation}

The Newton solver considers both the sizes of incremental changes to
the variables and the sizes of residual errors for the equations.  For
the energy equation, the residual used by the solver is defined to be
the timestep $\delta t$ times the difference between the left and
right sides of Equation~\eqref{e.energy} divided by
$e_{{\rm start},k}$; in other words, the residual is the error as a
fraction of the specific energy at the start of the step.  By
adjusting tolerances for the average and maximum size of residuals, we
force the Newton solver to take extra iterations to reduce the
residuals which will in turn reduce the total error in energy
conservation.

A second and related way to control energy conservation errors is to
use the average and maximum energy residuals to adjust the timesteps.
For example, if the maximum magnitude for an energy residual exceeds a
specified hard limit, then the proposed solution is rejected and the
step is retried with a smaller timestep.  If the maximum is smaller
than the hard limit but exceeds another specified limit, then there is
no forced retry for this step, but the next timestep is reduced by the
ratio of the limit divided by the maximum magnitude.  If the maximum
is smaller than both limits, then other factors determine the next
timestep.  Later in this section, we will show that these approaches
in combination with the intrinsic conservation form of the equations
yield a solution for a shock in an envelope that evolves with
reasonably large timesteps while conserving energy to a high degree of
accuracy.

\subsection{Limiting Acceleration of Convective Velocity}
\label{s.hydro-limit}

When using hydrodynamics, we often require timesteps that are so small
that we need to limit the increase in convective velocities as
calculated in the standard instantaneous mixing length theory (MLT) so that
they do not assume unphysically large accelerations.  If convection
velocities are allowed to adjust instantaneously, then our methods for
artificially creating shocks will fail since however rapidly we inject
energy over a limited region, convection will be able to transport the
energy away.  To be able to simulate shocks we need to have a way to
limit convection velocity acceleration.

The primary scheme we use for this is derived from
\citet{arnett_69_mlt} and \citet{wood_74}.  The MLT implementation in
\mesa\ has been extended to take as additional arguments the timestep
and the previous convection velocity at the same mass location
($v_{c,\mathrm{prev}}$).  It calculates a provisional convection
velocity ($v_{c0}$) using the standard instantaneous MLT, then defines
a convective timescale ($\taumlt$) as the local pressure scale height
($H$) divided by the sum of the provisional plus previous velocities.
If $\delta t$ is less than $\tau_{\rm MLT}$, then the next convection
velocity ($v_c$) is only incremented from the previous one by the
difference of the provisional minus the previous velocities times the
ratio of the timestep divided by the time scale
\begin{equation}
v_c = v_{c,\mathrm{prev}} + \min\left(1, \frac {\delta t} {\tau_{\rm MLT}}\right) ( v_{c0} - v_{c,\mathrm{prev}} )
\enskip,
\label{e.cv}
\end{equation}
where
\begin{equation}
\tau_{\rm MLT} = \frac {H} { ( v_{c0} + v_{c,\mathrm{prev}} ) }~.
\label{e.tau}
\end{equation}
As an alternative scheme for limiting convection acceleration, we also
allow the maximum rate of change of convection velocity to be set as a
fraction, $g_{\theta}$, of the local gravitational acceleration.  If
$v_{c0} > v_{c,\mathrm{prev}}$, then
\begin{equation}
v_c = \min(v_{c0},  v_{c,\mathrm{prev}} + \delta t\ g_{\theta}\ g)
\enskip,
\label{e.cv2}
\end{equation}
The final $v_c$ is used to recalculate the convection efficiency, which is
used to calculate the MLT temperature gradient.

These methods for limiting the acceleration of convective velocities
reduce the energy transport rate as well as the rate of compositional
mixing.  Both schemes seem to give at least qualitatively reasonable
results and avoid the problems of unphysically large accelerations
that are possible with standard instantaneous MLT.  Hopefully this ad
hoc solution will soon be replaced by a quantitatively correct
formulation.

\begin{figure*}[!htb]
\begin{center}
\includegraphics[width=3.15in]{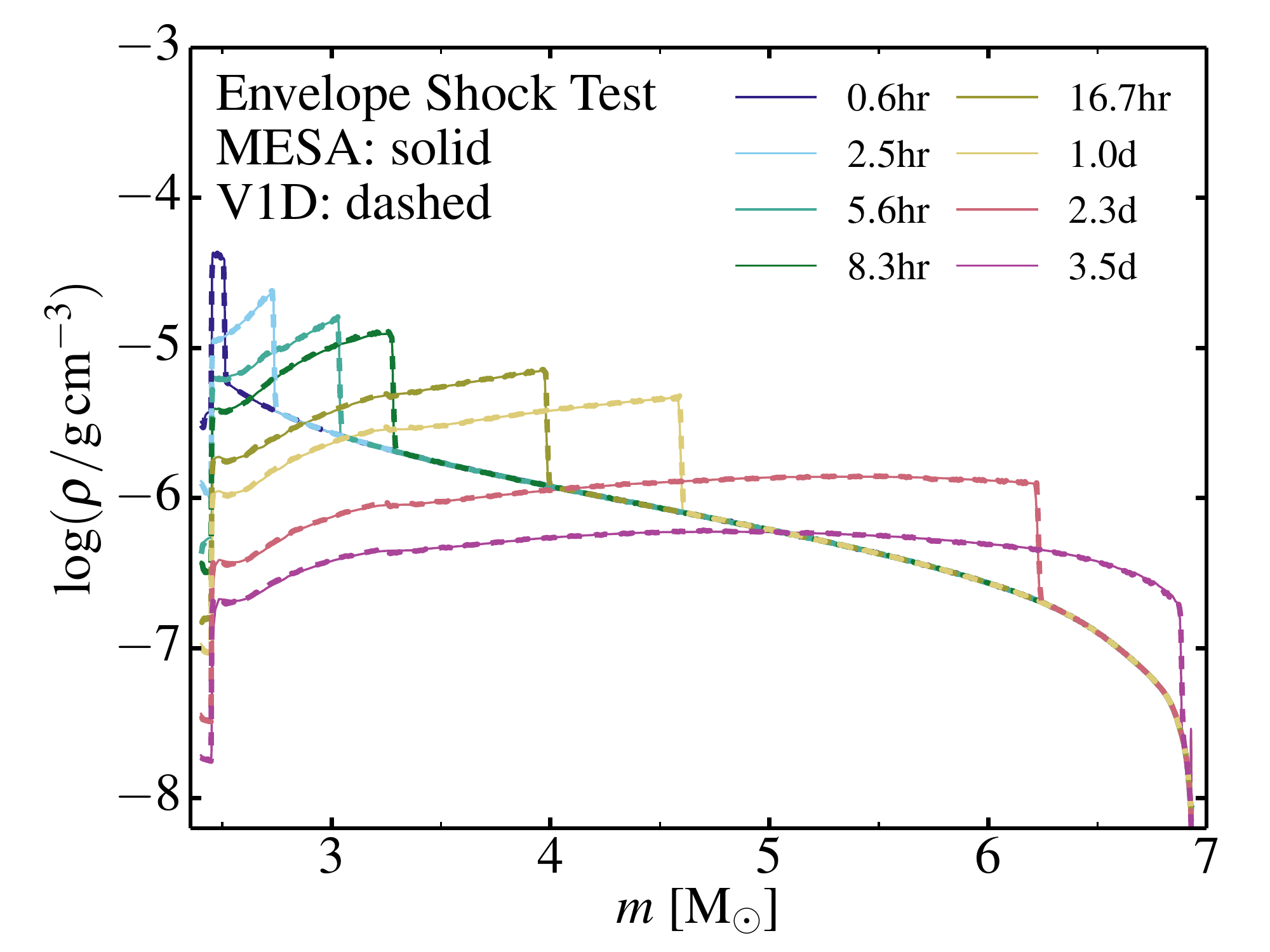}
\includegraphics[width=3.15in]{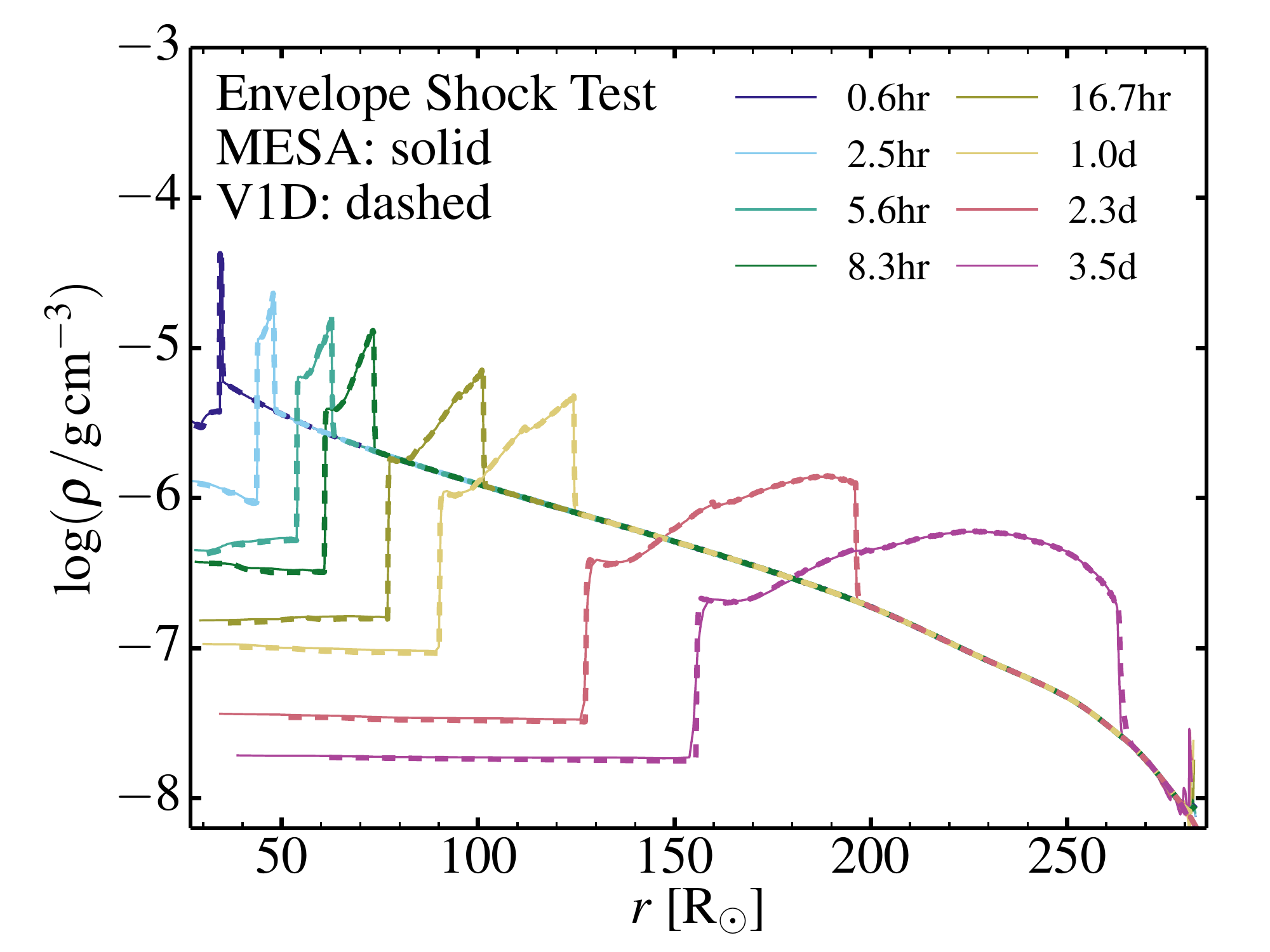}
\includegraphics[width=3.15in]{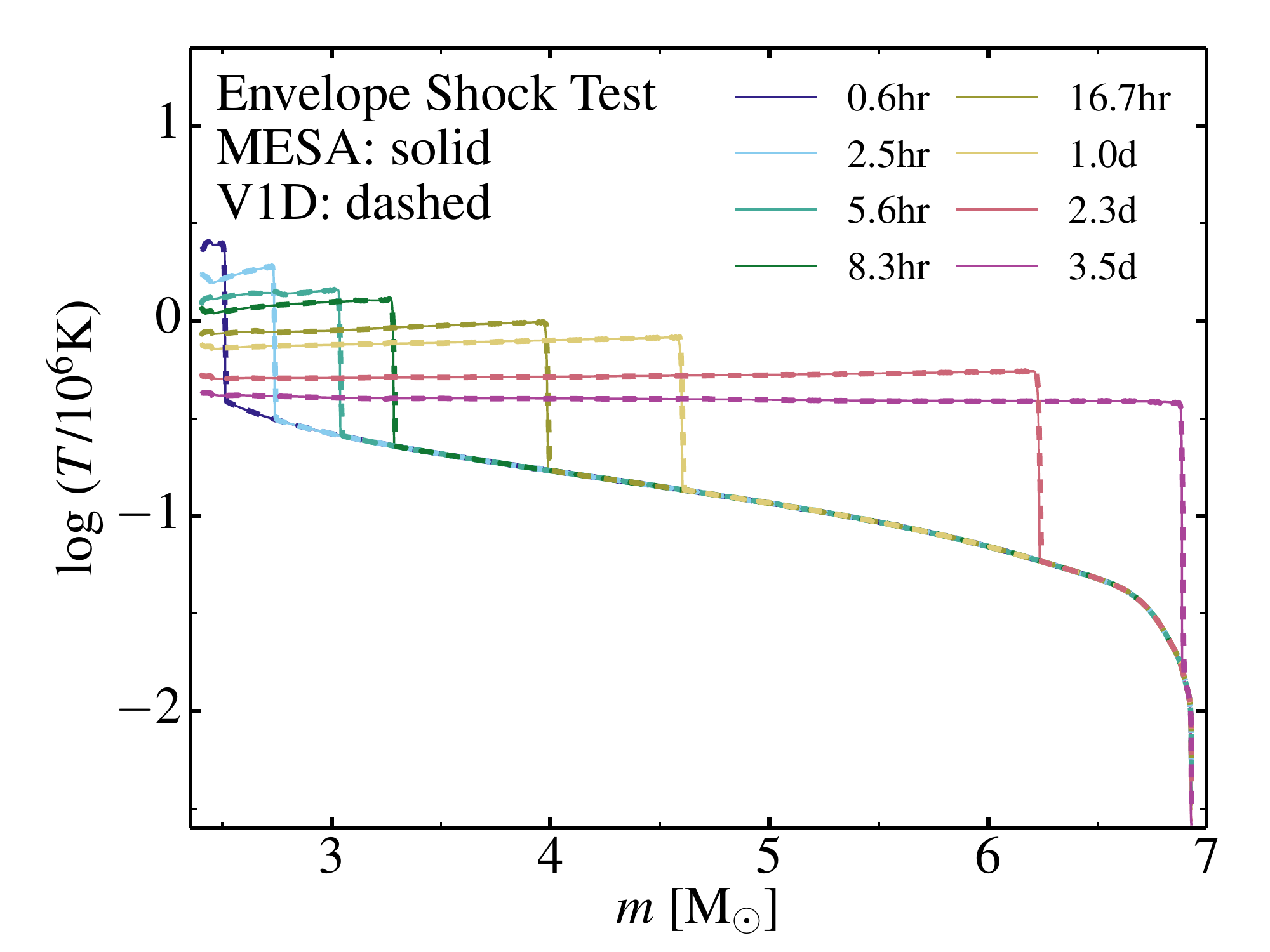}
\includegraphics[width=3.15in]{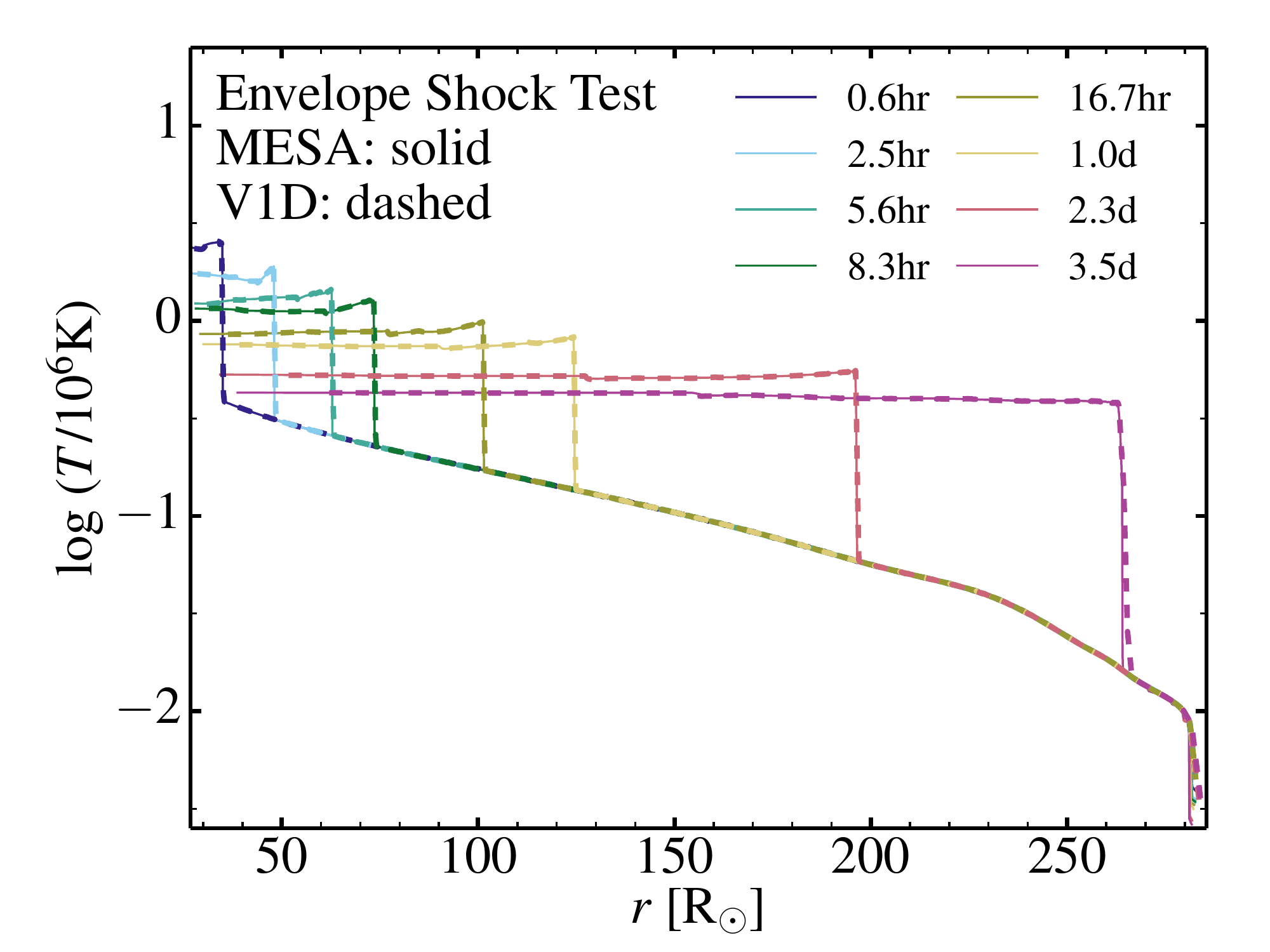}
\includegraphics[width=3.15in]{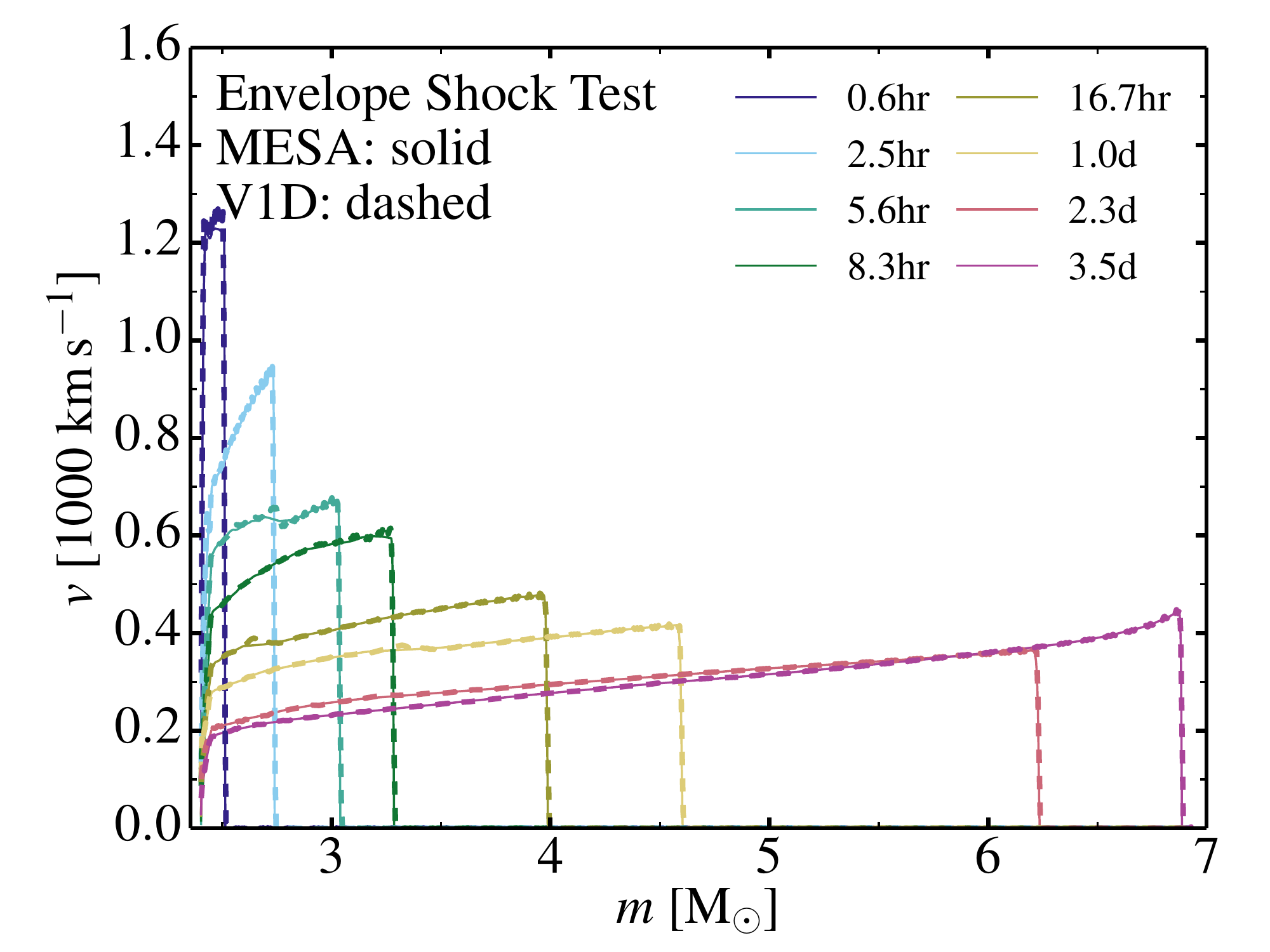}
\includegraphics[width=3.15in]{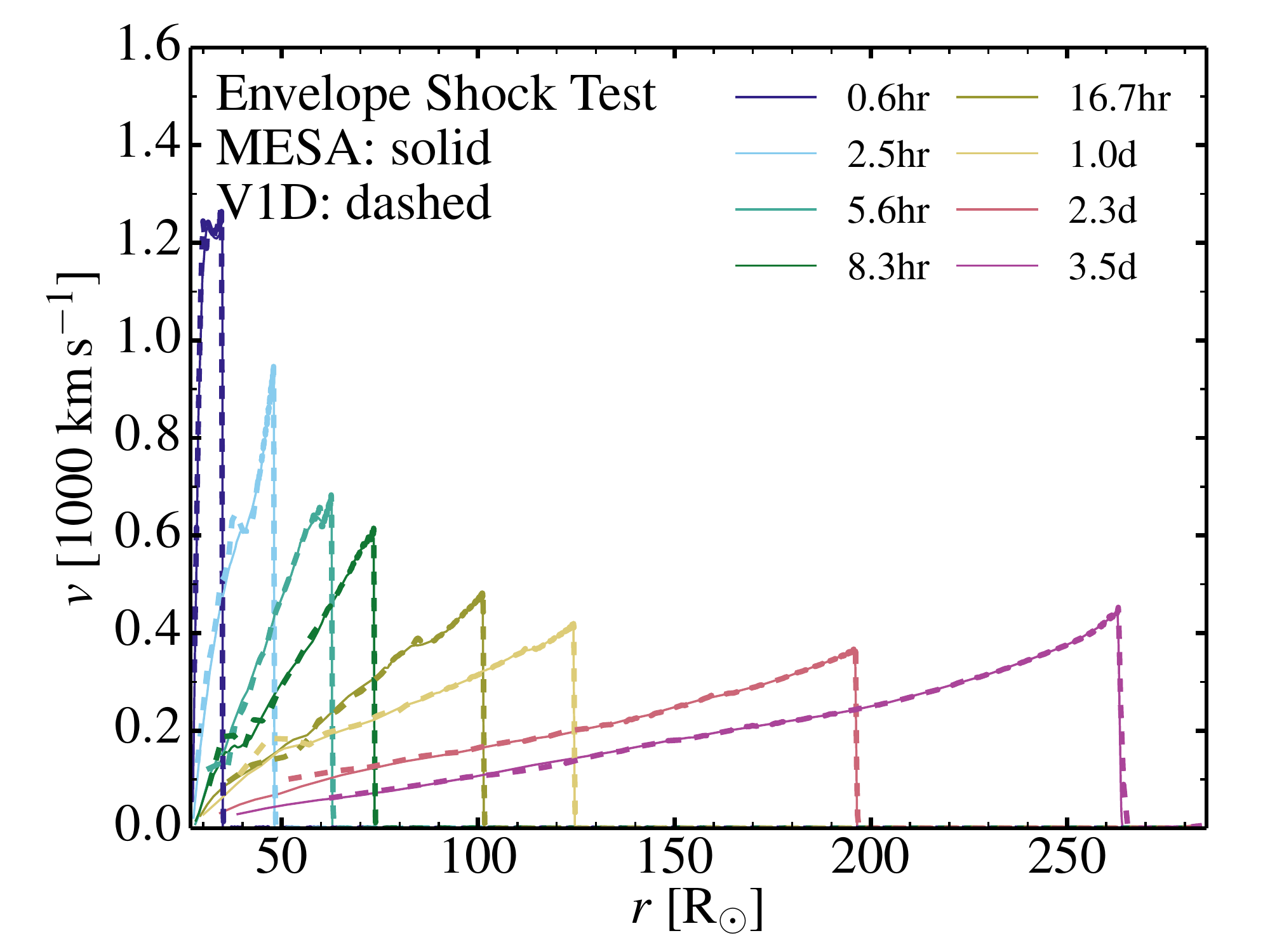}
\caption{Multi-epoch snapshots for the hydrodynamical simulation of a $10^{49}$\,erg shock
in the envelope of a $6.93M_\odot$
AGB star. We show the density (top row), temperature (middle row), and velocity (bottom row),
versus Lagrangian mass (left column) and radius (right column).
In each panel, the solid line refers to the \mesa\ results and the dashed line to the \voned\ results.
\label{f.env_slice}}
\end{center}
\end{figure*}

\subsection{Surface Boundary Conditions}\label{s.hydro_surface_bcs}

\MESA\ provides a variety of options for surface boundary conditions
(see, e.g., section 5.3 of \mesaone), and several more have been added
for use with hydrodynamics.  The simplest allow specification of a
particular value for the surface pressure, the surface temperature, or
the $\Teff$ if the surface is not at the photosphere.  In the case of
a given fixed surface pressure, the corresponding surface temperature
is set using the surface luminosity and radius based on the usual
black body relation. For the second case, where the surface
temperature is fixed, the surface pressure is set to the corresponding
radiation pressure.  For both of these, if the surface is not at the
photosphere, $\Teff$ is set using the Eddington $T$-$\tau$ relation.
Finally, for specified $\Teff$ when the surface is not at the
photosphere, the corresponding surface temperature is also derived
using the Eddington $T$-$\tau$ relation, and the surface pressure is
set to the radiation pressure for that temperature.

For computations involving shocks at the surface, there is an option
to use boundary conditions that specify a vanishing gradient for
compression at the surface and a temperature corresponding to black
body radiation. The outermost cells ($k=1,2$) satisfy the equation
\begin{equation}
\frac {1}{\rho_1} - \frac{1}{\rho_{{\rm start},1}} = \frac {1}{\rho_2} - \frac{1}{\rho_{{\rm start},2}}
\enskip,
\label{e.surfbc}
\end{equation}
which represents the vanishing of the surface compression gradient.

Finally, for computations involving interior shocks but low velocities
at the surface, there is an option to use the surface pressure from
the selected atmosphere prescription with the momentum equation
relating the surface velocity to the surface pressure gradient.  This
form for the surface boundary conditions is used in the shocked
massive star example in Section~\ref{s.ccsn} and in the following
envelope shock test.

\subsection{Shock Test}\label{s.shocks}

To test the implementation, we shock the extended envelope of a
6.93\,\msun\ asymptotic giant branch (AGB) star evolved from a 7\,\msun\
main-sequence star without rotation and an initial metallicity of
0.001.  This case is chosen because of the uniform properties of the
extended envelope (i.e., small density range, smooth density, and
uniform composition).  Our interest is to study the propagation of the
shock, the properties of the shocked material, and the magnitude of
energy conservation errors.  In Section~\ref{s.ccsn}, we present
results that mimic core-collapse supernovae.

Explosion simulations with \mesa\ start from a converged model.  The
core is excised by removing inner shells of the model and setting new
inner boundary conditions for mass, radius, velocity, and luminosity.
For the current test, we remove the center just above the helium core
at a mass of 2.40\,\msun\ which corresponds to an inner radius of
27.2\,\rsun. The stellar surface lies at a radius of 282.7\,\rsun.
During the following evolution, the excised region is treated as a
point mass and is linked to the above layers by the inner boundary
conditions which can be changed at each timestep to simulate various
core events.  The model grid was adjusted at each step to give higher 
resolution in the vicinity of the shock. The total number of cells stayed at about 1000,
with cell masses dropping to about $10^{-4}$ of the total in approximately 100 cells around the shock.

In \MESA, the artificial explosion that creates a shock can be
produced in three ways: a piston, a luminosity flash, or a thermal
bomb:
\begin{itemize}
\item The first option changes the inner boundary conditions for
  velocity and radius to mimic a piston.  A core-collapse supernova
  can be simulated by moving the inner radius inwards (collapse) at a
  free-fall speed and then violently outwards (bounce and explosion).
  The parameterization for the piston-driven explosion is the same as
  that described in \citet{ww95}, and includes the infall piston time,
  the final inward piston radius, the initial outward piston speed,
  and the final piston radius.
\item The second option increases the inner boundary luminosity over a
  specified time in order to deliver the desired total energy. In this
  approach, the inner boundary radius is fixed at all times and
  becomes a zero-flux inner boundary once the flash is over.  In this
  approach, the inner boundary radius is fixed (zero velocity) and we
  inject the energy within the first zone of the domain.
\item The third option deposits energy at a constant rate during a
  specified time and in a region bounded by two specified
  Lagrangian-mass coordinates.  As in the second option, the inner
  boundary radius is fixed at all times.
\end{itemize}
The differences amongst these three options can alter the properties
of the shocked envelope.

To benchmark \mesa\ for these shock tests, we have used the explicit
radiation-hydrodynamics code \voned\ \citep{livne93,DLW10b,DLW10a}.
Options 1 and 3 are implemented in \voned.  For the present envelope
shock test, and subsequently for the explosion tests, we use option 3
in both codes.  We initiate the explosion by depositing a total of
10$^{49}$\,erg at a constant rate over 10\,s between the Lagrangian
mass coordinates of 2.40 and 2.45\,\msun.  This energy deposited is
well in excess of the initial binding energy, which is approximately
$-2\times$10$^{47}$\,erg.  Once the energy injection is over, we save
a model which is then used as initial conditions for a shock evolution
simulation.

Once the stellar core has been excised, the remaining envelope has a
smooth density profile, resembling a power law whose exponent is $-$1
at depth and decreases outwards to become about $-$10 at the
photosphere (top row panels of Figure~\ref{f.env_slice}).  Because
convective accelerations are limited, the energy deposited increases
the internal energy within the innermost 0.05\,\msun\ of the grid. The
pressure build-up leads to the sudden expansion of the innermost
layers and the formation of a mildly supersonic shock (Mach number
$\approx 2$).  The shock propagates at a velocity in excess of
1000\,\kms\ initially, but slows to a few 100\,\kms\ by the time it
reaches the stellar surface after 3$\times$10$^5$\,s.  The density
contrast across this somewhat weak shock is $\approx$\,6.  For a strong
shock, one expects a density jump of 4 for an ideal gas with an
adiabatic index of 5/3 and a value of 7 for a radiation-dominated gas
($\gamma=$\,4/3).

This simulation is analogous to a shock-tube test. However, in the
stellar context (realistic stellar envelope, realistic equation of
state, spherical expansion), there is no analytical solution for
comparison.  We thus run the same simulation with the code \voned\ and
include the results in Figure~\ref{f.env_slice}.  The results agree at
multiple times spanning the progression of the shock towards the
stellar surface (the times used for comparison are the same to within
1\% and the grid resolution is comparable).  The sharpness of the
shock in the two simulations differs with time and location. In
\voned, the artificial viscosity has a physical spread of two grid
zones, irrespective of radius, while in this \mesa\ run, the spread is
set to 0.1\% of the local radius (see Section~\ref{s.hydro_visc}).

\begin{figure}
\begin{center}
\includegraphics[width=3.5in]{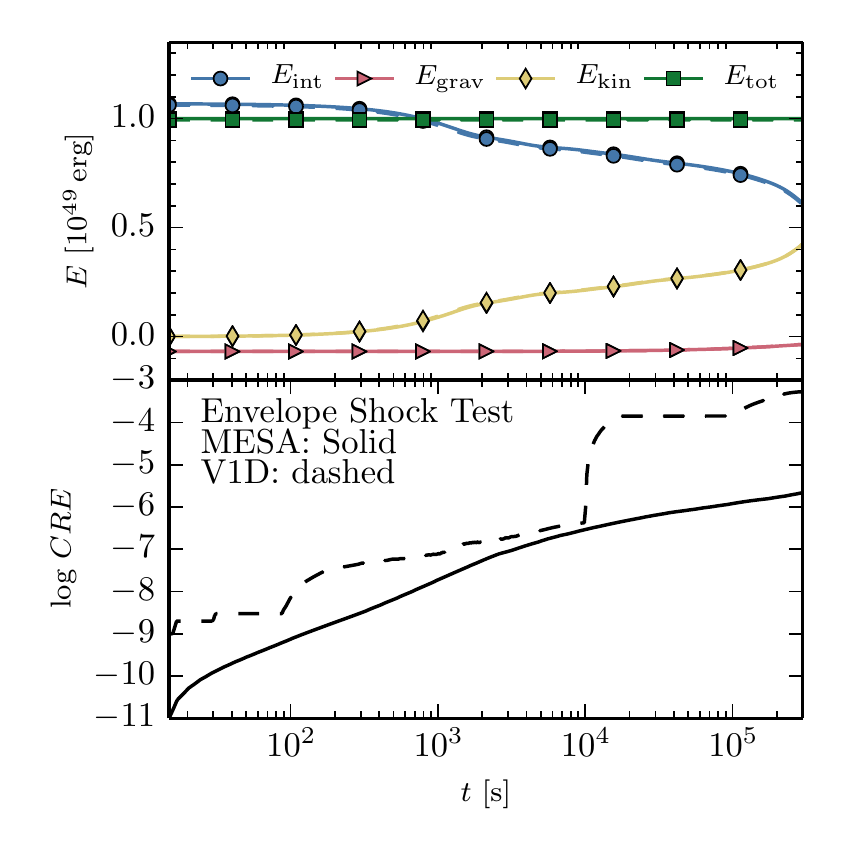}
\caption{
{\it Top:} Evolution of internal energy $E_{\rm int}$, gravitational energy $E_{\rm grav}$,
kinetic energy $E_{\rm kin}$, and their sum $E_{\rm tot}$
for the envelope shock test simulated with \mesa\ and \voned.
{\it Bottom:} Log of cumulative relative error $CRE$ (Equation~\ref{e.cre}) of the total energy $E_{\rm tot}$.
We neglect sources (nuclear burning) and sinks (radiation losses), which are negligible.
\label{f.env_energy}}
\end{center}
\end{figure}

\begin{figure}[!htb]
\begin{center}
\includegraphics[width=3.5in]{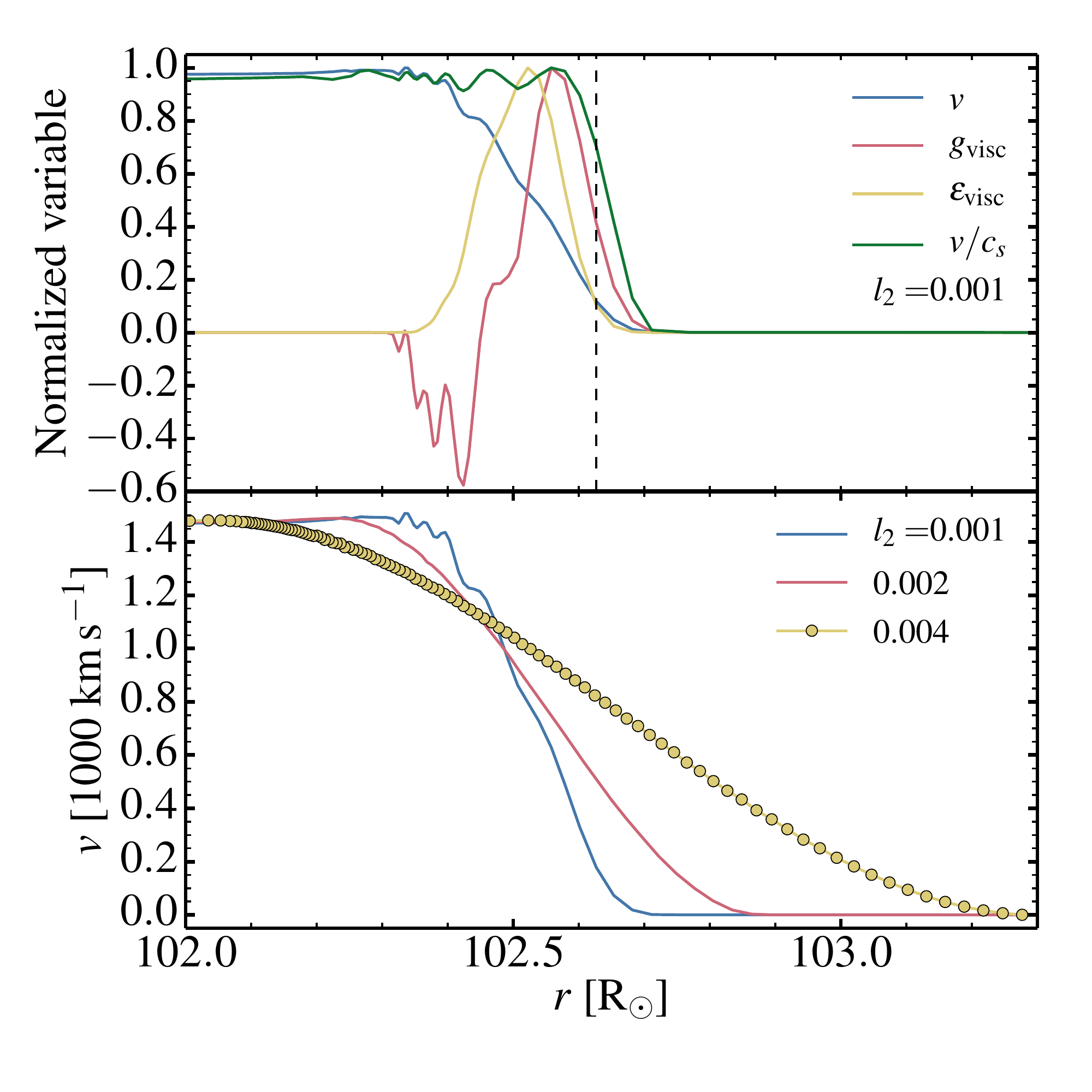}
\caption{
{\it Top:} Normalized values of  velocity,  artificial acceleration $g_{\rm visc}$,
artificial viscous heating $\epsilon_{\rm visc}$, and Mach number in the vicinity of the shock. The dashed vertical line marks where the Mach number is unity.
{\it Bottom:} Dependency of the shock morphology on changes in the viscosity parameter $l_2$.
The dots shown for the model with $l_2=$\,0.004 denote the location of the \mesa\ grid points at that time.
For all these tests, we deposit an energy of 10$^{50}$\,erg at a constant power over 1\,s.
\label{f.env_visc}}
\end{center}
\end{figure}

Since the explosion is started as a thermal bomb, the bulk of the
energy is initially internal, see Figure~\ref{f.env_energy}.  As the
material expands and accelerates, the kinetic energy increases,
mirroring the decrease in internal energy (essentially no energy is
used to unbind the envelope).  At the time of shock emergence, the
internal and kinetic energies are comparable.

In the present case, we can preserve good accuracy while still
allowing time steps an order of magnitude greater than the Courant
time.\footnote{The Courant time, equal to the minimum sound crossing
  time through a grid zone. In this envelope test, it is of the order
  of 10\,s initially, increasing progressively to 40\,s prior to shock
  emergence.} The error in energy conservation at timestep $i$ is
\begin{equation}
\delta E_{\mathrm{error}, i} = E_{i} - E_{i-1} - E_{\mathrm{sources}, i}
\end{equation}
where $E_{\mathrm{sources}, i}$ is the right hand side of
Equation~\eqref{e.econserv} multiplied by $\delta t$.  The cumulative
relative error in energy at a timestep $n$ is
\begin{equation}
CRE(t_n) = \frac{1}{E_n} \sum_{i=1}^{n} \delta E_{\mathrm{error},i}~.
\label{e.cre}
\end{equation}
In the test case, after about 15,100 timesteps when the shock reaches
6.6 \msun, the cumulative relative error has grown to about
$-1.4\times 10^{-6}$, corresponding a roughly linear growth rate of
about $-1\times 10^{-10}$ per step (bottom panel in
Figure~\ref{f.env_energy}).  Note that at this stage of evolution the
shock is nearing the outer edge of the envelope but has not actually
broken out through the surface.  Issues of shock break out are beyond
the scope of the current implementation.
Using the parameters selected for the test, the energy conservation
with \voned\ is not as good as with \mesa\ (the jumps in cumulative
error correspond to times when the limit on the time step are
loosened); comparable accuracy can be obtained with \voned\ by
reducing the explicit time step well below the Courant limit.


Finally, to illustrate the effects of artificial viscosity, we vary
the quadratic term $l_2$ that controls the spread of the shock in
response to compression (see Equation~\ref{e.etaviscquad}), with the
explosion energy increased to 10$^{50}$\,erg in order to produce a
stronger shock, and otherwise the same parameters and initial
conditions.  The top panel of Figure~\ref{f.env_visc} shows the
artificial acceleration ($g_{\rm visc}$) and energy
($\epsilon_{\rm visc}$) terms that enter the momentum and the energy
equations for $l_2=$\,0.001.  The acceleration term is positive
ahead of the shock, causing a pre-acceleration of the unshocked
material, and negative behind the shock causing a deceleration of the
post-shock material.  The energy corresponding to those changes in
momentum is balanced by the extra term for artificial viscous heating
in the energy equation ($\epsilon_{\rm visc}$).  The lower panel of
Figure~\ref{f.env_visc} shows the expected increase in the width of
the shock as we raise the parameter $l_2$. For the model with
$l_2=$\,0.004, dots locate grid cells.  Note that with the smallest
value ($l_2=$\,0.001), the velocity is showing small oscillations
(``ringing'') behind the shock indicating that we have reached a
practical lower bound for the shock spread given the other parameter
choices and the nature of the specific problem.



\section{Advanced Burning}\label{s.advburn}

For the advanced stages of stellar burning, we show here that more
accurate summations yield more efficient time integrations.  This
development allows \MESA \ to use large in-situ reaction networks. It
offers an improvement by providing a single solution methodology that
avoids the challenges of stitching together different solution methods
such as nuclear statistical equilibrium (NSE) or co-processing a reaction
network.  We discuss this development and apply it to the
evolution of an X-ray burst on a neutron star.  In Section~\ref{s.ccsn} we discuss the
pre-supernova progenitors and combine the new capability for advanced
burning with the implicit hydrodynamics module to discuss the
explosion of core-collapse supernovae.

The equations that describe the continuum limit
of reacting nuclei are
\begin{align}
\begin{split}
\label{e.nucode}
{\dot Y_i} & = \sum_j c_i \lambda_j Y_j +
\sum_{j,k} \frac {c_i}{|c_j|! |c_k|!} \lambda_{j,k} Y_j Y_k \\
& \quad \; + \sum_{j,k,l} \frac {c_i}{|c_j|! |c_k|! |c_l|!} \lambda_{j,k,l} Y_j Y_k Y_l
\enskip,
\end{split}
\end{align}
where $Y_i$ is the abundance of isotope $i$, $\lambda$ is a reaction rate,
and the three sums are over reactions which
produce or destroy a nucleus of species $i$ with 1, 2, and 3 reacting
nuclei, respectively \citep[e.g.,][]{meyer_1998_aa, hix_2006_aa, guidry_2013_aa, longland_2014_aa}.
The positive or negative stoichiometric coefficients $c_i$ account for
the numbers of nuclei created or destroyed in a reaction.  The
factorials in the denominators avoid double counting of identical
particles.

\begin{figure}[!htb]
\begin{center}
\includegraphics[width=\apjcolwidth]{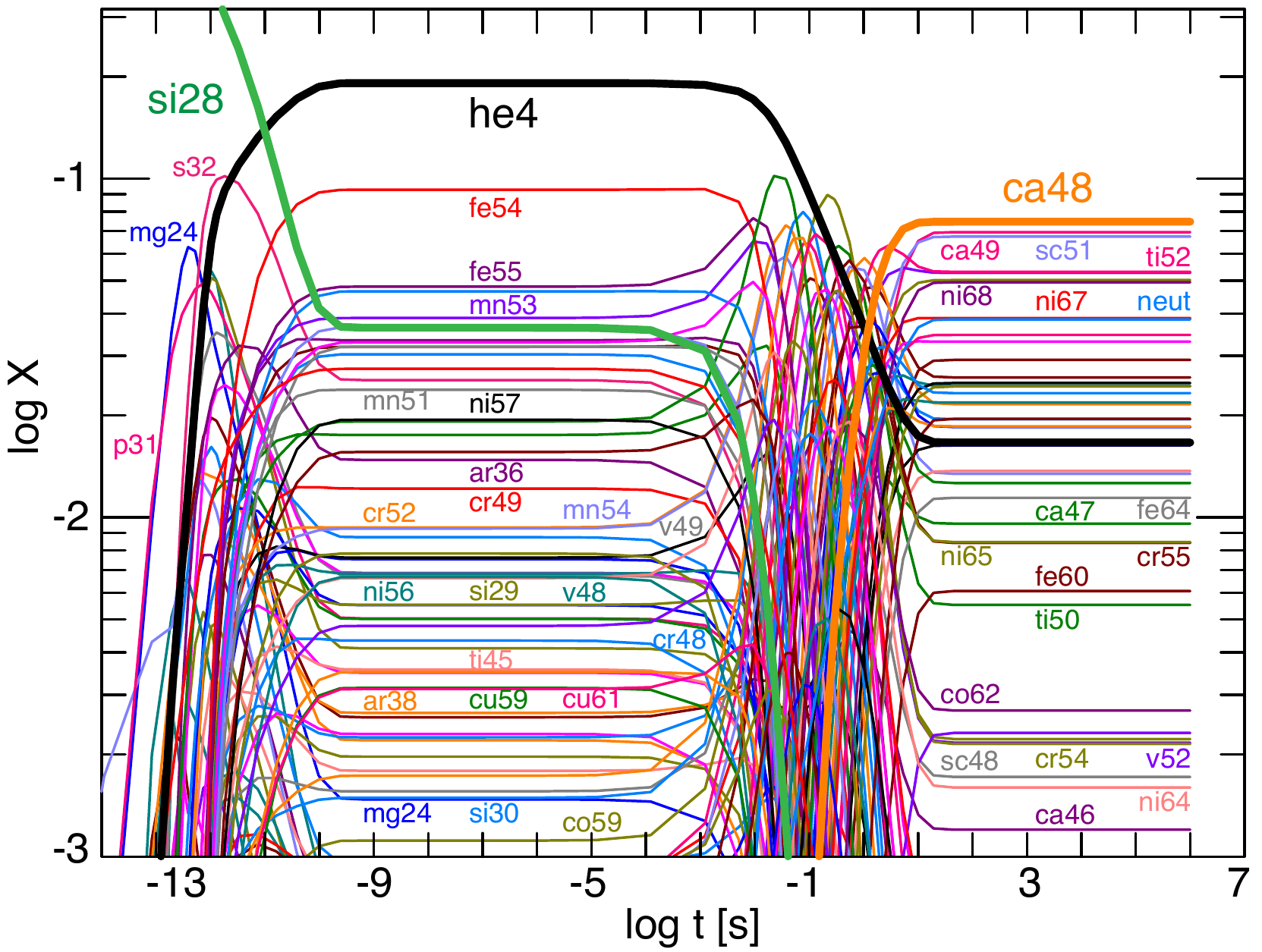}
\caption{
Evolution of the composition 
for a one-zone burn at constant T=9.6$\times$10$^9$ K and
$\rho$=6.0$\times$10$^9$ g cm$^{-3}$ for 10$^6$ s starting with a pure
$^{28}$Si composition. The calculation uses the {\tt mesa\_204.net}
isotope listing (see Table \ref{t.netlist}), the most abundant isotopes 
are drawn with thick lines, and several
isotopes are labeled. The initial composition is quickly erased as
NSE for $\Ye \approx$ 0.5 is established
by $\approx$ 10$^{-8}$ s. Several orders of magnitude in time pass before
weak reactions drive a second period of rearrangement. By $\approx$ 10 s
a second NSE quiescent period with $\Ye \approx$ 0.403
is established. 
}
\label{f.m204evol}
\end{center}
\end{figure}

Figure \ref{f.m204evol} shows the evolution of the mass fractions for
a \mesa\ one-zone burn at constant $T$=9.6$\times$10$^9$ K and
$\rho$=6.0$\times$10$^9$ g cm$^{-3}$ for 10$^6$ s starting with a pure
$^{28}$Si composition. The 204 isotope network, {\tt mesa\_204.net},
used in the calculation is listed in Table \ref{t.netlist}, and
includes the isotopes identified in \citet{heger_2001_aa} as important
for $\Ye$ in core-collapse models.  The thermonuclear reaction rates
are from JINA reaclib version V2.0 2013-04-02 \citep{cyburt_2010_aa}.
Implementation of reaction rates and associated quantities are
described in \mesaone \ and \mesatwo.

The thermodynamic conditions used in Figure \ref{f.m204evol} are
representative of the central regions of massive stars during
the advanced stages of evolution.  At such temperatures the initial
composition of pure \silicon[28] undergoes a rapid
readjustment.  The timescale for the initial $\Ye$ $\approx$ 0.5
composition to relax to an NSE composition is roughly \hbox{$\tau_{{\rm nse}} \approx \rho^{1/5}
  \exp(179.7/T_9 - 40.5)$ s} = $3\times 10^{-8}$ s
\citep{khokhlov_1991_aa,calder_2007_aa}, commensurate with the first
burning phase in Figure \ref{f.m204evol}.  Between $\approx$ 10$^{-8}$
s and $\approx$ 10$^{-4}$ s the isotopes $^{4}$He and $^{54}$Fe
dominate the $\Ye$ $\approx$ 0.5 NSE composition.  Since $T$ and
$\rho$ are constant, only changes to $\Ye$ can change the abundances.
A second period of intense rearrangement begins at $\approx$ 10$^{-4}$
s and ends at $\approx$ 10 s. This activity is driven primarily by
$p(e^-,\nu)n$ and $n(e^+,{\bar \nu})p$ and other weak reactions that
change $\Ye$.  Beyond $\approx$ 10 s the isotopes $^{48}$Ca,
$^{49}$Ca, and $^{51}$Sc dominate the $\Ye \approx$ 0.403 NSE
composition.

Table \ref{t.yelist} shows the sensitivity of the final $\Ye$ in this
calculation to the number of isotopes in the network. Each
successively larger network encompasses the previous smaller network
and was crafted to yield approximately the same final $\Ye$ value as
given by the largest network.  The 204 isotope network used in Figure
\ref{f.m204evol} is in the regime where larger networks give the same
final $\Ye$ to 3 significant figures.

\begin{deluxetable}{cccccc}
\tablecolumns{6}
\tablewidth{0pt}
\tablecaption{204 Isotope Network Listing.\label{t.netlist}}
\tablehead{\colhead{Element} & \colhead{$A_{{\rm min}}$} & \colhead{$A_{{\rm max}}$} &
           \colhead{Element} & \colhead{$A_{{\rm min}}$} & \colhead{$A_{{\rm max}}$} }
\startdata
n  &    &    &  S  & 31 & 37 \\
H  &  1 &  2 &  Cl & 35 & 38 \\
He &  3 &  4 &  Ar & 35 & 41 \\
Li &  6 &  7 &  K  & 39 & 44 \\
Be &  7 & 10 &  Ca & 39 & 49 \\
B  &  8 & 11 &  Sc & 43 & 51 \\
C  & 12 & 13 &  Ti & 43 & 54 \\
N  & 13 & 16 &  V  & 47 & 56 \\
O  & 15 & 19 &  Cr & 47 & 58 \\
F  & 17 & 20 &  Mn & 51 & 59 \\
Ne & 19 & 23 &  Fe & 51 & 66 \\
Na & 21 & 24 &  Co & 55 & 67 \\
Mg & 23 & 27 &  Ni & 55 & 68 \\
Al & 25 & 28 &  Cu & 59 & 66 \\
Si & 27 & 33 &  Zn & 59 & 66 \\
P  & 30 & 34 &     &    &    \\
\enddata
\end{deluxetable}

\begin{deluxetable}{cccc}
\tablecolumns{4}
\tablewidth{0pt}
\tablecaption{Final $\Ye$ for Figure \ref{f.m204evol}\label{t.yelist}}
\tablehead{\colhead{\# of Isotopes} & \colhead{$\Ye$} & \colhead{$Z_{{\rm max}}$} & \colhead{$A_{{\rm max}}$} }
\startdata
75   & 0.4093 & Ni & 68 \\
125  & 0.4065 & Ni & 68 \\
160  & 0.4032 & Ni & 68 \\
204  & 0.4032 & Zn & 66 \\
368  & 0.4035 & Zn & 77 \\
833  & 0.4029 & Sn & 125 \\
3298 & 0.4039 & At & 211 
\enddata
\end{deluxetable}

\begin{deluxetable*}{cclcrr}
\tablecolumns{6}
\tablewidth{0pt}
\tablecaption{Results of Summation Experiments.\label{t.sumexp}}
\tablehead{\colhead{IEEE}         & \colhead{Maximum} & \colhead{Strategy} & \colhead{Minimum}                          & \colhead{Number of}                  & \colhead{Ratio of} \\
           \colhead{Arithmetic}   & \colhead{Digits Compared}  &                    & \colhead{Correct Digits \tablenotemark{a}} & \colhead{Timesteps\tablenotemark{b}} & \colhead{CPU Times\tablenotemark{c}}}
\startdata
$\tau_{\rm int}$=10$^{-4}$  \quad $y_{\rm scale}$=10$^{-3}$ & & & & \\
  64-bit  & 16 & in order given      & 6  & 3062 & 31.7 \\
  64-bit  & 16 & sorted, ascending   & 7  & 2614 & 24.4 \\
  64-bit  & 16 & sorted, Kahan sum   & 8  & 1141 & 13.1 \\
  128-bit & 32 & in order given      & 21 & 55 &    1.0 \\
  128-bit & 32 & sorted, ascending   & 22 & 55 &    1.0 \\
$\tau_{\rm int}$=10$^{-6}$  \quad $y_{\rm scale}$=10$^{-5}$ & & & & \\
  64-bit  & 16 & in order given      & 6  &  10081 & 156 \\
  64-bit  & 16 & sorted, ascending   & 7  &  7972  & 123 \\
  64-bit  & 16 & sorted, Kahan sum   & 8  &  7674  & 112 \\
  128-bit & 32 & in order given      & 21 &    88  & 1.0 \\
  128-bit & 32 & sorted, ascending   & 22 &    88  & 1.0 \\
\enddata
\tablenotetext{a}{Relative to the 100 digit sum by the {\tt MP} and {\tt MPf90} multiple precision packages.}
\tablenotetext{b}{For a Bader-Deuflard integrator in IEEE 64-bit arithmetic.}
\tablenotetext{c}{For a single thread on one 2.7 GHz Intel Xeon E5 core with the Intel 15.0.1 Fortran compiler,
and relative to the execution time for the integration with 128-bit summations with terms in the order given.
}
\end{deluxetable*}

\subsection{More Accurate Summations Yield More Efficient Integrations} \label{s.summation}

We now test different summation methods for Equation (\ref{e.nucode})
and demonstrate that improved accuracy of the summations reduces the number
of time steps with a commensurate reduction in the execution time $-$
while producing the same answers to within the specified integration
accuracy.

When the summations in Equation (\ref{e.nucode}) are accumulated in IEEE
64-bit arithmetic (16 significant figures, \code{real*8} precision in Fortran
on most architectures; more specifically binary64 with round to nearest 
and round ties to even) the integration in Figure \ref{f.m204evol} takes
3062 time steps using a variable-order Bader-Deuflhard integrator with
a specified accuracy of $\tau_{\rm int}$=10$^{-4}$ and a scaling
value $y_{\rm scale}$=10$^{-3}$.  The specified accuracy
$\tau_{\rm int}$ limits the maximum error over one time step for any
isotope. Other potential, but less demanding, choices for the meaning
of $\tau_{\rm int}$ include limiting the average or root-mean-square
error over one time step for all isotopes.  When an abundance is
greater than $y_{\rm scale}$ a relative error is calculated, while for
abundances smaller than $y_{\rm scale}$, the absolute error is
calculated \cite[e.g.,][]{press_1992_aa}.  In essence, only abundances
greater than $y_{\rm scale}$ can exert control on the size of the time step.

When the summations are accumulated in IEEE 128-bit arithmetic (32
significant figures, \code{real*16} precision in Fortran on most
architectures), the same integration takes only 55 time steps, a
factor $\approx$50 improvement in the number of time steps, and a
factor of $\approx$30 less execution time.  Both calculations returned
the same answers to within the specified integration error
tolerances. For tighter integration tolerances of $\tau_{\rm int}$=10$^{-6}$ 
and $y_{\rm scale}$=10$^{-5}$, the evolution with
summations in IEEE 64-bit arithmetic takes 10,081 time steps while the
evolution with summations in IEEE 128-bit arithmetic takes 88 time
steps. This is a factor of $\approx$100 improvement in the number of
time steps, a factor of $\approx$150 in execution time, with both
calculations again producing the same abundances to within the
specified integration error tolerances.  Both sets of integration
tolerances are practical, everyday usage tolerances; they are not
extreme cases of hypothetical interest only.  Using low-order
Rosenbrock and first-order Euler integrators also showed similar
improvements in the number of time steps when the summations were
performed in IEEE 128-bit arithmetic instead of IEEE 64-bit
arithmetic.  We achieve a reduction in the number of timesteps and
execution times regardless of the number of isotopes, choice of
integrator, integration tolerances, or linear algebra solver.  This
improvement in efficiency is fundamentally driven by a reduction in
the numerical noise of the function being integrated.

At temperatures larger than $\approx$ 5$\times$10$^9$ K, integrating
Equation (\ref{e.nucode}) 
can be challenging as terms in the summation usually
become large and opposite in sign. As shown above, the classic symptom
during an integration under these thermodynamic conditions is the
integrator taking an excessive number of very small time steps in
order to satisfy the specified integration accuracy criteria.
The traditional workaround to this numerical problem is abandoning a
network integration at elevated temperatures and deploying equilibrium
solution methods.  This switching of methods raises its own numerical
issues when used within the larger context of multi-dimensional
simulations or stellar evolution models (see Section~\ref{s.nonse}).

Unless precautions are taken the summation of
large sets of numbers can be very inaccurate due to the accumulation
of rounding errors. Methods for accurate summation within the bounds
of a given arithmetic remain an active field of research
\citep[e.g.,][]{demmel_2003_aa, mcnamee_2004_aa, ogita_2005_aa,
  rump_2008_aa, graillat_2012_aa, collange_2014_aa}.  These summation
discrepancies also worsen on heterogeneous architectures $-$ such as
clusters with NVIDIA GPUs or Xeon Phi accelerators $-$ which
combine programming environments that may obey various floating-point
models and offer different precision results.

The summations in Equation (\ref{e.nucode}) for the neutron, proton,
and $\alpha$-particle abundances are especially prone to inaccuracies
because every isotope in a network reacts with these three particles.
We report on the summation
errors for these three isotopes.  Each term in the summations of
Equation (\ref{e.nucode}) is calculated using IEEE 64-bit arithmetic
and then copied into a IEEE 128-bit variable using the
Fortran promotion rules.  Each IEEE 128-bit term is then imported into
the {\tt MP} \citep{brent_1978_aa} and {\tt MPf90}
\citep{bailey_1995_aa} multiple precision packages.  All other aspects
of the integration were executed in IEEE 64-bit arithmetic.  The
summations are then accumulated with
\begin{itemize}[nolistsep]
\item IEEE 64-bit: terms in the order as given.
\item IEEE 64-bit: terms sorted \add{by their absolute value} in ascending order.
\item IEEE 64-bit: terms sorted \add{by their absolute value} in ascending order and the \citet{kahan_1965_aa} algorithm,
which reduces the numerical error in summation by retaining a separate variable to accumulate
the errors.
\item IEEE 128-bit: terms in the order as given.
\item IEEE 128-bit: terms sorted \add{by their absolute value} in ascending order.
\item {\tt MP} and {\tt MPf90} 100 digits: terms \add{sorted by their absolute value} in ascending order.
\end{itemize}
There are many summation methods and alternative multiple precision packages
we did not deploy in these studies \citep[e.g.,][]{knuth_1997_aa,
higham_2002_aa, li_2002_aa, muller_2010_aa, collange_2014_aa}.

Table \ref{t.sumexp} summarizes these summation
experiments.  We confirm that 100 digits are sufficient to prevent
errors in our multiple precision sums.  Column 4 gives the minimum
number of correct digits in a summation.  There are three time periods
in the evolution of Figure \ref{f.m204evol} where summations performed
in IEEE 64-bit arithmetic greatly increase the number of time steps
taken by the integration.  One is during the first rearrangement into
the NSE state ending around 10$^{-8}$ s, another is during the second
rearrangement around 10$^{-1}$ s and the third time period is when the
abundances do not change much (${\dot Y} \approx$ 0) and reaction
rates reach equilibrium. It is during these equilibrium periods where
a summation in IEEE 64-bit arithmetic with the terms summed in the
order given may yield only 6 accurate digits. As a result, 3062 time
steps are needed to complete the integration (e.g., row 2 of Table
\ref{t.sumexp}). It is important to note that this strategy and choice
of arithmetic is commonly used by nuclear reaction networks
\citep[e.g.,][]{timmes_1999_ab} $-$ and the most inaccurate choice.
Row 4 of Table \ref{t.sumexp} is an important case, sorted plus Kahan
summation, because it demonstrates that a marginal improvement in the
accuracy of the summation (8 minimum correct digits) has a major
reduction on the number of time steps (1174 time steps) and execution
time (a factor of $\approx$2.5 smaller).  This establishes the general
trend that improved accuracy of the summations reduces the number of
time steps with a commensurate reduction in the execution time $-$
while producing the same answers to within the specified integration
accuracy.

The left hand side of Equation (\ref{e.nucode}) for $\dot Y_i$ is a IEEE
64-bit array to be filled with one of the summations. Setting $\dot
Y_i$ equal to one of the IEEE 128-bit summations (at least 22 digits
of accuracy) or the multiple precision package summations gives
the most efficient integration (55 time steps, rows 5 and 6 in
Table \ref{t.sumexp}) because the Fortran precision demotion rules
assure $\dot Y_i$ is accurate to the limit of IEEE 64-bit arithmetic.
The next best strategy, but a distant second, is setting $\dot Y_i$
equal to the sorted, Kahan summation. The worst case is setting $\dot
Y_i$ equal to the 64-bit arithmetic sum with the terms in the
order they appear $-$ which is a common approach \citep[e.g.,][]{timmes_1999_ab}.

\begin{figure}[!htb]
\includegraphics[width=\apjcolwidth]{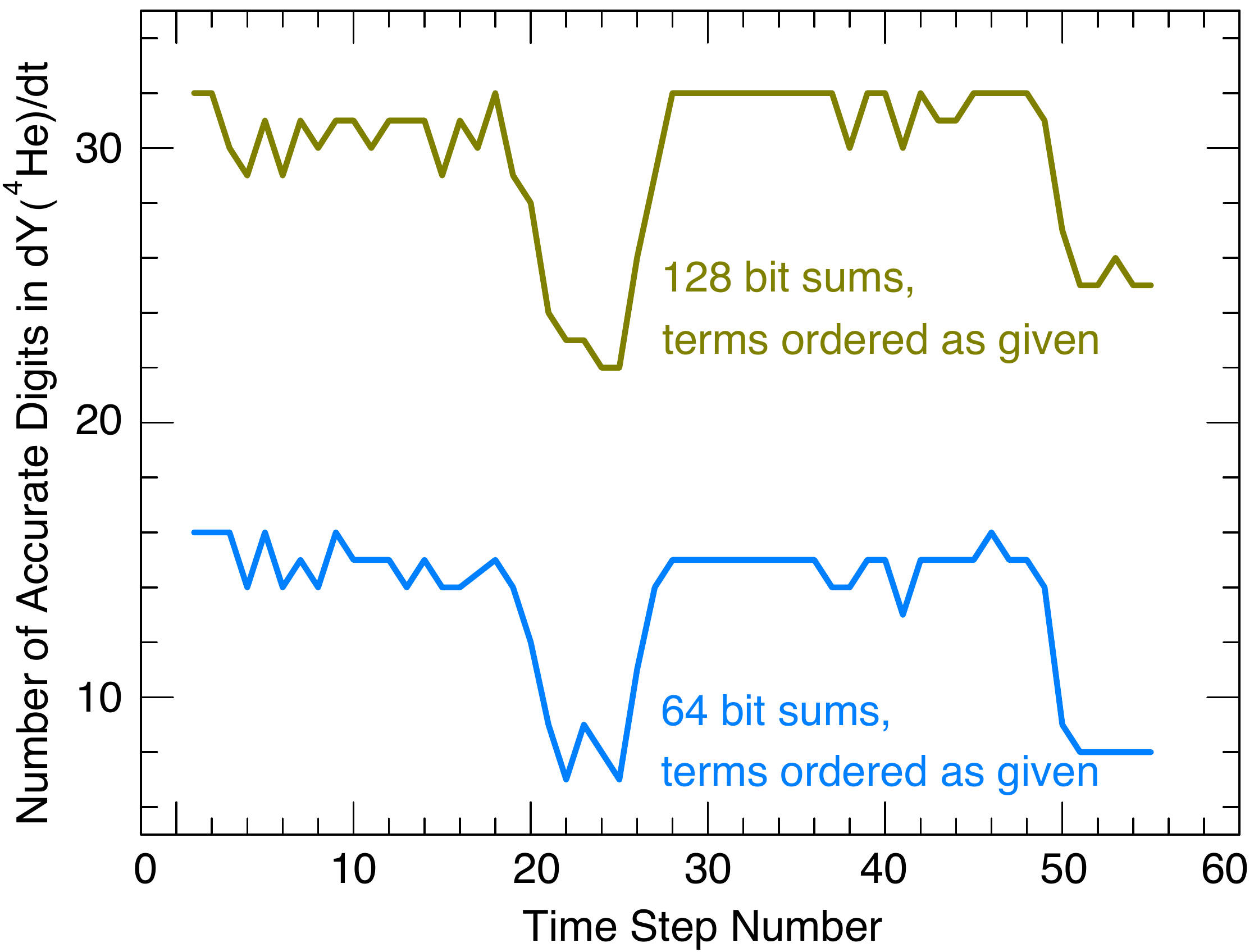}
\caption{
Number of accurate digits in 64-bit and 128-bit summations for
$\dot{Y}(^4\rm He)$ as measured by the 100 digit sum calculated by the
multiple precision packages.  The x-axis gives the time step number
for the integration done with IEEE 128-bit summations.  The number of
accurate digits in $\dot Y(\rm p)$ and $\dot Y(\rm n)$ are within a
few digits of $\dot{Y}(^4\rm He)$.
}
\label{f.r16mr8}
\end{figure}

Figure \ref{f.r16mr8} shows the number of correct digits in 64-bit and
128-bit summations for $\dot{Y}$($^4$He) with the terms
accumulated in the order they are given. The number of correct digits
is measured against the 100 digit sum calculated by the multiple
precision packages {\tt MP} and {\tt MPf90}.
  The
choices for the integrator and integration tolerances are the same as
in Figure \ref{f.m204evol}.
Figure \ref{f.r16mr8} shows that the minimum number of accurate digits
is usually within a few digits of the limit of IEEE 64-bit arithmetic,
but degrades to 6 digits (see row 1 of Table \ref{t.sumexp}) during a
time period of intense isotope rearrangement.  These relatively large
inaccurate summations cause the right hand side of Equation
(\ref{e.nucode}) to be poorly defined in IEEE 64-bit arithmetic.  As a
direct result, the integration of Equation (\ref{e.nucode}) with IEEE
64-bit summations takes 3062 time steps to complete.  Sorting the
terms in the sums in ascending order and using the Kahan summation
algorithm results in an accurate digit pattern that is very similar
except the number of accurate digits is improved by one or two (see
row 3 of Table \ref{t.sumexp}). As a direct result of the improved
accuracy of the summations, the number of timesteps is reduced from
3062 to 1141.

For the IEEE 128-bit sum relative to the 100 digit sum in Figure
\ref{f.r16mr8}, the minimum number of accurate digits is usually
near the limit of IEEE 128-bit arithmetic, but degrades to
21 digits (see row 5 of Table \ref{t.sumexp}) during the second period
of intense isotope rearrangement.  Relative to the IEEE 64-bit
summations the number of correct digits is improved by at least 15,
consistent with our conversion of the IEEE 64-bit terms in the sum to
IEEE 128-bit using the Fortran promotion rules.  As a direct result of
the improved accuracy of the summations the integration that used the
IEEE 128-bit summation took only 55 timesteps to complete the
evolution.

We found no notable improvements by increasing the accuracy of the
summations for the Jacobian matrix used by the stiff ordinary
differential equation integrators. For this problem, it is evidently
more important to better define the function $-$ the right hand side
of Equation (\ref{e.nucode}) $-$ than the Jacobian matrix holding the
derivatives of the function.

Based on these experiments we currently choose to improve the accuracy
of the summations by converting the terms in the order they appear
from IEEE 64-bit to IEEE 128-bit using the Fortran promotion rules and
adding the terms in IEEE 128-bit arithmetic.  IEEE 128-bit precision
is presently almost always implemented in software by a variety of
techniques (e.g., double-double methods), since direct hardware
support for IEEE 128-bit precision is presently rare.  However, Table
\ref{t.sumexp} shows the reduction in the number of time steps from
accumulating the sums in  IEEE 128-bit arithmetic far exceeds the
extra computational cost per addition.

\subsection{A Uniform Solution Method for Nuclear Burning in Stellar Evolution} \label{s.nonse}

At high temperatures, the traditional workaround
for the
numerical problem of inaccurate summations in IEEE 64-bit arithmetic
is to forgo using a reaction network integration to evolve the
abundances and nuclear energy generation rate and to replace it with
equilibrium solution techniques.  An example of such an equilibrium
calculation is NSE, where a root-find for the neutron and proton
chemical potentials is performed. Once these two chemical potentials
are known, all the abundances can be determined from nuclear Saha
equations
\citep[e.g.,][]{clifford_1965_aa,hartmann_1985_aa,meyer_1998_aa,nadyozhin_2004_aa,seitenzahl_2008_aa,odrzywolek_2012_aa}.
Equilibrium solution methods by themselves are efficient, robust, and
inexpensive.

However, combining reaction networks and equilibrium solution methods creates its own
numerical issues, especially when the temperature and density are
spatial and time dependent.  For example, the temperature of a cell
may start relatively low, move into quasi-static
equilibrium (QSE) range above 3$\times$10$^9$ K, and then
move into NSE range above 5$\times$10$^9$ K.
Ad-hoc decision trees
must be created for
switching between a network integration, QSE solutions, and NSE
solutions.
These switches can introduce
unphysical discontinuities in the abundances either from one timestep
to the next or in the abundance spatial profiles from one
cell to the next.

Furthermore, cells near the transition between a
network integration and an equilibrium method can be unstable in the
sense that the equilibrium solution can evolve a cell to lower
temperatures pushing the cell into using a network integration, while
the solution from the network integration can evolve the cell towards
higher temperatures evolving the cell back towards using the
equilibrium solution.  Moreover, the reaction network used for
the time integration is different (usually smaller) than the isotope
listing used for the equilibrium methods. This necessitates crafting a
delicate mapping between two abundance vectors, which may also
introduce unphysical discontinuities. In addition, care must be taken
to assure the reaction rate screening corrections used in the time
integration are properly taken into account in the equilibrium
solution method, otherwise a fundamental incompatibility exists
between the abundance vectors.

Finally, equilibrium methods determine the composition at a fixed
electron fraction $\Ye$.  It then becomes necessary to solve an
ordinary differential equation for $\dot \Ye$ based on weak reaction
rates in order to advance the abundance solution with a time varying
$\Ye$  \citep[][also see Section \ref{s.weak}]{mclaughlin_1996_aa,townsley_2009_aa,arcones_2010_aa}.
Switching between integration and equilibrium methods mid-stream is a
liability, not a positive asset.

The need for traditional workarounds forced by limited accuracy of the
summation is now avoided.  The summation experiments in Section
\ref{s.summation} demonstrate that network integration can be robust
and efficient, even at very high temperatures, when the accuracy of
the summations is improved.  We stress this is not just a solution to
issues of limited accuracy.  It also offers an improvement in
\mesa\ by providing a single solution methodology, network
integration, that avoids the challenges of stitching together
different solution methods.


\subsection{X-ray Burst Models and Adaptive Nets} \label{s.xrb}

The new capabilities described above allow \MESAstar\ to use large in-situ reaction networks
(i.e.\ fully coupled to the stellar evolution rather than uncoupled co-processing).
A demonstration is Type 1 X-ray bursts,
a class of objects with unstable nuclear burning 
on the surface of a neutron star (NS).
These bursts are sensitive functions of accretion rate \citep{chen_1997_aa}, 
accretion composition \citep{galloway_2006_aa}, the spatial
distribution of burning on the surface of the NS \citep{bildsten_1995_aa}, the 
type of burning that occurs between bursts \citep{galloway_2008_aa} as well as 
possibly other
conditions, for instance ``superbursts'' where carbon, rather than H/He, burns 
\citep{cumming_2001_aa}.
Here we focus on a simplified model of constant
accretion rate, where the burning 
occurs over the whole surface of the NS.
GS 1826-24 \citep{tanaka_1989_aa}, also known
as the ``clocked burster'' \citep{ubertini_1999_aa}, provides an example of such 
a system due to its regular Type 1 X-ray bursts.

\begin{figure}[!htb]
\begin{center}
\includegraphics[width=\apjcolwidth]{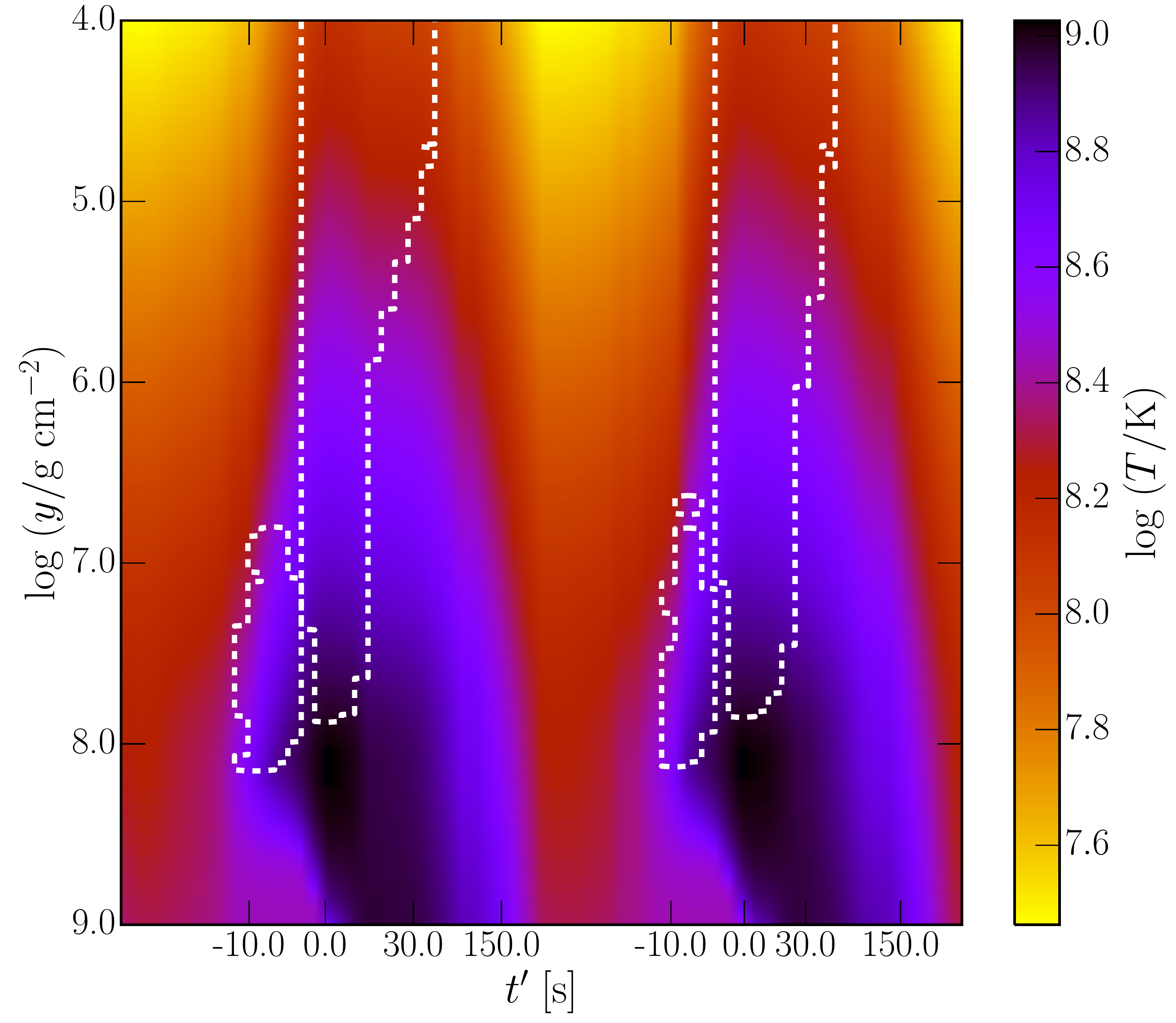}
\caption{
Kippenhahn plot during two X-ray bursts for the {\tt rp\_305} net with the solar 
metallicity accretion model. The x-axis values are times relative to the
peak of each burst, note the non-linearity of the scale. The y-axis values are 
the column depth and the color coding shows the
temperature of the NS envelope. The dashed contours show the extent of the 
convective regions.\label{f.xray_profile}}
\end{center}
\end{figure}

\begin{deluxetable*}{lccc}
\tablecolumns{4}
\tablewidth{0pt}
\tablecaption{Recurrence times of X-ray bursts.\label{t.charxray}}
\tablehead{\colhead{Model}        &
           \colhead{Accretion rate ($10^{-9}\,\mdotyr$)} &
           \colhead{Composition}         &
           \colhead{Recurence time (hrs)}       }
\startdata
GS 1826-24 &  &                & $4.0750 \pm 0.0003$\\
{\tt rp\_53}      & 3.00 & 2\% metals     & $1.5 \pm 0.10$\\
{\tt rp\_153}     & 3.00 & 2\% metals     & $3.3 \pm 1.80$\\
{\tt rp\_305}     & 3.00 & 2\% metals     & $3.2 \pm 0.07$\\
{\tt rp\_305}     & 3.00 & 2\%  \nitrogen & $3.0 \pm 0.07$\\
{\tt rp\_305}     & 2.40 & 2\% metals     & $4.1 \pm 0.30$\\
\cite{heger_2007_aa} & 1.17 & 2\% \nitrogen & $5.4\pm0.10$
\enddata
\smallskip
\end{deluxetable*}

\begin{figure}[!htb]
\begin{center}
\includegraphics[width=\apjcolwidth]{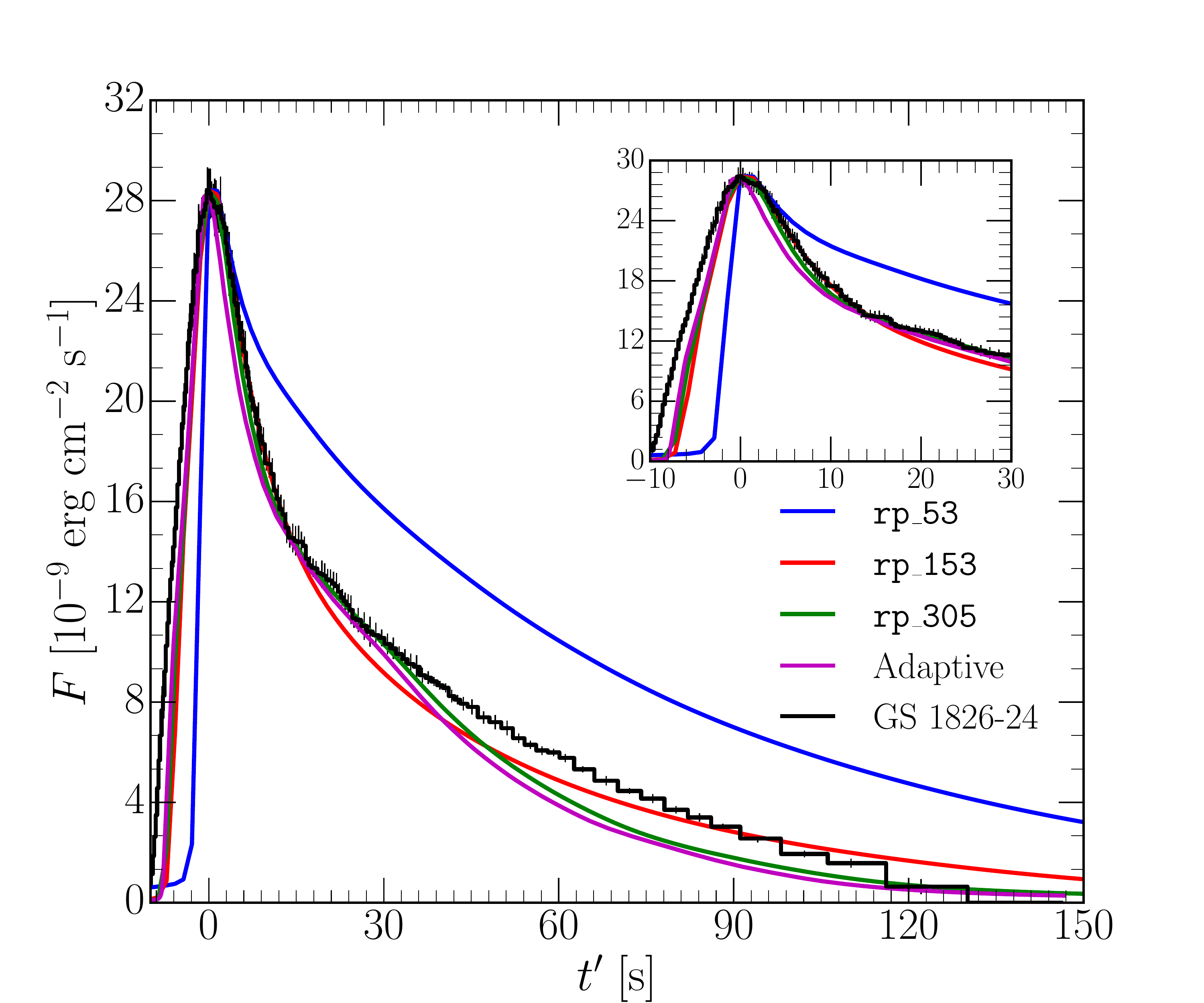}
\caption{
The folded burst profiles for the different nuclear networks as compared to
GS 1826-24 for an accretion rate of $3 \times 10^{-9}\,\mdotyr$ with
2\% metals with a solar composition, Three {\tt rp} network models are shown and 
 one of the adaptive net models. The insert shows a zoom in of the first $30\,\rm{s}$ during the burst.
 \label{f.xray_rp}}
\end{center}
\end{figure}

As material is accreted at the surface of a NS
it is compressed and heats the underlying material.
The accreted hydrogen (from a low mass main sequence (MS) star 
\citep{chen_1997_aa}) burns via the hot CNO cycle. However, 
with high enough accretion rates
the hydrogen will be accreted faster than the hot CNO cycle, which is limited by 
the $\beta$-decay timescale (of order minutes), can process the material.
The accreted helium ignites unstably in a hydrogen
rich environment, allowing rapid proton ({\tt rp}) captures onto seed nuclei
\citep{wallace_1981_aa}.
This process forms nuclei along the proton drip line up to and 
beyond the iron group \citep{schatz_1999_aa}, peaking at $^{107}$Te, when $\alpha$-decays 
prevent heavier elements from being formed \citep{schatz_2001_aa,fisker_2008_aa}. Once 
the burst begins, convection will commence, mixing the freshly burnt material with the ashes of 
previous burning episodes \citep{weinberg_2006_aa}.

GS 1826-24 has been studied
by the \textit{Rossi X-Ray Timing Explorer} (RXTE) over several years \citep{galloway_2004_aa,galloway_2008_aa}.
The bursts showed a decrease in the
recurrence time between bursts, from 4.1 \hour \ in 2000 to 3.56 \hour\ in 2002,
though during each observational epoch the bursts were consistent with each other.
Based on the ratio of the burst energy to the persistent flux, it is assumed
that the bursts are powered by hydrogen burning of solar metallicity material.


\begin{figure}[!htb]
\begin{center}
\includegraphics[width=\apjcolwidth]{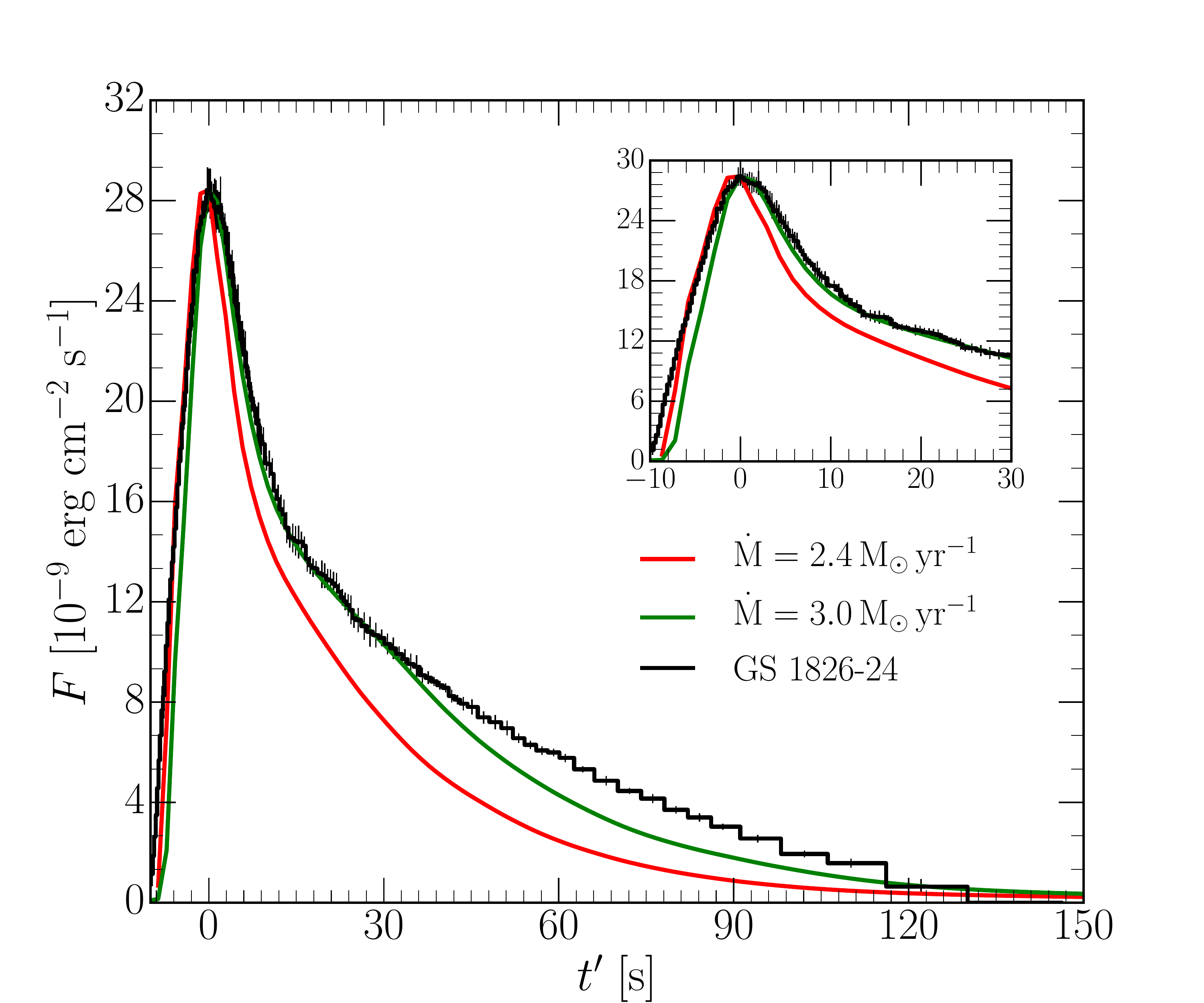}
\caption{
Folded burst light curve for the {\tt rp\_305} net, with a solar metallicity 
accretion composition, shown for  $\dot M=2.4\times10^{-9}\,\mdotyr$  and $\dot M=3\times10^{-9}\,\mdotyr$, normalized
to the peak flux measured for GS 1826-24.
The insert shows the first $30\,\rm{s}$ of the burst. 
\label{f.xray_acc_rate}}
\end{center}
\end{figure}



We model the NS envelope using inner boundary conditions for mass and radius of
$M_{\rm c}=1.4\,$\msun\ and $R_{\rm c}=11.2\,$\kilo\meter\ \citep{heger_2007_aa},
implying a gravitational redshift of $1+\grredshift=1.26$.
The base of the envelope is composed of  an inert
layer that does not undergo reactions. The luminosity 
at the base of
the envelope is set to $L=1.6\times 10^{34}\, \ergspersecond$ \citep{woosley_2004_ab}.
We base our nuclear networks on  the 304 species {\tt rp.net} network of \citet{fisker_2008_aa}, which 
includes proton rich isotopes up to $^{107}$Te. Isotopes above \zinc[66], which 
is the peak isotope in the
{\tt mesa\_204.net}, are included due to the proton captures possible on 
high-$Z$ isotopes during the peak of the burst \citep{fisker_2006_aa}.
We also include the effects of rotational mixing by setting  a minimum amount of mixing 
in the NS envelope. This
mixing,  while having a physical motivation \citep{piro_2007_aa,keek_2009_aa}, is there 
primarily to improve the convergence of \mesa \ models by smoothing out the
compositional gradients that form in the ashes of previous bursts.
We include the post-Newtonian correction to 
correct the local gravity in
each cell for GR effects.
During the burst we allow the accretion to continue.

Our results are compared to the RXTE observations of GS 1826-24 over bursts 9-20. 
Time resolved spectra were binned during the
bursts' rise time and decay \citep{galloway_2008_aa,zamfir_2012_aa}.
Data output by \MESA\ is not GR time corrected, thus we 
set the burst times to be $\tprime=t(1+\grredshift)$,
and average multiple bursts to produce a scaled light-curve.


%

Figure \ref{f.xray_profile} shows the temperature profile during two
X-ray bursts, for the {\tt rp\_305} net, accreting solar metallicity material at a rate of $3.0\times10^{-9}\,\mdotyr$.
At $\tprime\approx -10\,\rm{s}$ the envelope ignites material and drives the 
formation of the first convection zone. This zone expands outwards
in the envelope mixing the ashes from the burning at the base of the envelope 
outwards to lower pressures \citep{weinberg_2006_aa}. As the burst decays the convection zone recedes outwards and 
by $\tprime\approx 150\,\rm{s}$ the envelope returns to its pre-burst temperature profile.
%


We test three reaction networks, {\tt rp\_53}, {\tt rp\_153} and {\tt rp\_305}, each a modified form of that in
\citet{fisker_2008_aa}. Table~\ref{t.charxray} shows that increasing
the number of isotopes in the reaction network increases the
recurrence time and that all (for $\dot M= 3.0\times10^{-9}\,\mdotyr$)
have recurrence times $\approx 1\,\rm{hr}$ less than that of GS
1826-24.

Figure~\ref{f.xray_rp} and its insert show the folded light curves
for each of the three {\tt rp} reaction networks plus the GS 1826-24
observations. The rise time is sensitive to the net, with the largest
net matching the observed slow rise. The observed decay profile is
also best matched by {\tt rp\_305}. Burst to burst variations of the
models decrease with increasing net size and can be further reduced by
increasing the temporal resolution of the models.  However, increasing
the size of the net reduces the variation without having to increase
the temporal resolution and also highlights the impact of \mesa\
capability to include large nuclear networks.

To achieve a better match to the GS 1826-24 recurrence time (see
Table~\ref{t.charxray}), we reduce the accretion rate to
$\dot M= 2.4\times10^{-9}\,\mdotyr$.  However,
Figure~\ref{f.xray_acc_rate} shows that the light curve comparisons
are not as good as for the higher $\dot M$.

GS 1826-24 was also modeled by \citet{heger_2007_aa} with accretion of
hydrogen, helium and 2\% \nitrogen.  For comparison, we run a model
with this same composition with the {\tt rp\_305} net.  Table
\ref{t.charxray} shows that the recurrence time decreases slightly
when accreting 2\% \nitrogen\, rather than 2\% metals. The model with
metal accretion is in better agreement with both the light curve rise
and decay.


We now explore adaptive nets \citep{woosley_2004_ab}, where we
allow \mesa\ to determine which isotopes (and reactions) are necessary
by assessing the available reaction pathways for the most abundant
isotopes. The network is constructed by first finding those isotopes
with an abundance above a threshold, $X_{\rm{keep}}$, and then
introducing those isotopes which are connected by adding or removing
protons, neutrons, or $\alpha$ particles. That determination is made
via the additional parameters $X_{\rm{n}}$ (i.e. neutron reactions)
and $X_{\rm{p}}$ (i.e. proton and $\alpha$ reactions) potentially
re-adding isotopes removed with the initiating $X_{\rm{keep}}$
threshold.


Accreting solar composition material at
$\dot M=3.0\times10^{-9}\,\mdotyr$ we follow the model to the second
burst, finding a recurrence time of $3.1\,\rm{hrs}$, comparable to
that from the {\tt rp\_305} net (Table~\ref{t.charxray}).  The
adaptive net has a better rise time profile than the {\tt rp} nets,
while the {\tt rp\_305} net has a better fit to the decay.  This gives
us confidence that the {\tt rp\_305} net includes all relevant
isotopes which drive the X-ray burst and thus is a useful
approximation.  For suitable values for the sensitivity of the
adaptive net, the net limits itself to $\approx400$ isotopes between
bursts, which increases to $\approx600$ isotopes during the
burst. Variations of a factor 100 in the threshold parameters only
change the isotope count by at most 50 isotopes and do not affect the
final results.

\def\foe{10$^{51}$\,erg}

\section{Core-Collapse Supernovae}
\label{s.ccsn}

The capability of using large, in-situ reaction networks without the
need for equilibrium or co-processing techniques was described in
Section \ref{s.advburn} and applied to X-ray burst models. We
extend our demonstration of this capability by first considering
pre-supernova models. We then combine the advanced burning
development with the implicit treatment of shocks discussed
in Section \ref{s.hydro} to core-collapse supernovae models.

\begin{figure}[!htb]
\begin{center}
\includegraphics[width=\apjcolwidth]{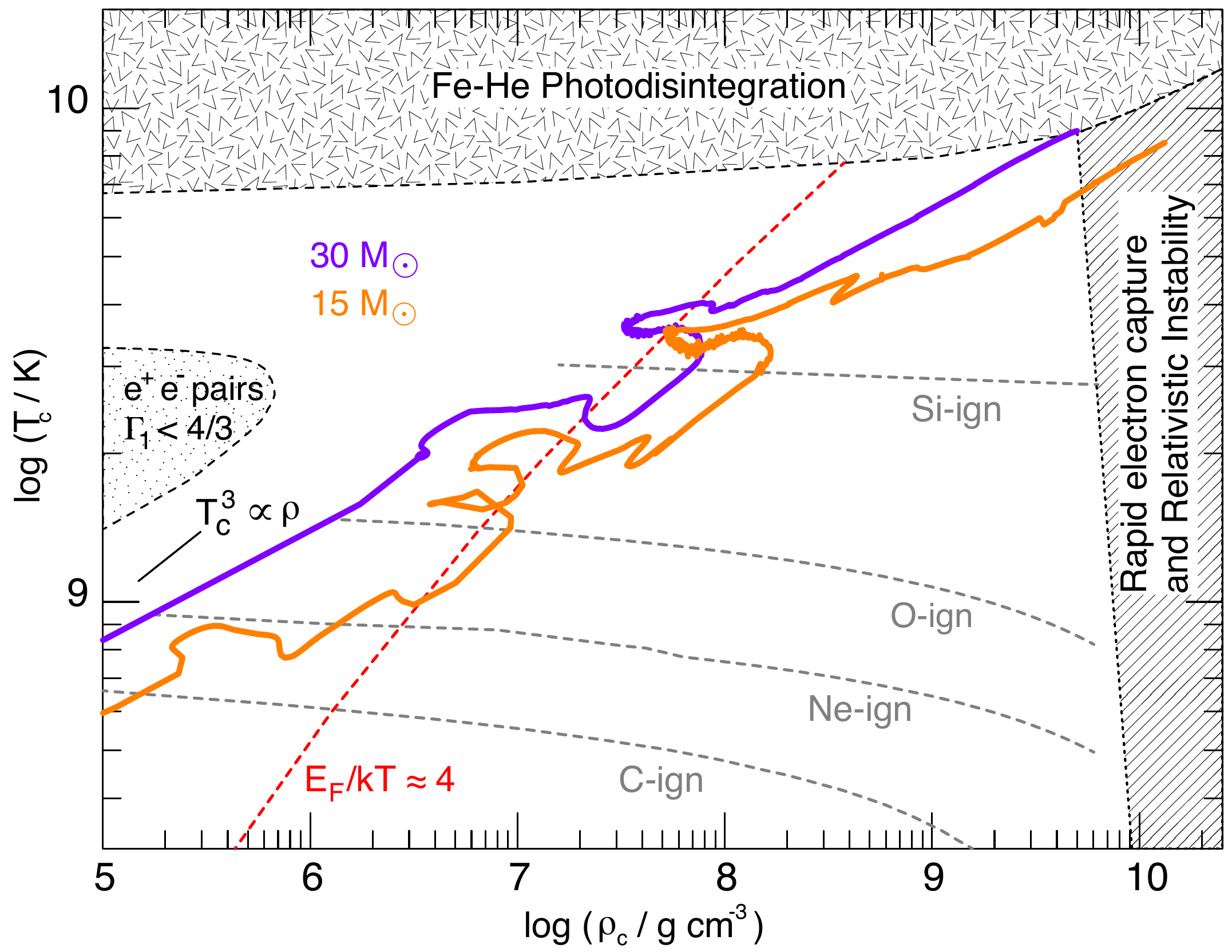}
\caption{Evolution of \Tc\ and \rhoc\ in solar metallicity,
non-rotating $M_i$ = 15 and 30 \msun \ pre-supernova models.
The curves are calculated using an in-situ 204 isotope reaction network.
Locations of the core carbon, neon, oxygen, and silicon
ignition are labeled, as is the scaling relation $\Tc^3 \propto \rhoc$,
and the $\EF/\kB T \approx 4$ electron degeneracy curve.
Regions dominated by electron-positron pairs, photodisintegration,
and rapid electron capture are shaded and labeled.
}
\label{f.tcdc}
\end{center}
\end{figure}

\begin{figure}[!htb]
\begin{center}
\includegraphics[width=\apjcolwidth]{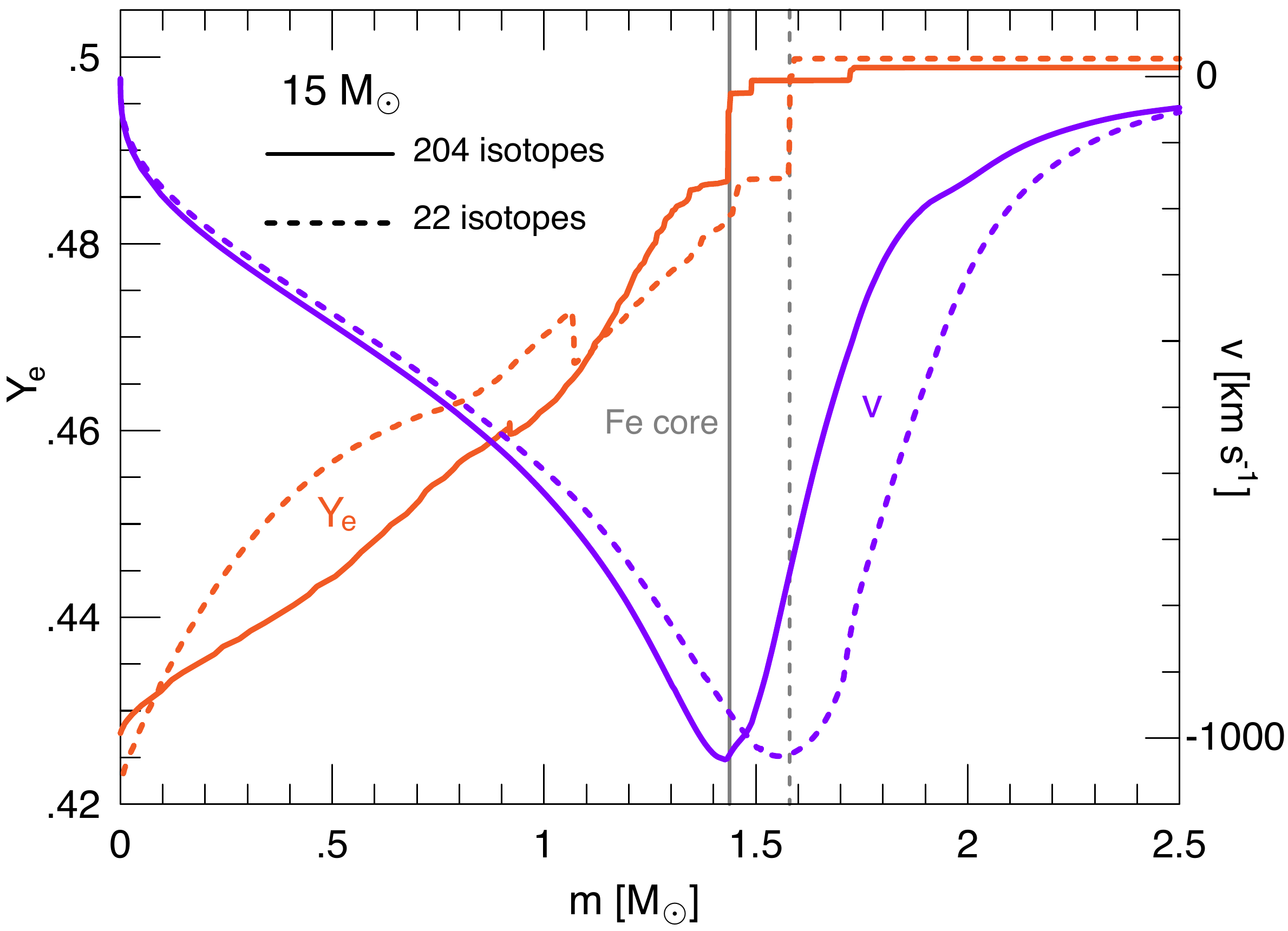}
\caption{
Radial velocity and $\Ye$ profiles at the onset of core collapse
for the $M_i$ = 15 \msun \, model. Dashed curves show the results
using a 22 isotope network and solid curves show the results
using a 204 isotope network. Both models are evolved from the pre main-sequence
to the onset of core collapse with their respective reaction network.
The vertical gray lines mark the mass of the iron core as defined by the $\Ye$ jump.
}
\label{f.ye_vel_profiles}
\end{center}
\end{figure}

\begin{figure}[!htb]
\includegraphics[width=\apjcolwidth]{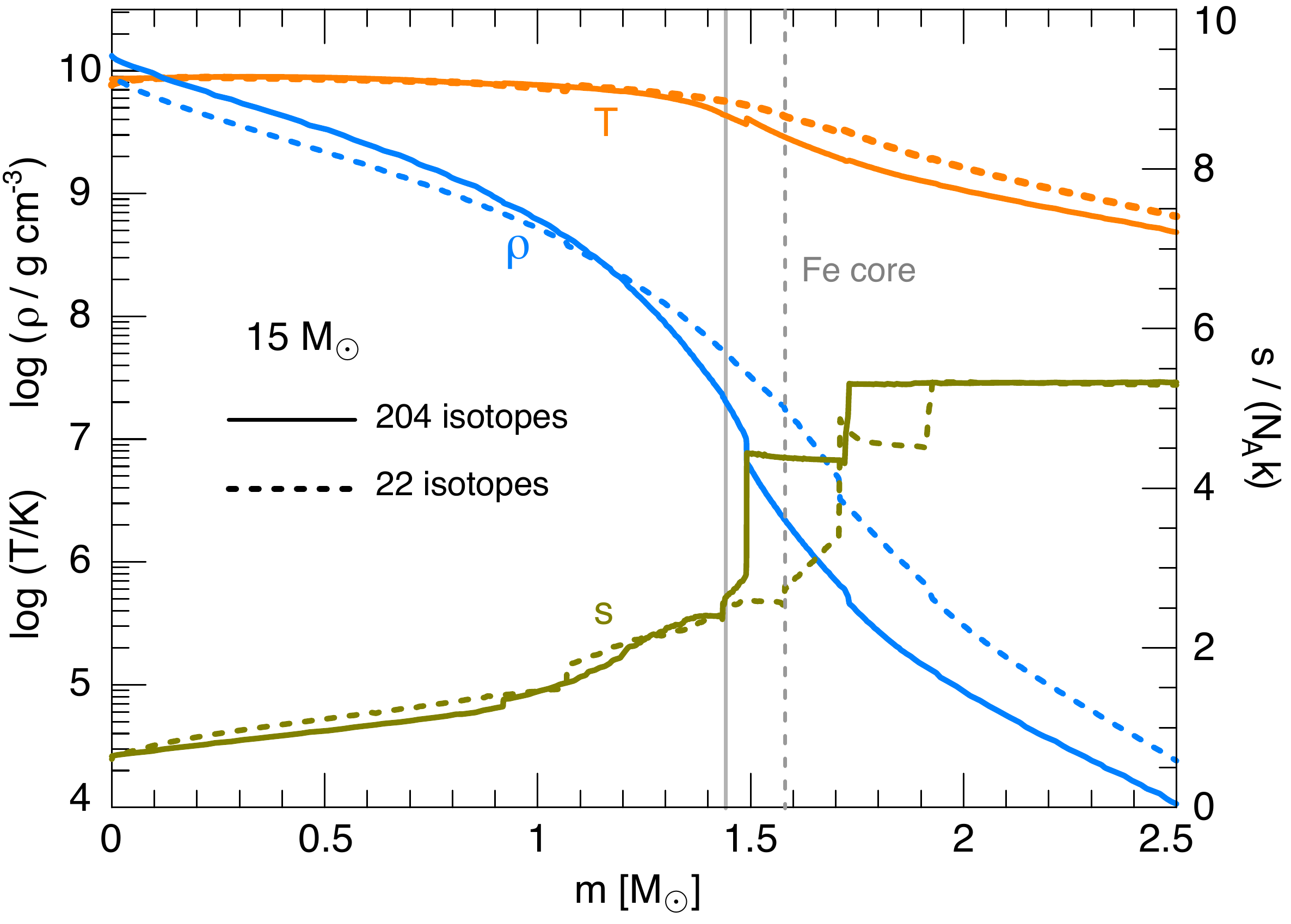}
\caption{
Thermodynamic profiles at the onset of core collapse for the $M_i$=15
\msun \ model.  Dashed curves show the results using a 22 isotope
network and solid curves show the results using a 204 isotope
network. Both models are evolved from the pre main-sequence to the onset of
core collapse with their respective reaction network. The
vertical gray lines mark the mass of the iron core as defined by the $\Ye$ jump.
}
\label{f.trho_profiles}
\end{figure}

\begin{figure}[!htb]
\includegraphics[width=\apjcolwidth]{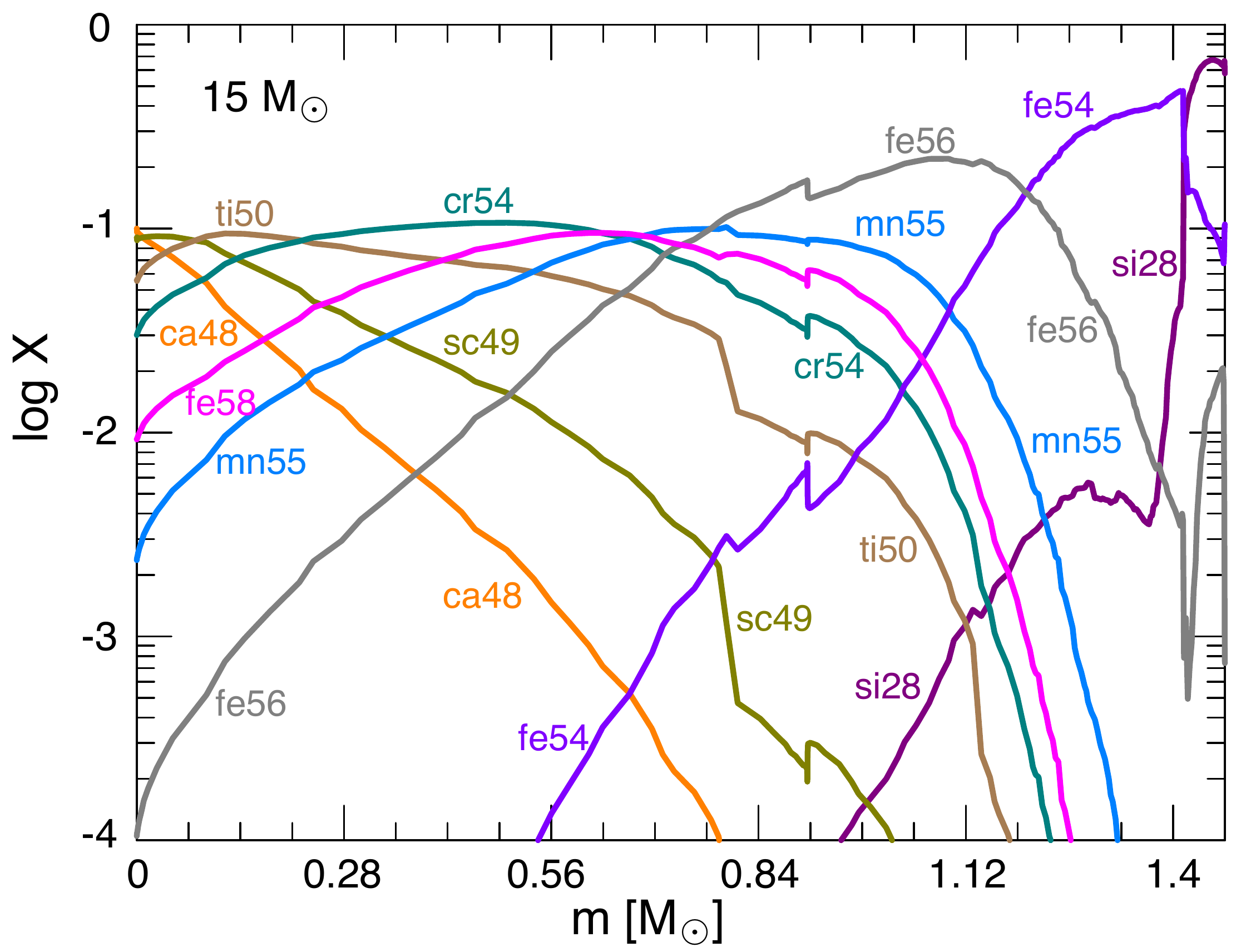}
\caption{
Mass fraction profiles of the ten most abundant isotopes within the
iron core at the onset of core collapse for the $M_i$ = 15
\msun\ model evolved with the 204 isotope network.  The entire iron
core is in NSE and the mass fractions adapt to the changing temperature,
density and $\Ye$.
}
\label{f.mostabund}
\end{figure}

\subsection{Pre-Supernova Evolution without QSE or NSE}
\label{s.presn}

Figure \ref{f.tcdc} shows the $\Tc - \rhoc$ evolution of $M_i$ = 15
and 30\,\msun \ models from the onset of carbon burning until
iron-core collapse.  These non-rotating, solar metallicity models used
the 204 isotope reaction network described in Section \ref{s.advburn}
and \MESAstar's ``Dutch'' mass loss prescription with $\eta$=0.8.
These models have $\approx$\,2200 cells on the main-sequence,
$\approx$\,3500 cells as the star becomes a red supergiant,
and $\approx$\,2300 cells at the onset of core collapse.  At core
collapse the final masses are $M_f$ = 13.0 and 15.2\,\msun. The curves
fall below the $\Tc^3 \propto \rhoc$ scaling relation because the core
becomes partially electron degenerate, as indicated by tracks crossing
the Fermi energy $\EF/(\kB T) \approx$\,4 curve.  Evolution towards
lower density at nearly constant temperature signals ignition of a
nuclear fuel.

Figure \ref{f.ye_vel_profiles} shows the radial velocity and $\Ye$
profiles at the onset of core collapse for the $M_i$ = 15 \msun
\ model.  Dashed curves show the results using a 22 isotope network
and solid curves show the results using a 204 isotope network. Both
models are evolved from the pre main-sequence to the onset of core
collapse with their respective reaction network.  The vertical gray
lines mark the mass of the iron core as defined by the $\Ye$ jump,
which is $m \approx$\,1.43\,\msun\, for the 204 isotope model and $m
\approx$\,1.59\,\msun\, for the 22 isotope model.  The infall speed
has reached $\approx$ 1000 km s$^{-1}$ just inside these iron core
locations.

Figure \ref{f.trho_profiles} shows the thermodynamic profiles at the
onset of core collapse for the $M_i$=15 \msun \ model.  Dashed curves
again show the results using a 22 isotope network and solid curves
show the results using a 204 isotope network.  The vertical gray lines
again mark the mass of the iron core as defined by the $\Ye$ jump in
Figure \ref{f.ye_vel_profiles}.  The impact of these differences
remains to be explored.

Figure \ref{f.mostabund} shows the mass fraction profiles of the ten
most abundant isotopes within the iron core at the onset of
core collapse for the $M_i$ = 15 \msun\ model evolved with the 204
isotope network.  Each isotope shown dominates the NSE composition at
some location within the iron core, although we stress that no NSE or
QSE approximation was used; the same 204 isotope reaction network was
used throughout the entire model from the pre-main-sequence to the
onset of core collapse.

The most abundant isotopes in an NSE distribution generally have an
individual $\Ye$ that is within a small range of the local $\Ye$.  A
small spread usually exists due to nuclear structure effects.  For
example, the dominant isotopes at the center in Figure
\ref{f.mostabund} are $^{49}$Sc and $^{48}$Ca. These isotopes have
individual $Y_{\rm e}$ of 0.429 and 0.417, respectively; commensurate
with the central \Ye\,$\approx$\,0.428 shown in Figure
\ref{f.trho_profiles}.  The dominant isotope changes as the NSE
distribution adapts to the rapidly decreasing density profile and
increasing $Y_{\rm e}$ profile. All the isotopes in the iron core
eventually become part of the compact remnant after the
explosion. However, the thermodynamic and composition profiles near
the mass cut depend on the profiles interior to the mass cut.

\begin{figure*}[!htb]
\begin{center}
\includegraphics[width=\apjcolwidth]{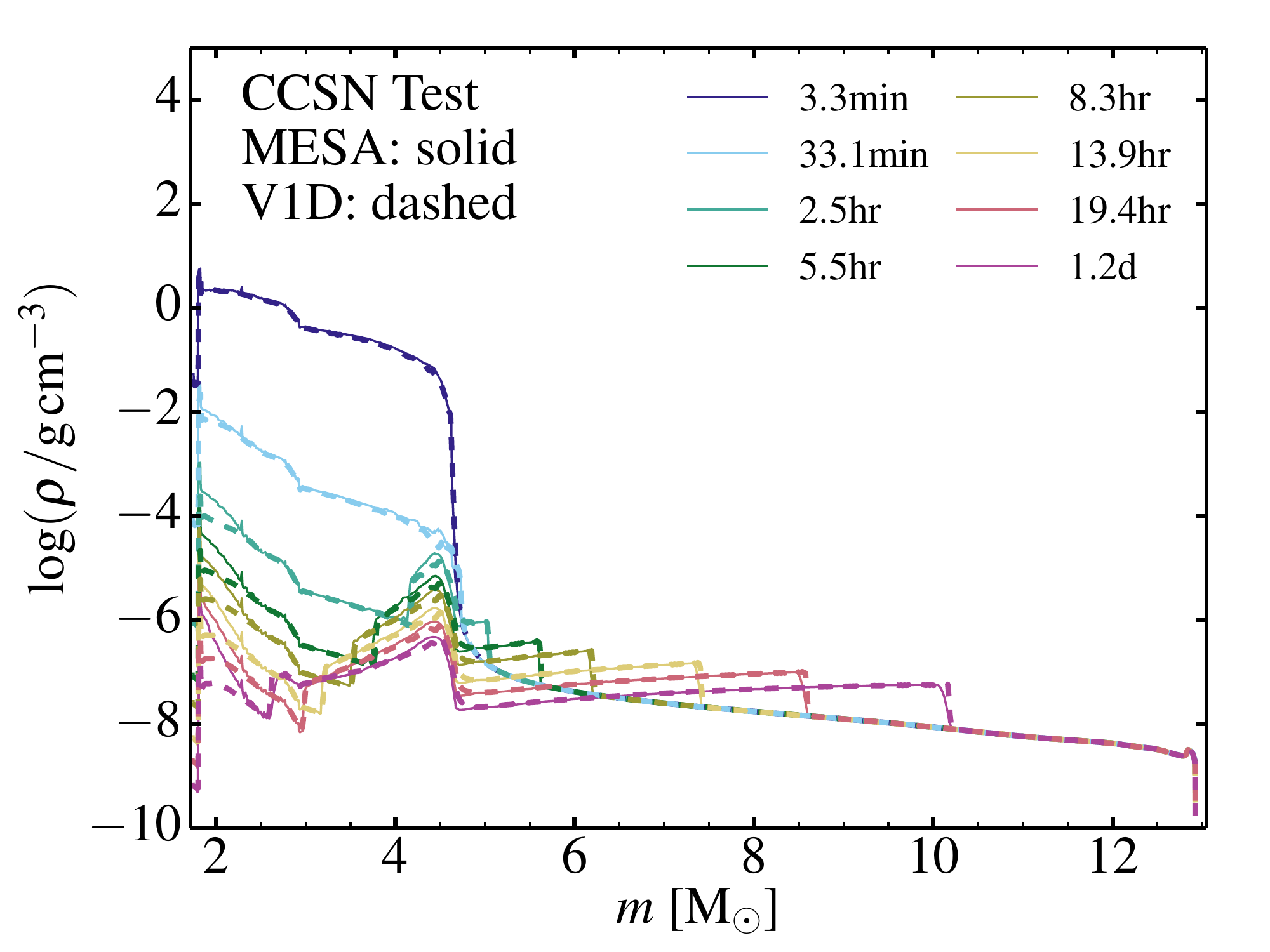}
\includegraphics[width=\apjcolwidth]{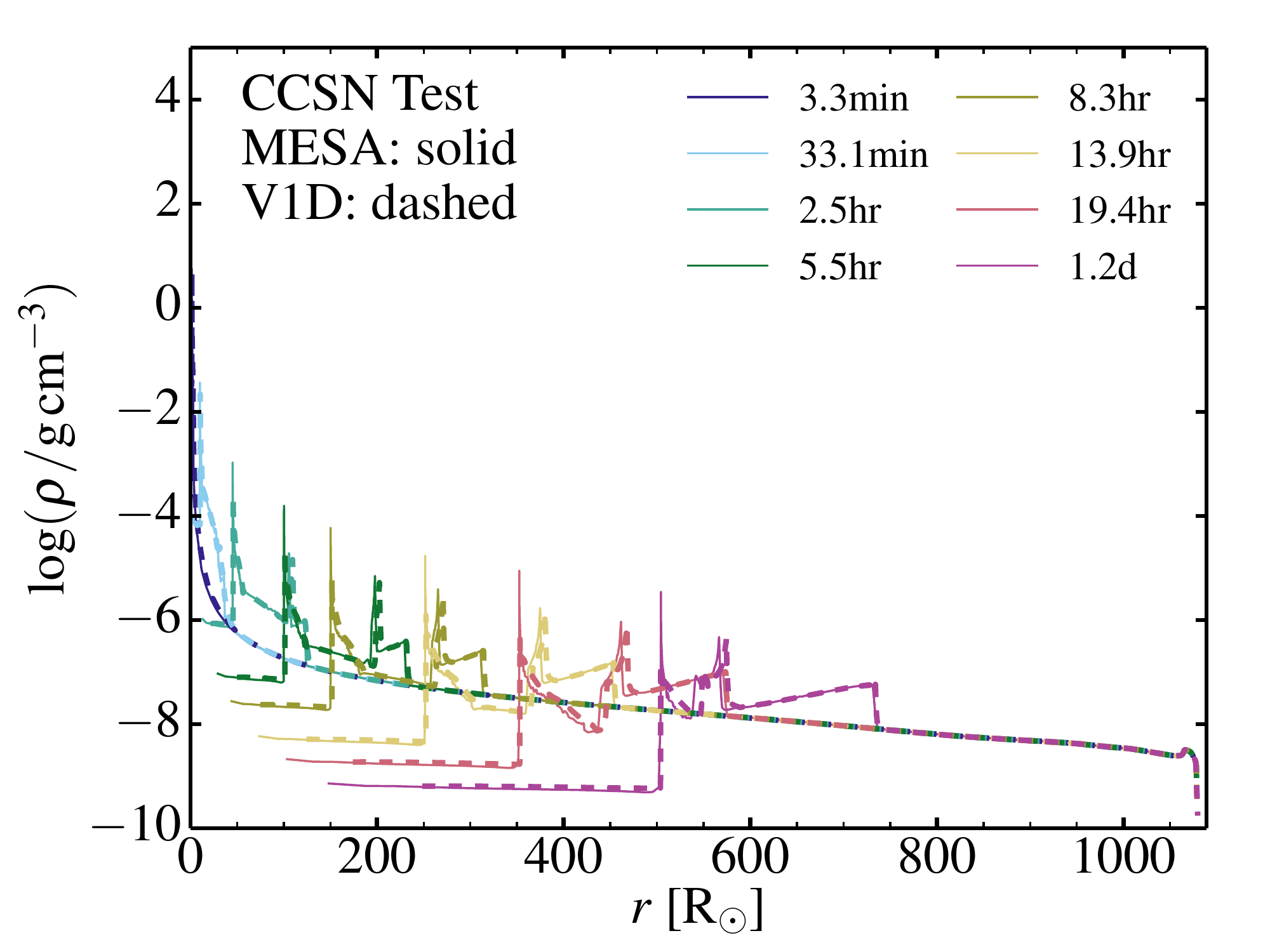} \\
\includegraphics[width=\apjcolwidth]{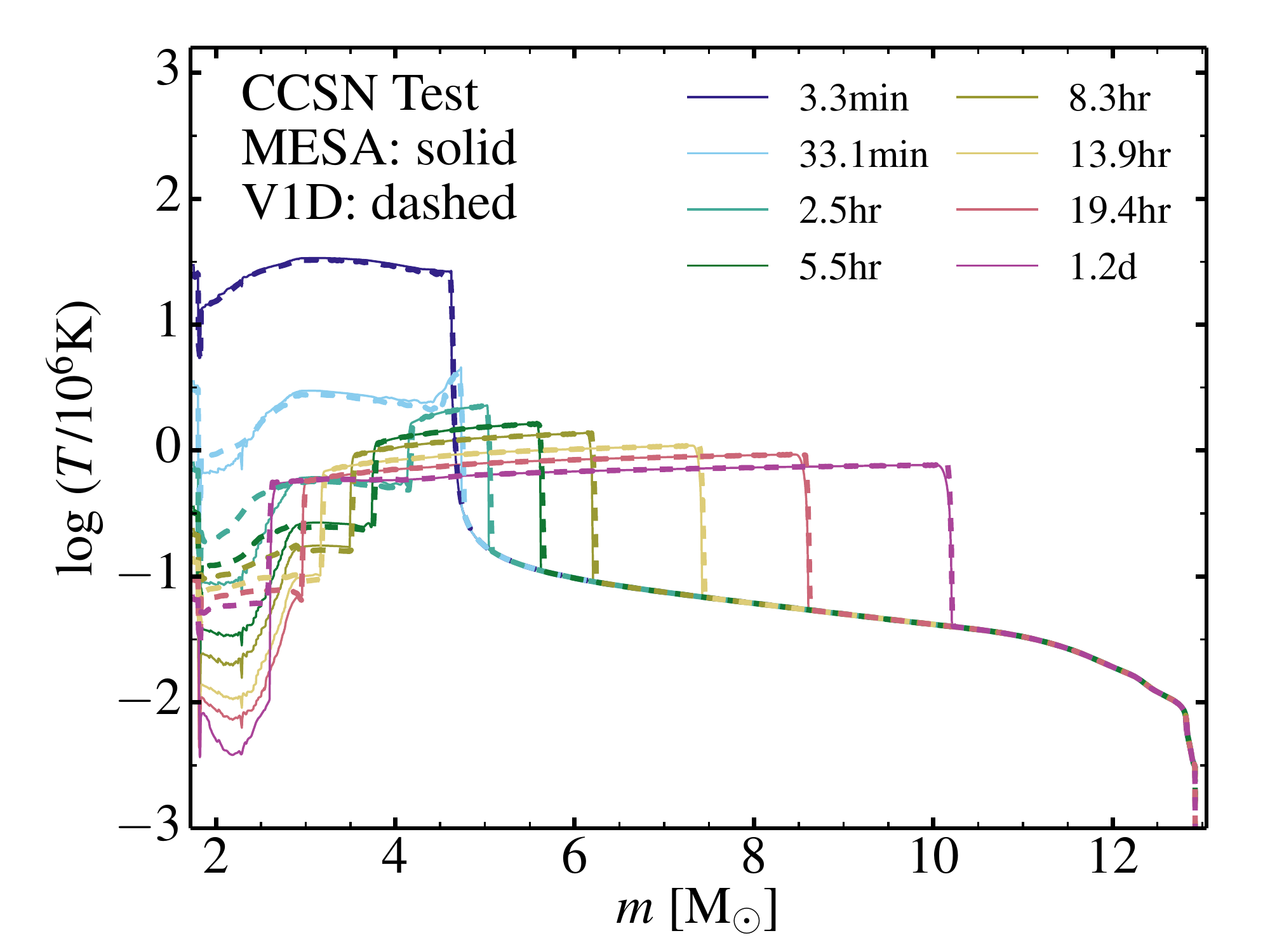}
\includegraphics[width=\apjcolwidth]{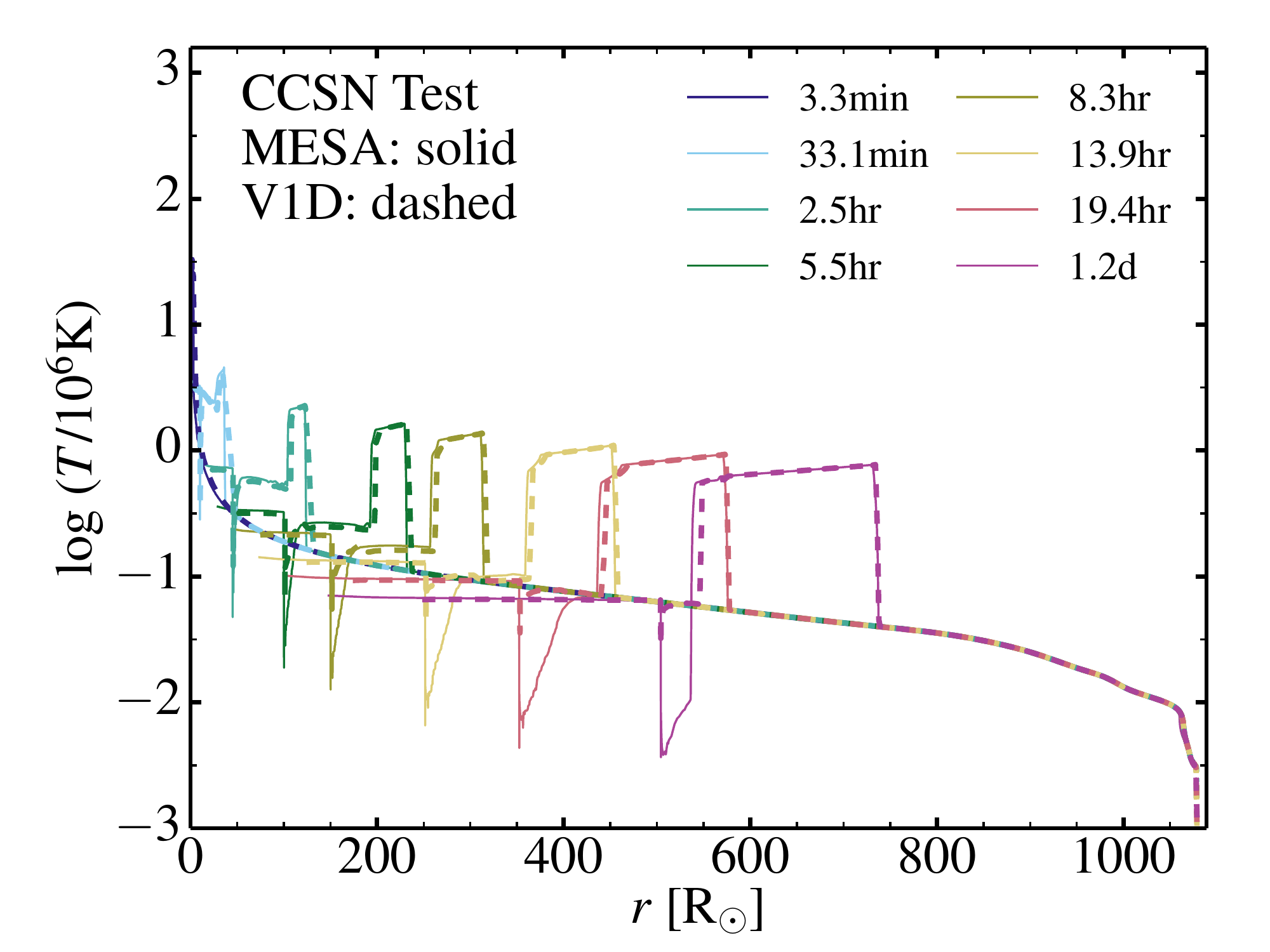} \\
\includegraphics[width=\apjcolwidth]{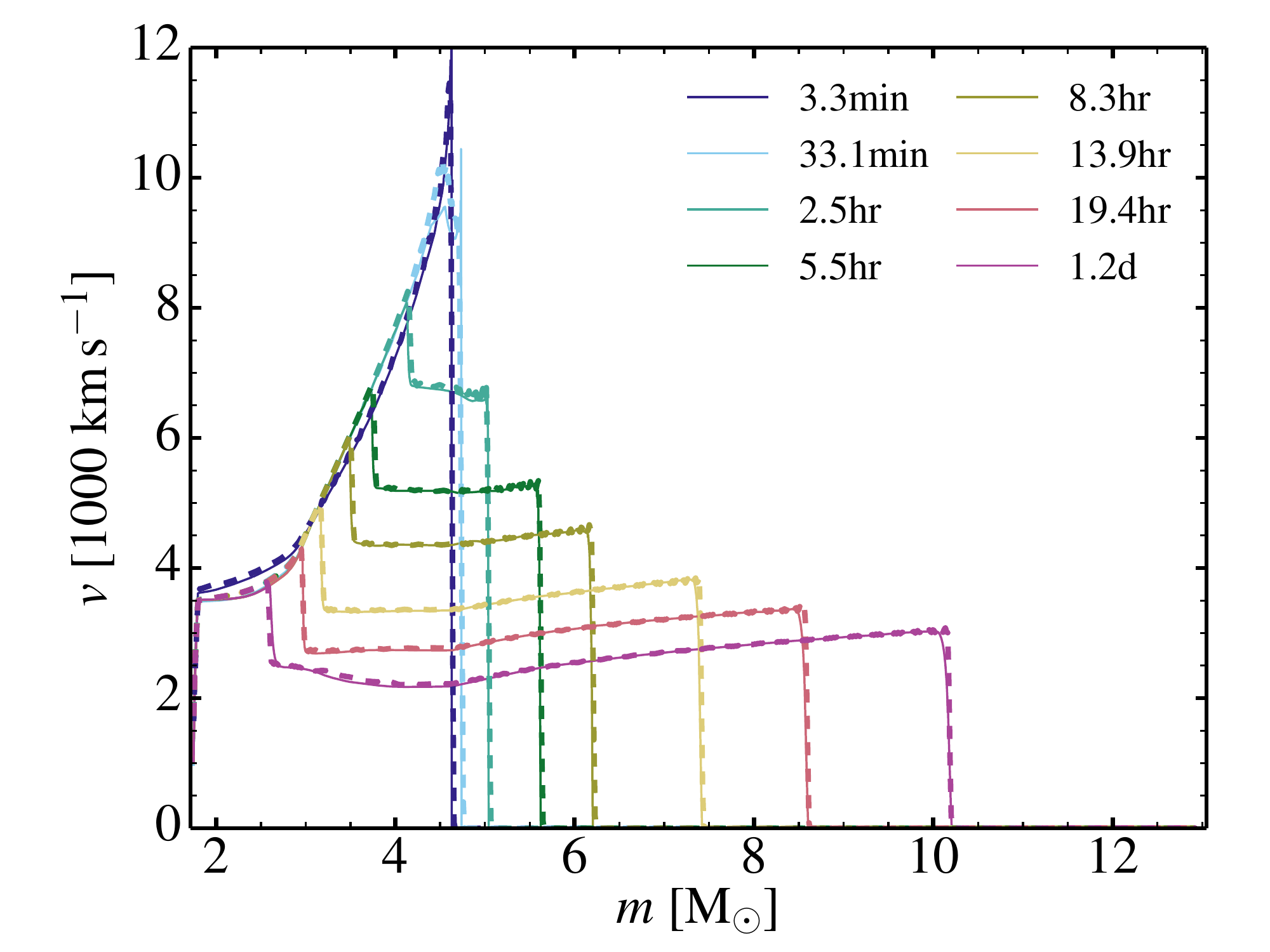}
\includegraphics[width=\apjcolwidth]{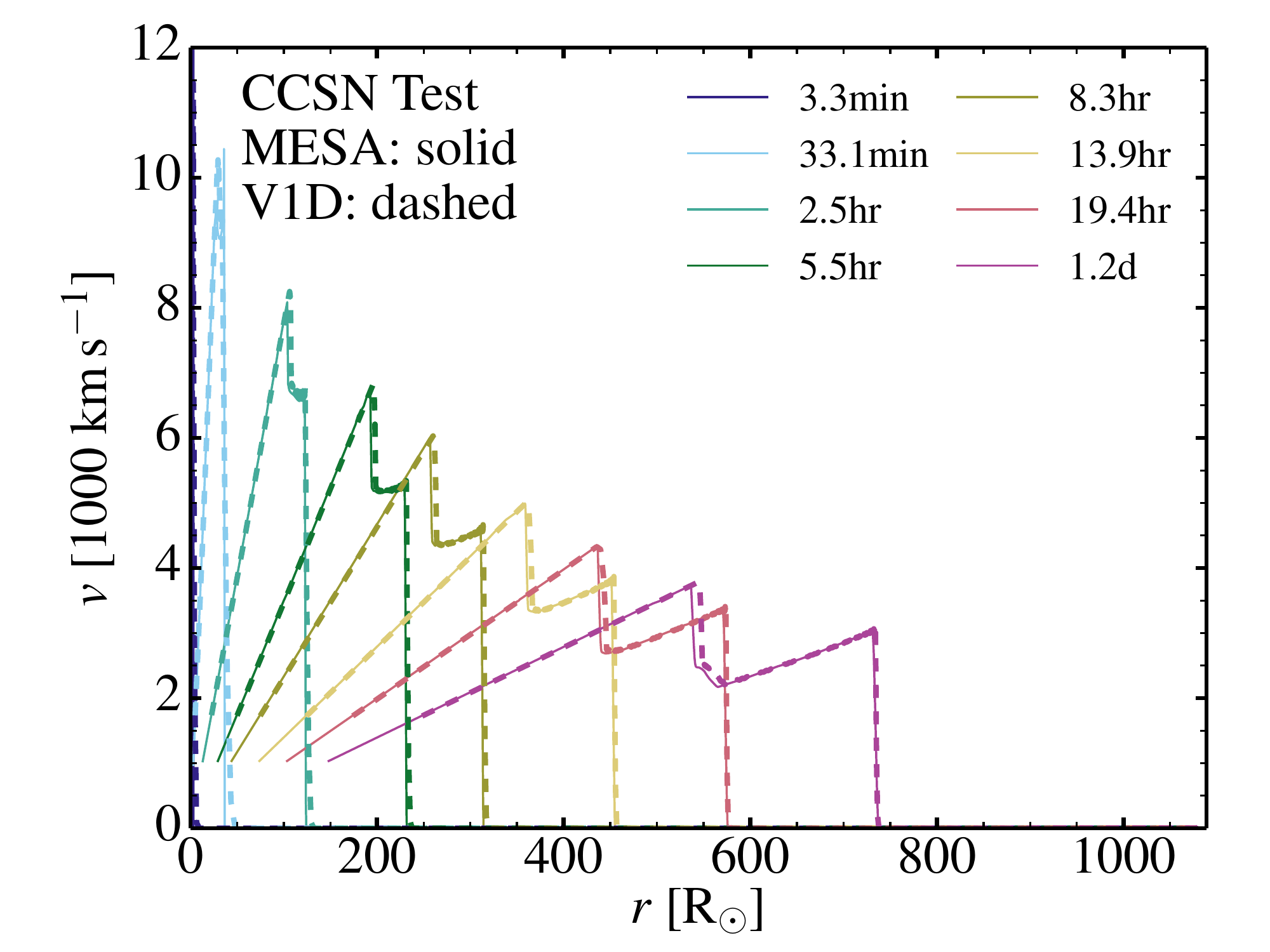}
\caption{
Multi-epoch snapshots of the hydrodynamical simulation from energy injection mimicking
core-collapse supernova.
The initial model is a 15\,\msun\ star at solar metallicity, evolved with mass loss
but no rotation, and employing a nuclear network of 22 isotopes.
We show the density (top
row), temperature (middle row), and velocity (bottom row), versus
Lagrangian mass (left column) and radius (right column).  In each
panel, the solid line shows the \mesa\ results and the dashed line
the \voned\ results.
\label{f.m15_slice}}
\end{center}
\end{figure*}

\subsection{Core-collapse Supernova Explosions}
\label{s.ccsn_expl}

The envelope shock tests described in Section~\ref{s.shocks} show that
the hydrodynamic solver in \mesa\ meets the basic requirements for
shock propagation in a star. The AGB star model was
selected because of the well behaved conditions of its envelope --- a
density structure that is smooth and monotonically declining towards
the stellar surface, and a uniform composition.

Here we explore the more challenging conditions associated with a
strong shock born deep in the stellar interior of a massive star.  We
study the dynamics of such a supernova shock and the explosive
nucleosynthesis that takes place in the wake of the shock during the
first second.  The yields from explosive nucleosynthesis depend on
both the energy and the power (characteristic energy deposition
timescale), while the dynamics of the shock are primarily dependent on
the total energy deposited.

The starting conditions for the explosion simulations are the two
15\,\msun\ pre-supernova models discussed above; one for the
approximate 22 isotope network and one for the 204 isotope network.

\subsubsection{Explosion Dynamics}

Since the focus here is on the dynamics rather than nucleosynthesis,
we expedite the \MESA\ simulation by using the 22 isotope network for
both the pre-supernova and the explosion phases.  Before triggering
the explosion, we remove the iron core of the red-supergiant star by
placing the inner boundary of the grid at a Lagrangian mass of
1.75\,\msun.
We trigger the explosion by depositing 1.52$\times$10$^{51}$\,erg at a
constant rate during 1\,s. The artificial viscosity is raised
during the energy deposition phase (i.e., $l_2=$\,0.01), when the shock is at small radii,
and lowered in the subsequent evolution until shock breakout (i.e., $l_2=$\,0.003).
Since the binding energy of the envelope to be
shocked is $-$3.2$\times$10$^{50}$\,erg at the time we trigger the
explosion, this choice of energy deposition yields a total energy at
the end of the deposition phase of 1.2$\times$10$^{51}$\,erg. This is
generally considered a standard value for a core-collapse supernova.

Figure~\ref{f.m15_slice} shows that the development of the explosion
is analogous to the tests using the (low-density) envelope of an AGB
star in Section \ref{s.shocks}, but with significant quantitative
differences.  Here, the shock born at the edge of the iron core first
travels through the dense CO-rich core. At the outer edge of the
He-rich shell, the shock traverses a steep density gradient to enter
the low-density H-rich envelope. Hence, the shock crosses regions with
densities ranging from $\sim$10$^{6}$\,g\,cm$^{-3}$ at the edge of the iron
core down to 10$^{-10}$\,g\,cm$^{-3}$ at the stellar surface.

The radii of the innermost shells are initially very small since they
lie at the outer edge of the iron core.  Consequently, they suffer
considerable cooling from expansion. Figure~\ref{f.m15_slice} shows a
drop in temperature from a few 10$^9$\,K down to $\sim$\,10$^4$\,K at
$\sim$\,1\,day.  In addition, the supernova shock splits into a
reverse/forward shock structure when it encounters the density drop at
the transition between the He-rich core and the H-rich envelope.  The
reverse shock is the new feature, absent in the envelope shock test,
that causes a significant deceleration of He-core material.  The
conversion of kinetic energy into internal energy causes this inner
material to heat up, erasing the cooling effect from expansion. The
innermost layers, which travel the slowest, will be shocked
last. These innermost zones can evolve to temperatures $\sim$\,10$^4$
K.  It is in these innermost regions at late times that the
differences between \mesa\ and \voned\ are the largest. The
offset occurs in a region of relatively high density
($\sim$\,10$^{-7}$\,\grampercc) and low temperature
($\sim$\,10$^4$\,K).
The offset in temperature between the \mesa\ and \voned\ simulations at late times
stems from a difference in the equation of state for metal-rich regions.
\mesa\ accounts for ionization through the OPAL equation of state table for
metallicities z$<$\,0.04.
For higher metal abundances where OPAL tables are unavailable, \mesa\ currently
assumes full ionization while \voned\ solves for the ionization state of the gas.
Note that such density/temperature regimes are
normally not encountered in stellar interiors.  For other quantities
and/or locations/times, the agreement between \mesa\ and \voned\ is
excellent.

\add{We also note that in the \mesa\ simulation,
two small spikes appear in the temperature and density profiles at $<$\,2.5\,\msun\ at
$\approx$1000\,s after the energy deposition phase. This feature is absent in \voned\ 
because \voned\ uses a much
larger viscous spread when the shock is in the helium-rich core ($R<R_{\odot}$).
One can reduce or eliminate such spikes by increasing the viscous spread, 
although this may visibly smear the shock when it crosses the H-rich envelope --- 
the current choice seems a suitable compromise.}

In contrast to the envelope shock test, this supernova explosion
configuration raises the temperature by a factor of about ten.
Consequently, because $P_{\rm rad}/P_{\rm gas} \propto T^3/\rho$, the
post-shock material becomes completely radiation dominated
($P_{\rm rad} \gg P_{\rm gas}$).  If we neglect the binding energy and
the kinetic energy of the post-shock material, the post-shock energy
is of the order of the explosion energy.  We indeed find a good
correspondence between the post-shock temperature computed by \mesa\
and the temperature obtained from $(E_0/aV)^{1/4}$ (where $a$ is the
radiation constant, $E_0$ is a fitting parameter, typically of the
order of the explosion energy, and $V=4 \pi R_{\rm sh}^3/3$ is the
volume within the shock radius $R_{\rm sh}$).  As expected, we also
find that the shock accelerates (decelerates) in regions where
$\rho_{\rm sh} R_{\rm sh}^3$ decreases (increases) outward.

\subsubsection{Explosive Nucleosynthesis\label{s.expl_nucleo}}



Here we compare the shock nucleosynthesis results from the two
independent codes, \mesa\ and \voned.  The same initial 204 isotope
pre-supernova model was the starting point.  Our first test case is a
strong explosion triggered by injecting 1.57$\times$\foe\ for 0.05\,s
and within 0.02\,\msun\ of the mass cut, which is positioned at the
outer edge of the iron core at 1.5\,\msun.  The exact choice of
explosion energy, deposition time scale, and mass cut is not strictly
relevant.

\begin{figure}[!htb]
\begin{center}
\includegraphics[width=\apjcolwidth]{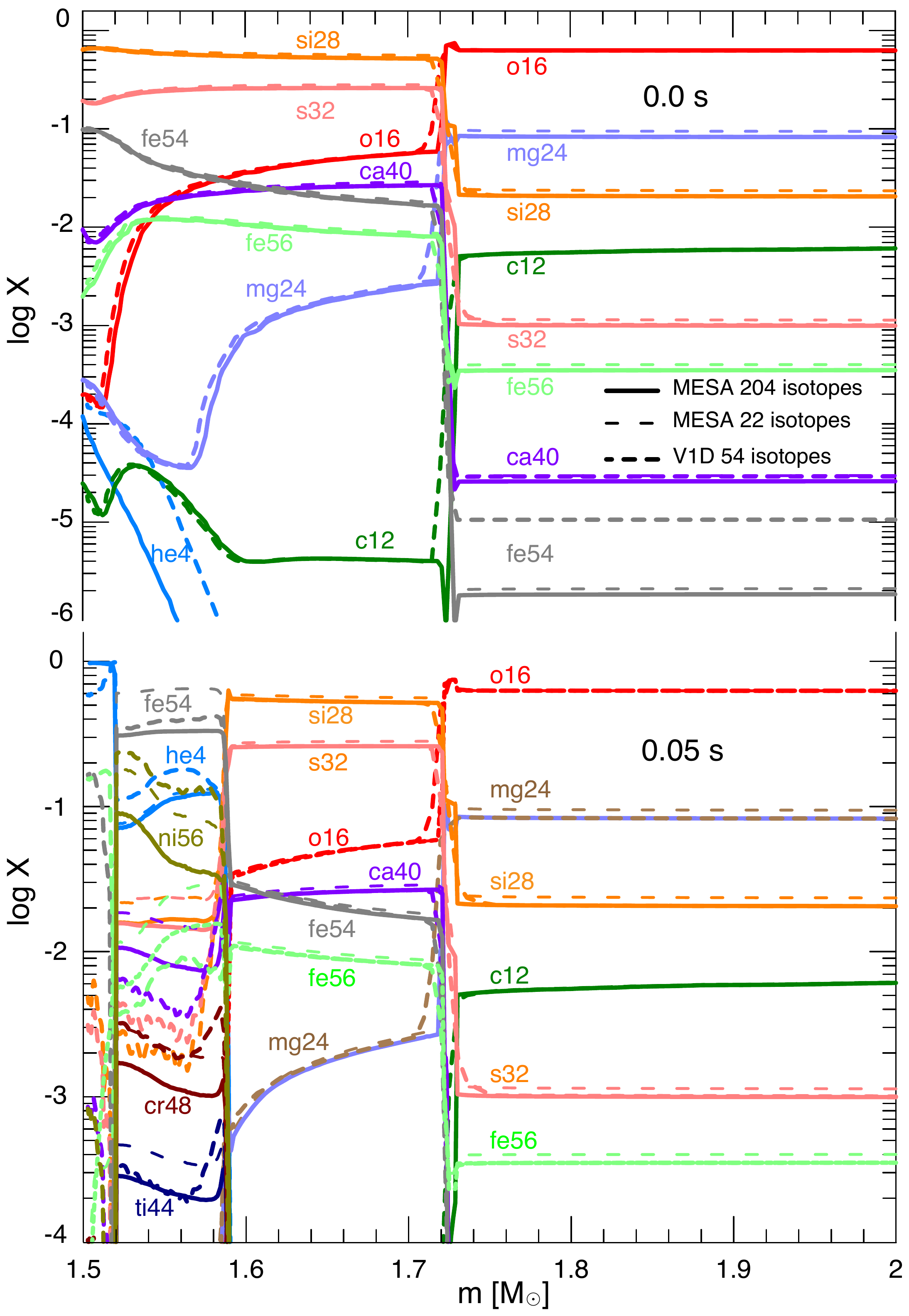}
\caption{
Nucleosynthesis profiles of selected isotopes for the
1.57$\times$\foe\ energy deposition test case at 0.0\,s (top) and 0.05\,s (bottom).
The dashed lines show the \MESA\ results with a 22 isotope network,
the solid lines  show the \MESA\ results with a 204 isotope network,
and the long dashed lines show the  \voned \ results with a 54 isotope network.
}
\label{f.comp_mesa_v1d}
\end{center}
\end{figure}

\begin{figure}[!htb]
\begin{center}
\includegraphics[width=\apjcolwidth]{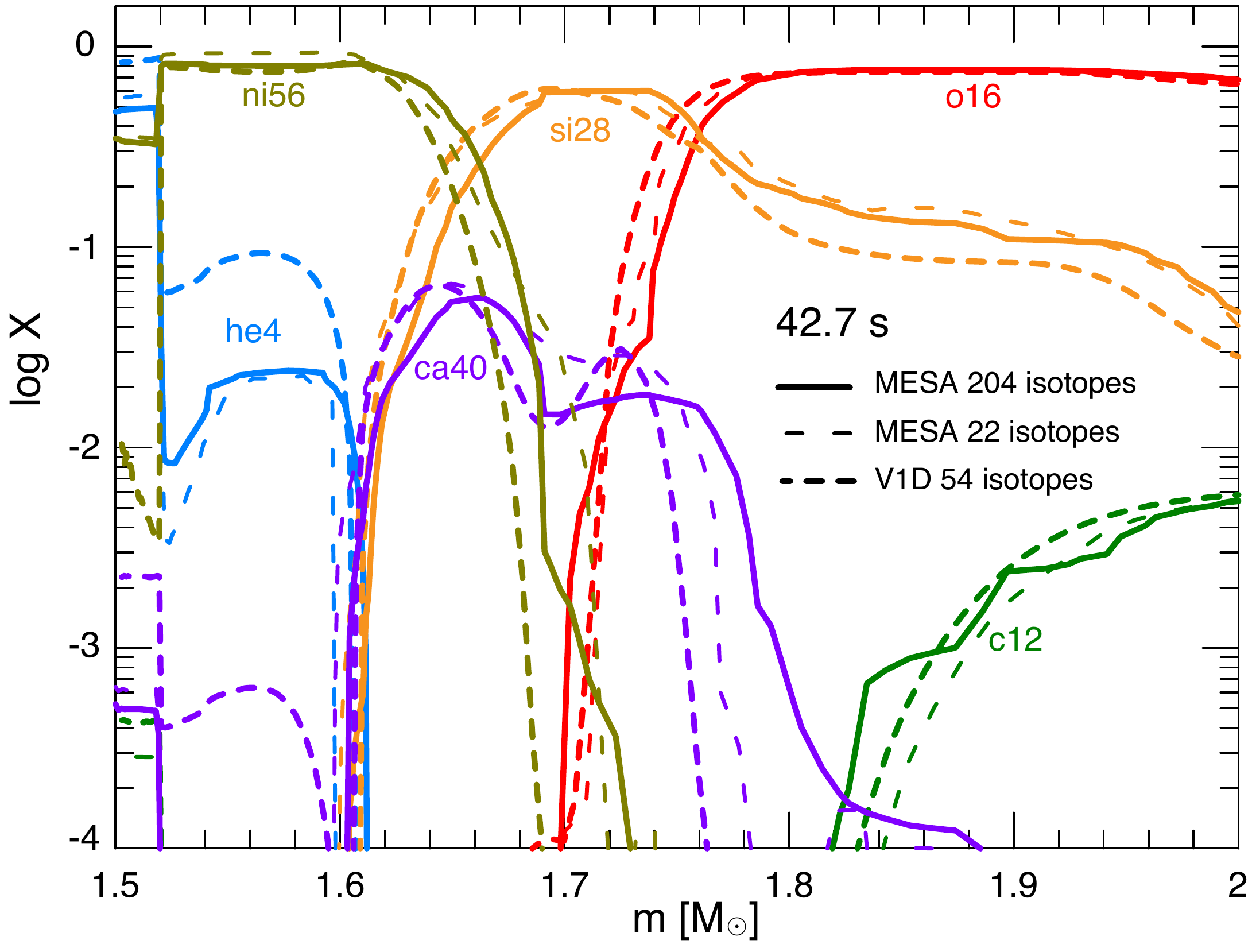}
\caption{
Same as Figure \ref{f.comp_mesa_v1d} but at a time after all nucleosynthesis
has completed.
}
\label{f.comp_mesa_v1d_final}
\end{center}
\end{figure}

Figures \ref{f.comp_mesa_v1d} and \ref{f.comp_mesa_v1d_final} compare
the mass fraction profiles of \MESA\ with a 22 isotope network, \MESA\
with a 204 isotope network, and \voned\ with a 54 isotope network. The
first comparison at 0.0\,s shows the impact of mapping from the
pre-supernova 204 isotope network to the networks used in the shock
nucleosynthesis test.  The next comparison at 0.05\,s is at the end of
the energy deposition phase.  The final comparison at 42.7\,s is after
explosive nucleosynthesis has completed.  In all cases, the
silicon-rich and oxygen-rich shells are strongly influenced by the
explosion; the former primarily for the production of \nickel[56] and
the latter primarily for the production of \silicon[28] and
\sulfur[32].
\add{The $^{56}$Ni yields at 42.7\,s are 0.092\,\msun\ for \voned,
0.087\,\msun\ for \mesa\ with 22 isotopes, and
0.096\,\msun\ for \mesa\ with 204 isotopes.}

Overall the agreement between \MESA\ and \voned\ on this strong
explosion is very good.  The small differences between \MESA\ and
\voned\ in Figures \ref{f.comp_mesa_v1d} and
\ref{f.comp_mesa_v1d_final} are due to the difference in mapping
procedures.

\MESA\ follows rules for mapping isotopes from one network to another
network: If an isotope present in the old network is also present in
the new network, then the abundance from the old network is copied to
the abundance for the new network.  Isotopes in the new network that
are not present in the old network are initially given a mass fraction
of zero. \MESA\ then separately renormalizes classes of isotopes to
have the same total mass fraction in the new network as in the old
network.  The classes are neutrons, hydrogen, helium, carbon,
nitrogen, oxygen, and other metals.  This procedure guarantees that
the sum of the mass fractions in a given class will be the same in the
new network as in the old network. \voned's mapping procedure for
isotopes is the following.  For any isotope present in the \voned\
network but absent in the \mesa\ input the mass fraction is set to the
solar metallicity value.  When an isotope included in the \mesa\ input
is absent in the \voned\ network, this isotope is left out in the
\voned\ simulation.  After completing the mappings, the resulting
composition is renormalized so that the sum of the mass fractions is
unity.


\begin{figure}[!htb]
\begin{center}
\includegraphics[width=\apjcolwidth]{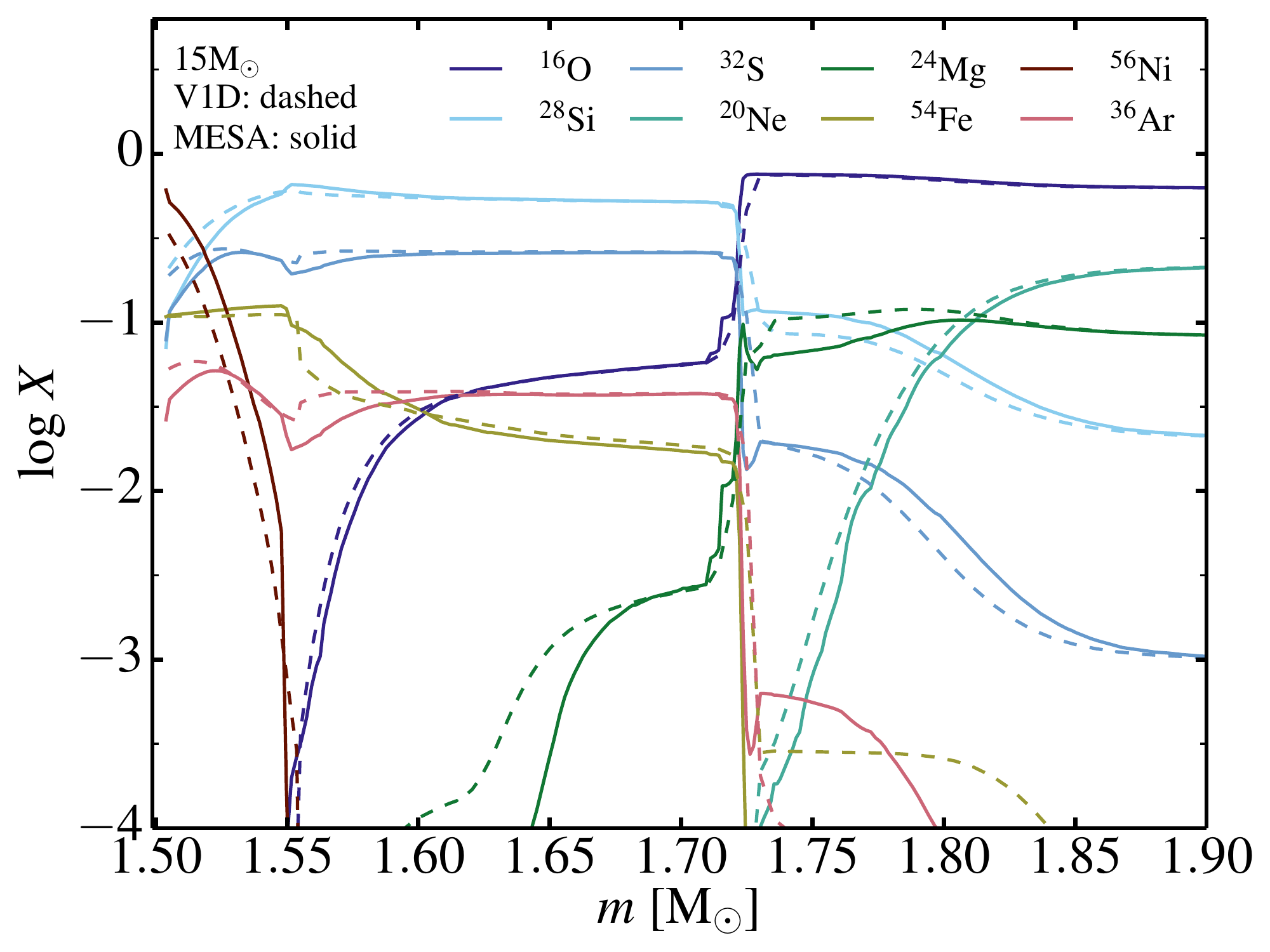}
\caption{ Composition profiles of the eight most abundant isotopes at
  1\,s in the inner 0.4\,\msun\ of the ejecta for \mesa\ (solid) and
  \voned\ (dashed) simulations of a \foe\ explosion in the 15\,\msun\
  model.  The Si-rich and O-rich shells have been influenced by
  explosive nucleosynthesis, the former primarily for the production
  of \nickel[56] and the latter primarily for the production of
  \silicon[28] and \sulfur[32].  }
\label{f.comp_mesa_v1d_1s}
\end{center}
\end{figure}

Our second test case is a lower power explosion.  We inject
1.326$\times$\foe\ in 1.0\,s over 0.05\,\msun\ of the mass cut, which
is also positioned at the outer edge of the iron core at 1.5\,\msun.
The total energy after the deposition phase is \foe.

In Figure~\ref{f.comp_mesa_v1d_1s}, we show the composition profiles
for the eight most abundant isotopes in the inner
$\approx$\,0.4\,\msun\ at the end of the energy-deposition phase
(i.e., at 1\,s).  The correspondence between \MESA\ and \voned\ is
again very good.  The \nickel[56] mass fraction approaches unity -- it
would reach unity if we appreciably increased the power (see Figure
\ref{f.comp_mesa_v1d} for example). Some \nickel[58] is produced in
the same region, while \iron[54] is synthesized in the layers
immediately above.
\add{The $^{56}$Ni yields at 1.0 s are 0.0041 \msun \ for \voned,
and 0.011 \msun \ for \mesa \ with 204 isotopes.}


This work shows that the power of the explosion has a a significant
impact on the abundance profiles.  In the high power explosion, the
yield of \nickel[56] is $\approx$ 10 times larger and the \helium[4]
is several orders of magnitude more abundant.  The nucleosynthesis of
the low power explosion is completed at end of deposition phase at 1.0
s, while nucleosynthesis in the high power case continues for
$\approx$ 30 s.  This sensitivity suggests potentially observable
signatures between low and high power explosions.  In addition, the
explosive nucleosynthesis that takes place in core-collapse supernovae
is sensitive to the way the explosion is triggered.  With the approach
we use (fixed power during the energy deposition phase), we find that
increasing the explosion energy (at a given power), the power (at a
given explosion energy), or both alters the amount of mass burnt.
Moving the mass cut deeper into denser layers considerably enhances
the amount of burnt material but this material may fall back rather
than be ejected. Moving the mass cut further out into lower-density
regions may completely quench the production of \nickel[56], in favor,
for example, of \silicon[28].  It is thus important to keep in mind
that the piston or thermal explosion trigger is artificial and that
the yields from explosive nucleosynthesis bear significant
uncertainties.



\section{Improved Treatment of Mass Accretion}\label{s.accretion}

Adding mass to a star requires a way to accurately and efficiently
compute the thermal state of the freshly accreted material in the
outermost layers. This is simplified by a hierarchy of timescales.
For accretion at $\dot M$, there are two important
timescales at a given location, $m$, the thermal time
$\tau_{\rm th}\simeq (M-m)C_PT/L$, where $C_P$ is the specific heat
at constant $P$, and the time to accrete this same layer, $\tau_{\rm  acc}\simeq (M-m)/\dot M$.  
Near the surface, $L\gg C_PT\dot M$,
implying that $\tau_{\rm th}\ll \tau_{\rm acc}$, so that these layers
have ample time to relax to the thermal equilibrium configuration
fixed by $L$ \citep{NomotoSugimoto1977,Nomoto1982,TownsleyBildsten2004}.  

In cases where $L$ arises solely from compression of material, such as
a very rapidly accreting star of high $\dot M$ or an old, cold
accreting WD, then $L\sim C_PT_{\rm b}\dot M$, where $T_{\rm b}$ is
the temperature at the degenerate/nondegenerate transition in a WD
\citep{TownsleyBildsten2004} or of the core in a normal star. 
Even in these
cases, the outer layers have $T\ll T_b$, allowing the
inequality $\tau_{\rm th}\ll \tau_{\rm acc}$ to hold.  This also implies that the thermal state of the
arriving material is unimportant, allowing us to safely use the
approximation that material arrives with the same entropy as the
photosphere, since material relaxes toward this on the very short
$\tau_{\rm th}$ at the photosphere.\footnote{A possible exception to
this case is rapidly accreting pre-main sequence stars where
the accretion shock is so
optically thick that the material's entropy remains high and is
advected inward \citep{PallaStahler1990}.} 
Even when $\dot M$ varies on short timescales, using an averaged
accretion rate is a good approximation for computing the evolution of
the interior layers due to their long $\tau_{\rm th}$
\citep{PiroArrasBildsten2005,TownsleyGaensicke2009}.

The timescale hierarchy $\tau_{\rm th}\ll \tau_{\rm acc}$
implies that the outer regions evolve nearly homologously in the
fractional mass coordinate $q=m/M$ \citep{SugimotoNomoto1975}.  Hence,
the thermal profile (e.g. the run of $T$ with $P$ or $\rho$) of the
outer layer is nearly constant in time even as fluid elements are
compressed to higher pressures and have $T(m)$ increase. More
formally, $T(q)$ varies slowly in time near the surface, where
$(1-q)\ll1$.  
This motivates reformulating the Lagrangian based form of 
\begin{equation}
\label{eq:standardDsDt} 
\epsilon_{\rm grav} = -T\frac{Ds}{Dt} \equiv -T\left(\frac{\partial
s}{\partial t}\right)_m\  
\end{equation}
that is needed in the energy equation, $\partial L/\partial m =
\epsgrav + \epsnuc - \epsnu$
(\citealp{Sugimoto1970,SugimotoNomoto1975}; \mesaone),
to a version in the coordinate $q$,
\begin{equation}
\label{eq:DsDt}
\epsgrav 
= -T\left(\frac{\partial s}{\partial t}\right)_{q}
+ T\frac{d\ln M(t)}{dt}\left(\frac{\partial s}{\partial\ln q}\right)_t\ .
\end{equation}
\citet{SugimotoNomoto1975}, and later works based on it, denote the second
term on the right hand side the ``homologous'' term.  Physically it is
the local loss of entropy in the fluid element as it is compressed to
higher pressure. They label the first term on the right hand
side the ``non-homologous'' term. It arises from the much slower
departure of the outer layers from simple homologous evolution on a timescale $M/\dot M$.

\MESAstar\ includes the ability to have an inner inert core of mass
$M_c$.  In this situation $q = (m-M_c)/(M-M_c)$ rather than the more
typical $m/M$.  For simplicity here, we will use $M_c$=0.  Approximate
homology holds in either case for $1-q\ll 1$.

In \mesatwo, following the work of \citet{TownsleyBildsten2004}, only
the homologous term in Equation \eqref{eq:DsDt} was included in
$\epsgrav$ in and near regions of newly accreted material.  We also
described the huge advantage of such an implementation, as it allows
the mass added per timestep to be much larger than the smallest cell mass
near the surface, while maintaining accurate thermal profiles at low pressures.
However, leaving out the non-homologous term can create a discontinuity in 
$\epsgrav$ at the location where the standard Lagrangian derivative, 
Equation \eqref{eq:standardDsDt}, begins to be used. We now describe the
improvement we have made to \MESAstar\ so that it now includes both
the homologous and non-homologous terms. Hence, the two forms of
$Ds/Dt$ are physically equivalent and there is no longer any
discontinuity.

%

\subsection{Lagrangian and Homologous Regions}
\label{s.accretion_meshregions}

The independent coordinates used for writing the time-dependent
structure of the star are $m$ and $t$, and for fluid elements deep
within the star at both timesteps, the conventional form of Equation
\eqref{eq:standardDsDt} is adequate.  One numeric subtlety of
accretion is that the derivative at constant $m$ cannot be evaluated for
material that is not present in the star at the beginning of the
timestep.  However, the simplification available when $\tau_{\rm th}\ll \tau_{\rm acc}$, 
manifest in Equation \eqref{eq:DsDt}, enables $\epsgrav$ to be
evaluated in the outer regions.

When $T$ and $\rho$ are used as independent variables, we write 
\begin{equation}
\label{eq:epsgrav_lag_trho}
\epsgrav = -C_PT\left[ (1-\nabla_{\rm ad} \chi_T)\left(\frac{\partial \ln T}{\partial t}\right)_m
 -\nabla_{\rm ad}\chi_\rho\left(\frac{\partial \ln \rho}{\partial t}\right)_m\right]
\end{equation}
in the interior of the star, as in \mesaone, and near the surface we choose to write, using Equation \eqref{eq:DsDt},
\begin{equation}
\label{eq:epsgrav_hom}
\epsgrav = \epsilon_{\rm grav, nh} + \epsilon_{\rm grav, h}\ ,
\end{equation}
where
\begin{equation}
\epsilon_{\rm grav,nh} = -C_PT\left[ (1-\nabla_{\rm ad} \chi_T)\left(\frac{\partial \ln T}{\partial t}\right)_q
 -\nabla_{\rm ad}\chi_\rho\left(\frac{\partial \ln \rho}{\partial t}\right)_q\right]
\end{equation}
and
\begin{equation}
\epsilon_{\rm grav,h} = \frac{C_PTGm\dot M}{4\pi r^4P}\left(\nabla_{\rm ad}-\nabla_T\right)\ .
\end{equation}
Here $\nabla_T=d\ln T/d\ln P$ is the $T$-$P$ profile in the star, and we have used the thermodynamic derivatives $\nabla_{\rm ad}=(\partial \ln T/\partial \ln P)_s$, $\chi_T=(\partial P/\partial T)_\rho$, and $\chi_\rho=(\partial P/\partial\rho)_T$.
There is also a transition region 
where a weighted combination of these forms is used, with weights varying linearly in $m$.

\begin{figure}[!htbp]
\resizebox{\apjcolwidth}{!}{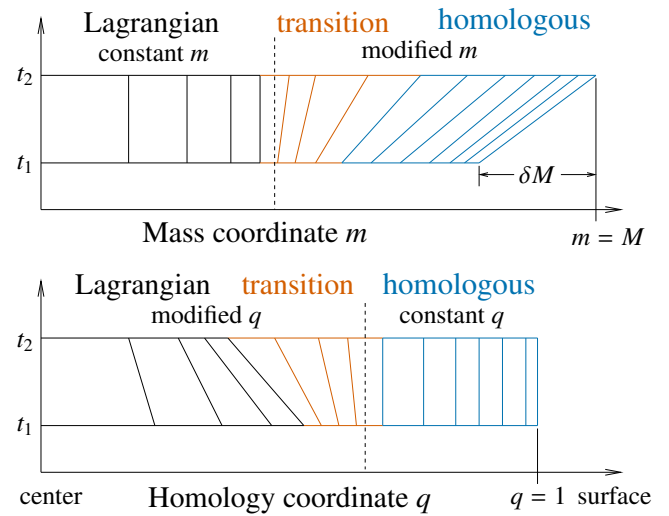}
\caption{\label{fig:mesh_regions}
Illustration of the \MESAstar\ mesh in both $m$ and $q$ for a timestep
in which mass is added to the star over the time interval $t_1$ to $t_2$.
Vertical and slanted lines indicate cell boundaries.  Cell size is
exaggerated; there can be many cells in the newly added material.
Three regions are chosen in the process of expanding the mesh for the
added material, an inner Lagrangian region, an outer homologous
region, and a transition region.
}
\end{figure}

Placement of the transition is related to the mesh.  The
\MESAstar\ mesh structure is unchanged from that discussed in \mesaone
\ and \mesatwo.  An illustration of the mesh regions and an indication
of the behavior of cell boundaries during accretion is shown in
Figure~\ref{fig:mesh_regions}.  In the case of mass loss, analogous
operations are performed; here we focus only on mass gain.

A mass $\delta M=\dot M(t_2-t_1)$ is added to the star from time $t_1$
to $t_2$.  Sizes are exaggerated for clarity; there are generally many
zones in each region, possibly hundreds in the newly added material.
The diagram is shown in both mass coordinate, $m$, and homology
coordinate, $q$.  Before each timestep, \MESAstar\ adjusts the initial
mesh resolution by splitting or merging cells based on local gradient
conditions, producing an adjusted resolution mesh, which we show at
$t_1$.  No simulation time elapses during that process. When
constructing the mesh that will represent the star at $t_2$,
\MESAstar\ divides the new mesh into three regions: an inner
Lagrangian region, in which the $m$ boundaries of each cell are
preserved during the timestep, a transition region, and an outer,
homologous region, in which the $q$ boundaries of each cell are the
same across the timestep.  The result of this operation is an expanded
mass domain shown at $t_2$ in Figure~\ref{fig:mesh_regions}.

Time derivatives appearing in the equations for physical evolution are
then estimated using first order differences.  In the Lagrangian mesh
region, the finite difference form of $(\partial /\partial t)_m$
involves a simple same-cell difference.  Similarly, in the homologous
mesh region, the finite difference form of $(\partial/\partial t)_q$
involves a same-cell difference.  In most cases, by design, these
same-cell differences are for values whose changes, e.g.\ $\delta \ln
T$, are directly available from the iterative solution of the new
structure, allowing us to avoid the numerical problems inherent in
subtracting two almost identical numbers.
In the transition region both $m$ and $q$ coordinates of cells have
been modified, so we cannot do a same-cell difference for either
$(\partial /\partial t)_m$ or $(\partial/\partial t)_q$.  Instead, we
interpolate values from the model at the start of the step to
corresponding locations in $m$ or $q$ at the end of the step.

A smooth and accurate value for $\epsgrav$ in the transition region is
important.  To ensure this, the location of the transition region is
selected to reduce the differences between the constant $m$ and
constant $q$ forms of the time derivative and maintain accurate finite
differencing.  As a simple mechanism to control these, we limit, in
units of cell size, the offset in the interpolation used to translate
locations from the beginning to the end of the timestep
\citep{Milesetal2015}.  Using the cell size implicitly takes advantage
of the limits imposed by mesh controls on the maximum possible
magnitude of cell-to-cell changes in key variables, including the
variables of interest for $\epsgrav$ time derivatives.

\subsection{Testing}
\label{s.accretion_testing}

In order to demonstrate that the interface between the outer
homologous region and the inner Lagrangian region provides a smooth
profile that is independent of timestep size, we have repeated the test
shown in \mesatwo\ section 5.3.  This test involves accretion of solar
composition material onto a WD at $10^{-10}M_\odot$~yr$^{-1}$.  We use
the same starting model as in \mesatwo, which was produced by
accreting hydrogen-rich material through several hydrogen shell
flashes on a $0.6M_\odot$ WD with an initial core temperature of about
$10^7$~K.  As accretion proceeds, the total accumulated accreted mass,
$M_{\rm acc}$, increases up to a maximum which causes the hydrogen
flash and nova runaway, $M_{\rm ign}$.

\begin{figure}[!htbp]
\centering
\includegraphics[width=\apjcolwidth]{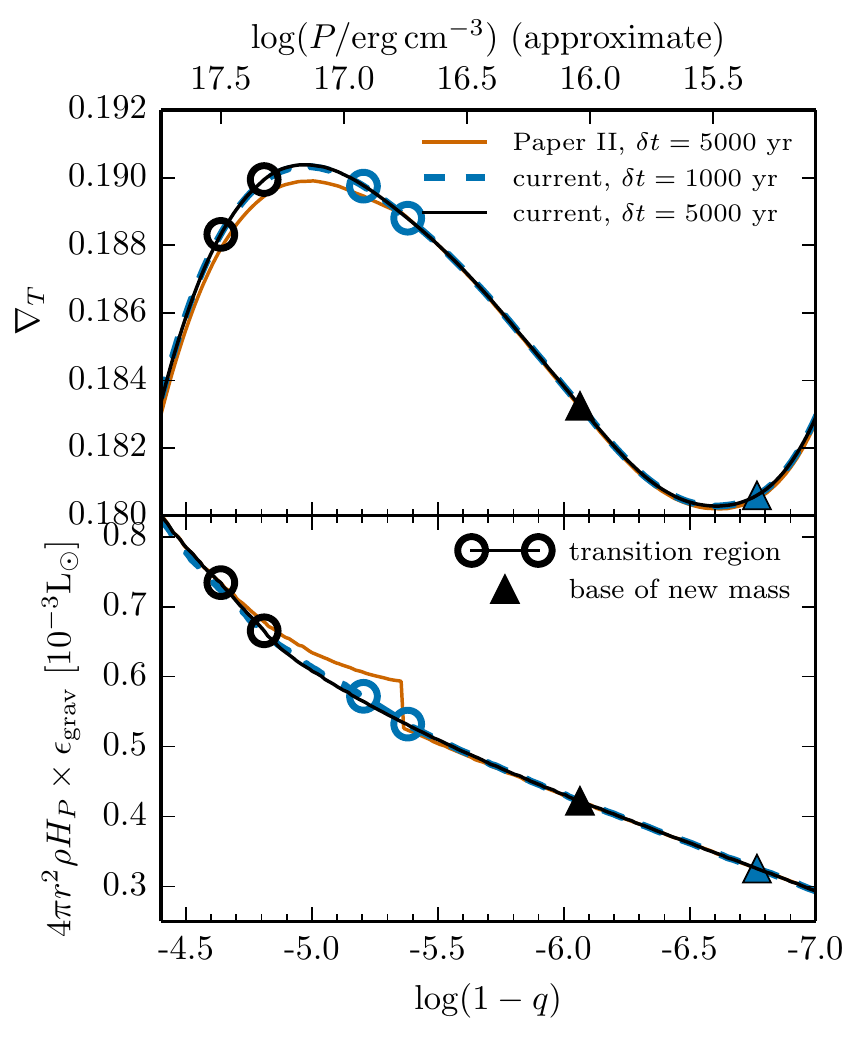}
\caption{\label{fig:eps_grav_profile}
Profiles spanning the Lagrangian-homologous grid transition region in a
$0.6M_\odot$ WD with a core temperature of $10^7$ K undergoing
accretion of solar composition material at a rate of $10^{-10}M_\odot$
yr$^{-1}$.  At the time shown $M_{\rm acc}/M_{\rm ign}=0.2$.  Shown
for comparison are the results of the treatment discussed here for a
5000~yr timestep and a 1000~yr timestep and the treatment discussed in
\mesatwo.  For the current treatment the
transition from homologous to Lagrangian  is indicated by two open circles at either end of
the region.  The location $\delta M$ in from the surface, the base of
material newly accreted this step, is indicated by the
triangle. 
}
\end{figure}

Profiles near the transitions region are shown in
Figure~\ref{fig:eps_grav_profile} at the time when $M_{\rm acc}/M_{\rm ign}=0.2$, 
which had the most severe discontinuity in \mesatwo.  The first panel
shows $\nabla_T$ and the second panel shows $4\pi r^2\rho
H_P\times \epsilon_{\rm grav}$, where $H_P$ is the pressure scale
height.  This is the amount of energy being released due to the
$T\,Ds/Dt$ term in the energy equation within a scale height, and has
units of luminosity.  We have chosen a timestep of $5,000$~yr, which
places the homology-Lagrangian transition region in a similar place to
the location of the discontinuity in \mesatwo.

The orange curve in Figure \ref{fig:eps_grav_profile} was computed
using \MESA\ version r4664, as used in \mesatwo.  This displays
the discontinuity due to using $\epsgrav$
from Equation \eqref{eq:epsgrav_lag_trho} and Equation
\eqref{eq:epsgrav_hom} with only the homologous term and a transition
point a factor of 5 deeper in pressure than $\delta M$.  
The black
curve shows the same simulation with the same timestep for the
treatment discussed here, in which $\epsilon_{\rm grav,nh}$ is
included and the transition region, indicated with solid dots at
either end, is placed as described in
Section~\ref{s.accretion_meshregions}.  We see that, away from the
transitions, $\epsgrav$ is unchanged from the values found in
\mesatwo, in which only the homologous term was used in the exterior.
We also show the result for a timestep of 1,000~yr is
indistinguishable on this plot; the $\epsgrav$ profile differs by a
fraction of a percent at the edges of the transition region, and less
elsewhere.  In the current treatment the profiles are independent of
timestep size.


\section{Weak Reactions} \label{s.weak}

\textcolor{red}{An erratum for this paper was published as \citet{Paxton16}.  The errors reported there have been corrected in this version of the paper.}

The \rates\ module provides weak reaction rates for hundreds of
isotopes.  By default, when atoms are fully ionized, these rates are
based (in order of precedence) on the tabulations of
\citet{Langanke00}, \citet{Oda94}, and \citet{Fuller85}.  These tables
span a wide range of density and temperature,
$1 \le \log (\rho \Ye/ \mathrm{g\ cm}^{-3}) \le 11$
and
$7 \le \log (T/\mathrm{K}) \la 10.5$,
but are relatively coarse, with 11 points in the $\rho \Ye$
dimension ($\Delta \log \rho \Ye = 1$) and 12 points in the $T$
dimension ($\Delta \log T \approx 0.25$).

These grids include the thermodynamic conditions where the electrons
are degenerate and relativistic, which are realized for example in
massive white dwarfs and cores of intermediate mass stars.  Under
these conditions, the rates of electron-capture and beta-decay
reactions are sufficiently sensitive to density and temperature that
they can change by tens of orders of magnitude between adjacent points
in these tables.  Linear or cubic interpolation cannot accurately
reproduce the value of the rate between the tabulated points.

The difficulty of interpolating in coarse rate tabulations
was discussed by \citet{Fuller85}, who proposed a physically-motivated
interpolation scheme, hereafter referred to as Fuller, Fowler, and
Newman (FFN) interpolation.  Their procedure assumes the rate has the
form given by a single transition between the parent and daughter
nuclear ground states.  However, the true rate may be dominated by
allowed transitions to or from excited states in the parent or
daughter nucleus.  This is almost always the case when the ground
state to ground state transition is highly forbidden.  The specific
transition that dominates the rate may change over the range of
thermodynamic conditions covered by the table.  The FFN interpolation
method does not account for these complications.

\begin{figure}
\centering
\includegraphics[width=\apjcolwidth]{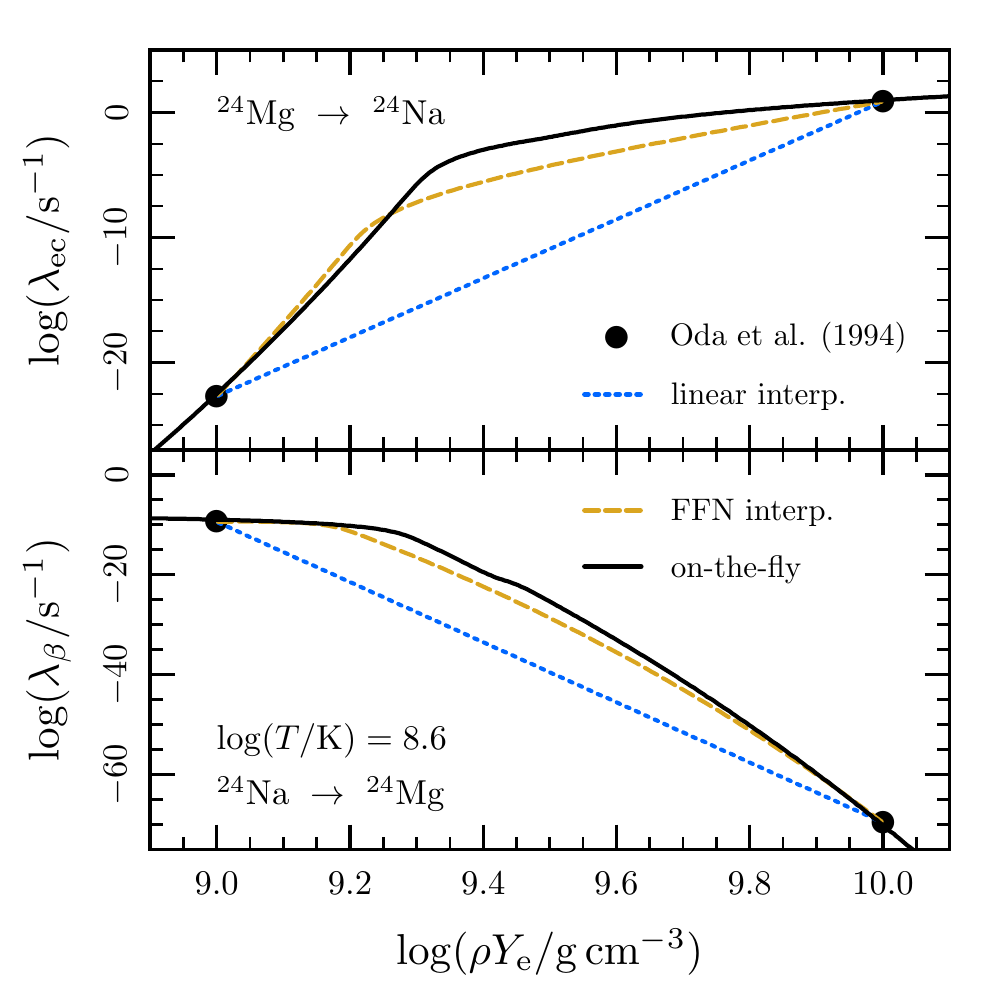}
\caption{
The top panel (bottom panel) shows the rate of electron capture on
$\magnesium$ (beta decay of $\sodium[24]$) as a function of density at
a fixed temperature of $\log(T/\mathrm{K})=8.6$.  The \citet{Oda94}
tabulated points are shown as black dots.  The dotted line shows the
result of using linear interpolation between the tabulated points.
The dashed line shows the result of using the physically-motivated
interpolation method suggested by \citet{Fuller85}. The solid line
shows the rate calculated using the on-the-fly rate calculation
capability of \mesa\ documented in this section.  Slight differences
between points and the line are due to differences in the input
nuclear data.}
\label{fig:interpolation}
\end{figure}

Figure~\ref{fig:interpolation} compares the results of the
interpolation methods described in the preceding paragraphs with the
on-the-fly approach that we have implemented in \mesa\ and will be
described here.  It shows the electron-capture rate on $\magnesium$
and beta-decay rate of $\sodium[24]$ at fixed temperature.  Linear
interpolation of these coarse tables fails to reproduce the rapid
variation in the rate.  The FFN interpolation method produces curves
with characteristic shapes more similar to the true rate, but because
the $Q$-value is that of the ground state to ground state transition
and not that of the transition that dominates the rate, the density
dependence is not correct in detail.

In recent years, a number of authors have discussed the importance of
well-sampled weak rates in capturing the influence of these processes
on stellar evolution.  This can be achieved by generating denser
tables for the specific reactions of interest or by using analytic
approximations to the rates \citep[e.g.,][]{Toki13, MartinezPinedo14}.
We now present a capability by which \mesa\ can
calculate weak reaction rates on-the-fly from input nuclear data.
This removes the potential for interpolation artifacts.  It also
enables easy experimentation in cases where the input nuclear data may
not be well-measured.  We begin with an
overview of how we calculate these weak rates and
illustrate their utility and a few applications.

\subsection{Calculation of Weak Rates}
\label{s.weak-details}

Consider two nuclei $A \equiv (Z,N)$ and $B \equiv (Z-1, N+1)$ that have
two states connected by an electron-capture transition
\begin{equation}
  \label{eq:electron-capture}
  A + e^- \to B + \nu_e
\end{equation}
and beta-decay transition
\begin{equation}
  \label{eq:beta-decat}
  B \to A + e^- + \bar{\nu}_e~.
\end{equation}
The energy difference between the ground states can be written as
\begin{equation}
  \label{eq:1}
  Q_\mathrm{g} =
 \begin{cases}
(M_{A} - M_{B})c^2 &\text{ for electron capture},\\
(M_{B} - M_{A})c^2 &\text{ for beta decay}.
\end{cases}
\end{equation}
where $M_A$ and $M_B$ are the \textit{nuclear} rest masses of the
ground states.  The total energy difference between any two states can
be written as
\begin{equation}
  \label{eq:q}
  Q_{ij} = Q_\mathrm{g} + E_{i} - E_{f}~,
\end{equation}
where $E_{i}$ and $E_{f}$ are the energies of the initial and final
states measured relative to the ground state.  For the transitions
that we consider here, $Q_\mathrm{g} < 0$ and $Q_{ij} < 0$ for electron capture
and $Q_\mathrm{g} > 0$ and $Q_{ij}>0$ for beta decay.

In this section, we use $J$ to represent the nuclear
spin. We work in the allowed approximation, which neglects all total
lepton angular momentum ($L = 0$).  This restricts us to Fermi
transitions, where the total lepton spin is $S = 0$, and therefore the
initial and final nuclear spins are equal ($J_i = J_f$), and
Gamow-Teller transitions, where $S = 1$, and therefore $J_i = J_f, J_f
\pm 1$ (excluding $J_i = J_f = 0$). In both cases, there is no parity
change: $\pi_i \pi_f = +1$ \citep[e.g.,][]{Commins73}.

The total rate of the process (electron capture or beta decay) is the
sum of the individual transition rates from the $i$-th parent state to
the $j$-th daughter state, $\lambda_{ij}$, weighted by the occupation
probability of the $i$-th parent state, $p_i$.
\begin{equation}
  \lambda = \sum_{i} p_i \sum_{j} \frac{\ln 2}{(ft)_{ij}} \Phi(\mue, T, Q_{ij}),
\end{equation}
where $(ft)$ is the comparative half-life and can be either measured
experimentally or calculated from theoretical weak-interaction
nuclear matrix elements. The $i$-sum is over all parent states and the
$j$-sum is over all daughter states.  The occupation probability is
\begin{equation}
  p_i = \frac{(2 J_i + 1) \exp\left(-\beta E_i\right)}{\sum_k (2 J_k + 1) \exp(-\beta E_k)}~,
\end{equation}
where we define $\beta = (\kB  T)^{-1}$.  The quantity $\Phi$ is a
phase space factor which depends on the electron chemical potential
$\mue$ (including the electron rest mass), on the
temperature $T$, and the energy difference $Q_{ij}$. The value of
$\Phi$ for electron capture is
\begin{equation}
\Phi_\mathrm{ec} = \frac{\exp({\pi\alpha Z})}{(\me c^2)^5} \int_{-Q_{ij}}^{\infty} \frac{E_{\rm e}^2 (E_{\rm e}+Q_{ij})^2}{1 + \exp[\beta (E_{\rm e} - \mue)]} dE_{\rm e}~,
\end{equation}
where $\alpha$ is the fine structure constant.  For beta decay it is
\begin{equation}
\Phi_\beta = \frac{\exp({\pi\alpha Z})}{(\me c^2)^5} \int_{\me c^2}^{Q_{ij}} \frac{E_{\rm e}^2 (E_{\rm e}-Q_{ij})^2}{1 + \exp[-\beta (E_{\rm e} - \mue)]} dE_{\rm e}~.
\end{equation}
Similarly, the total rate of energy loss via neutrinos is
\begin{equation}
  \varepsilon_\nu = \sum_{i} p_i \sum_{j}\frac{ \me c^2 \ln 2}{(ft)_{ij}} \Psi(\mue, T, Q_{ij})~,
\end{equation}
The value of $\Psi$ for electron capture is
\begin{equation}
\Psi_\mathrm {ec} = \frac{\exp({\pi\alpha Z})}{(\me c^2)^6} \int_{-Q_{ij}}^{\infty} \frac{E_{\rm e}^2 (E_{\rm e}+Q_{ij})^3}{1 + \exp[\beta (E_{\rm e} - \mue)]} dE_{\rm e}~,
\end{equation}
and for beta decay it is
\begin{equation}
\Psi_\beta = \frac{\exp({\pi\alpha Z})}{(\me c^2)^6} \int_{\me c^2}^{Q_{ij}} \frac{E_{\rm e}^2 (E_{\rm e}-Q_{ij})^3}{1 + \exp[-\beta (E_{\rm e} - \mue)]} dE_{\rm e}~.
\end{equation}

In order to implement these equations in \mesa, we rewrite the
integrals in terms of Fermi-Dirac integrals, following appendix A of
\citet{Schwab15}.  \mesa\ implements fast quadrature routines to
evaluate integrals of this form.  Each time a weak rate is
needed, it is calculated on-the-fly.  We discuss the computational cost
of this procedure in Section~\ref{s.weak-limitations}.

Assuming thermal equilibrium, the energy released by
weak reactions depends only on total reaction rate, total
neutrino loss rate, energy difference between the nuclei, and the
electron chemical potential.  Therefore, the total specific heating
rate from a reaction is
\begin{align}
  \label{eq:qec}
  \epsilon_\mathrm{ec} & = \frac{n_A}{\rho} \left[(Q_\mathrm{g} + \mue) \lambda_\mathrm{ec} - \varepsilon_{\nu, \mathrm{ec}}\right]~, \\
  \label{eq:qbeta}
  \epsilon_\beta & = \frac{n_B}{\rho} \left[(Q_\mathrm{g} - \mue) \lambda_\beta - \varepsilon_{\nu, \beta}~ \right]~,
\end{align}
where $n_A$ and $n_B$ are the number densities of the species undergoing electron capture and beta decay, respectively, and $\rho$ is the total mass density.

\begin{figure}
\centering
\includegraphics[width=\apjcolwidth]{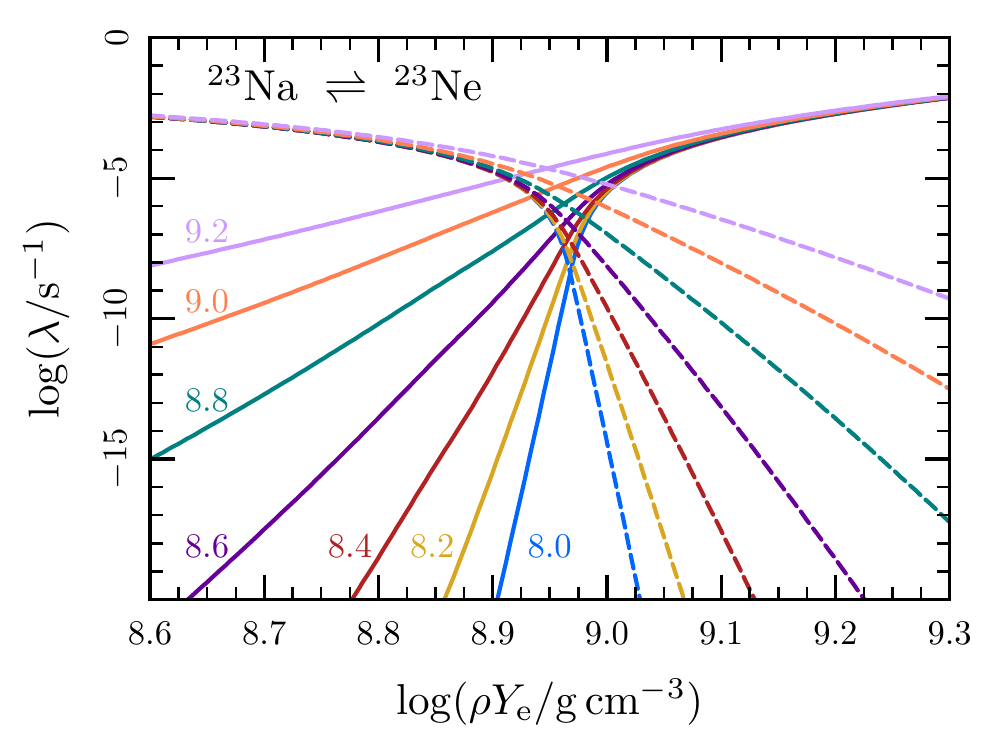}
\caption{
The electron capture (solid lines) and beta decay (dashed lines) rates
of the $\sodium[23]$--$\neon[23]$ Urca pair as calculated by \mesa,
using the on-the-fly methods described in this section.  The value of
$\log(T/\mathrm{K})$ is shown next to each electron capture line; the
beta decay line of matching color is at the same temperature.  The
rates vary rapidly, with both temperature and density, near the
threshold density, which is roughly in the center of the plot.}
\label{fig:a23-rates}
\end{figure}

Therefore, given a list of nuclear levels and the $(ft)$-values for
the transitions between them, \mesa\ can calculate the rates of
electron capture and beta decay and the corresponding energy
generation rates.  Typically only a few low-lying states and the
transitions between them are needed.  As an example,
Figure~\ref{fig:a23-rates} shows the rates for the
$\sodium[23]$--$\neon[23]$ Urca pair.

\subsubsection{Coulomb Corrections}
\label{s.coulombcorrections}

In a dense plasma, the electrostatic interactions of the ions and
electrons introduce corrections to the weak rates relative to those
which assume a Fermi gas of electrons and an ideal gas of ions.  Our
treatment of these effects, which is presented in appendix~B of
\citet{Schwab15}, is similar to appendix~A of \citet{Juodagalvis10}.

Since electron capture and beta decay change the ion charge, the
Coulomb interaction energy changes the energy difference between the
parent and daughter nuclear states.  To calculate this shift, we use
the excess
ion chemical potential $\mu_\mathrm{ex}$ from \citet{Potekhin09a}.  We
incorporate this effect by shifting the value of $Q_{ij}$, as defined in
Equation~\eqref{eq:q}, by an amount
$\Delta E = \mu_{\mathrm{ex,parent}} - \mu_{\mathrm{ex,daughter}}$.  This shift,
$Q_{ij}' = Q_{ij} + \Delta E$, enters the calculation of the phase space factors
and the energy generation rates.

The electron density relevant to the reaction rate is not the average
electron density, but rather the electron density at the position of
the nucleus.  This correction is accounted for as a
shift in the value of the electron chemical potential that enters the
phase space factor, $\mu'_{\rm e} = \mue + V_s$.  Values of $V_s$ have been
calculated by \citet{Itoh02}.  This correction does not enter the
energy generation rates because it has not changed the energy cost to
add or remove an electron.

\subsection{Applications}
\label{s.weak-applications}

When $\mue \la |Q|$, only the few electrons in the tail of the
Fermi-Dirac distribution have sufficient energy to overcome the energy
gap and capture on $A$ to form $B$.  Thus, the rate of
electron capture is small compared to the rate of beta decay, and so
isotope $B$ is favored in the equilibrium.  When $\mue \ga
|Q|$, there are only a few unoccupied states available to accept the
energetic electron from the beta decay. This final state blocking
means the rate of beta decay is small compared to the rate of
electron capture, and so isotope $A$ will be favored in the
equilibrium.

The shift in this equilibrium can have profound consequences when it
occurs in stellar interiors.  It modifies the composition, reduces
the electron fraction, and alters the thermal state of the plasma.  We
now discuss two applications of our on-the-fly treatment of the weak
rates: the Urca process and accretion-induced collapse.

\subsubsection{Urca Process}
\label{s.applications-urca}

\begin{figure}
  \centering
  \includegraphics[width=\apjcolwidth]{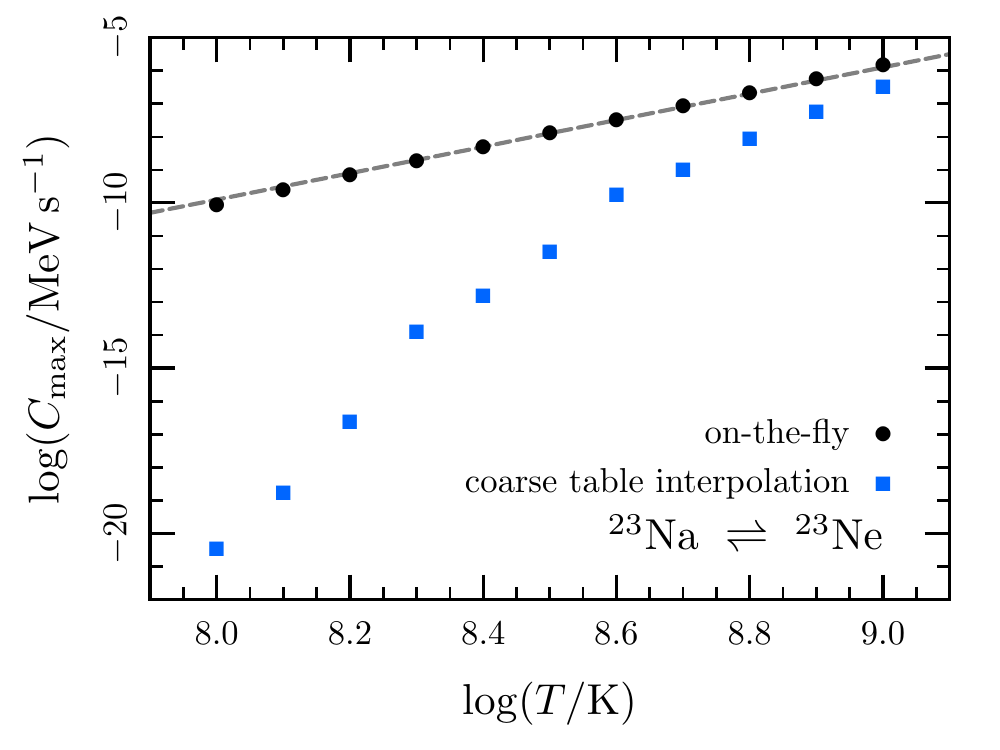}
  \caption{The effect of the interpolation method on the Urca process
    cooling rates.  The circles show the maximum value of $C$
    (Equation~\ref{eq:urca-C}) calculated using the on-the-fly methods
    discussed in this section; the squares show the results using the
    coarse tables of \citet{Oda94}.  Interpolation in these coarse
    tables severely underestimates the Urca cooling rates at low
    temperatures.  The dashed line shows the expected value of the
    cooling rate given by Equation~(\ref{eq:urca-Cmax}).}
  \label{fig:urca-23}
\end{figure}

When the ground state to ground state transition is
allowed (odd nuclei), the rates of electron capture and beta decay are
both significant when $\mue \approx |Q_\mathrm{g}|$.  Since each
reaction produces a neutrino which free-streams out of the star, this
can lead to significant cooling.  With a total number density of an
Urca species $n_U = n_A + n_B$, assuming the abundances are given by
the detailed balance condition
$n_A \lambda_\mathrm{ec} + n_B \lambda_\beta = 0$, the volumetric
neutrino cooling rate will be $n_U C$, where
\begin{equation}
  \label{eq:urca-C}
  C = \frac{\varepsilon_{\nu, \mathrm{ec}} \lambda_\beta + \varepsilon_{\nu, \beta} \lambda_\mathrm{ec}}{\lambda_\beta + \lambda_\mathrm{ec}}~.
\end{equation}
In the limit $\kB T \ll |Q|$, the maximum value of the Urca cooling rate at a given temperature has
a simple form \citep[e.g.,][]{Tsuruta70}
\begin{equation}
  C_\mathrm{max} = \frac{ 7 \pi ^4 \ln 2}{60}
  \left(\frac{\me c^2}{(ft)_\beta + (ft)_{\mathrm{ec}}}\right) 
  \left(\frac{\kB T}{\me c^2}\right)^4
  \left(\frac{Q}{\me c^2}\right)^2 \exp({\pi\alpha Z})~.
  \label{eq:urca-Cmax}
\end{equation}
Well-sampled rates such as those shown in Figure~\ref{fig:a23-rates}
are necessary to reproduce the correct Urca cooling rates.  We
illustrate this in Figure~\ref{fig:urca-23}, which shows
$C_\mathrm{max}$ for the $\sodium[23]$--$\neon[23]$ Urca process for
temperatures $10^8$ -- $10^9$ K.  The circles
show the results using the on-the-fly treatment described in this
paper; the squares show the results using the coarse tables of
\citet{Oda94}.  The dashed line shows the cooling rate expected
from Equation~(\ref{eq:urca-Cmax}) which is in excellent agreement
with the results of the on-the-fly method.  The Urca cooling rates
calculated from interpolating in coarse tables severely underestimate
the true cooling rate when $\kB T \ll |Q|$.

Thus, when the Urca process is important, well-resolved weak rates are
necessary to correctly capture the temperature evolution of the core
\citep{Toki13,Jones13}.
  \citet{Jones13} used MESA r3709 along with a denser table described
  in \citet{Toki13} to do their work.  The \citet{Toki13} table is not
  publicly available, so to reproduce the results of \citet{Jones14}
we save a model of an 8.8 $\Msun$ star at
$\log(\rhoc/\mathrm{g\,cm^{-3}}) = 8.95$ from our run with \mesa\
(version r3709) using the \citet{Jones14} inlists.
We then load this model into a newer \mesa\ version (r7503) that has
access to the on-the-fly weak rates and evolve this model using a
network with only the Urca process reactions.  During this phase other
nuclear reactions are not important to the central evolution.

\begin{figure}
  \centering
  \includegraphics[width=\apjcolwidth]{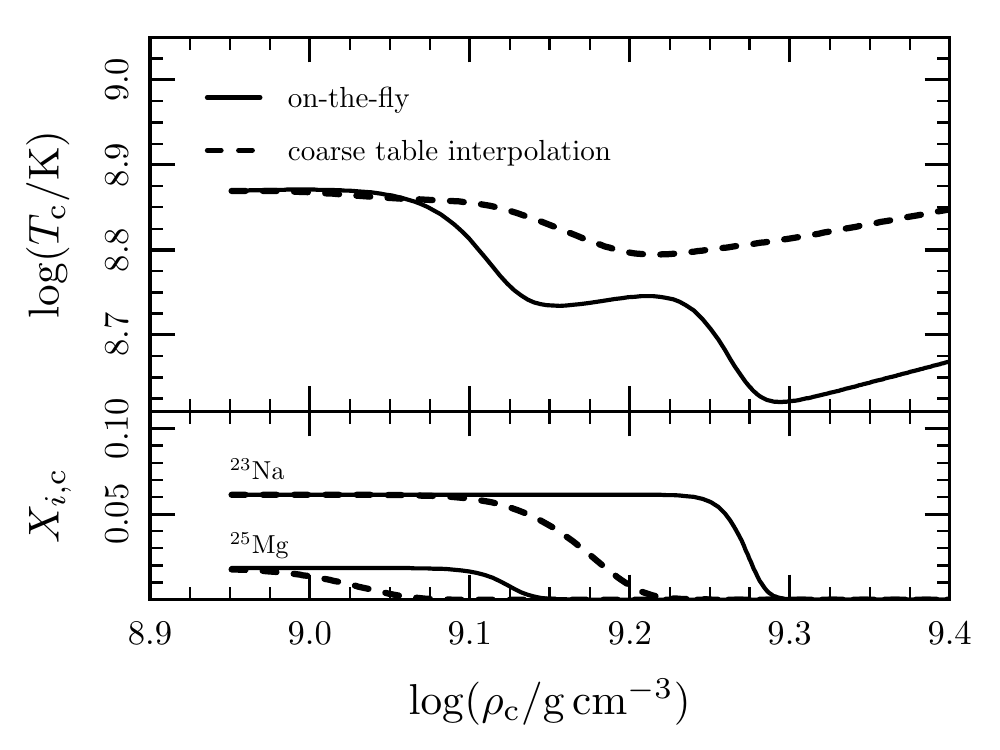}
  \caption{The top panel shows the evolution of $\Tc$ and $\rhoc$ in
    an 8.8 \msun\ star.  The bottom panel shows the central
    $\magnesium[25]$ and $\sodium[23]$ mass fractions.  The solid
    lines show the evolution using the on-the-fly rates, the dashed
    lines show the results when interpolating in coarse tables. The
    locations of the changes in mass fraction match the locations of
    cooling in the top panel. This demonstrates the importance of
    densely-sampled weak rates to the evolution of the core.}
  \label{fig:urca-star}
\end{figure}

Figure~\ref{fig:urca-star} shows the central temperature and density
of the core.  The solid lines show the evolution using the on-the-fly
rates, the dashed lines show the results when interpolating in coarse
tables.  The drops in temperature at
$\log(\rhoc/\mathrm{g\,cm^{-3}}) \approx 9.1$ and $\approx 9.25$
correspond to cooling from the $\magnesium[25]$--$\sodium[25]$ and
$\sodium[23]$--$\neon[23]$ Urca pairs, respectively.  The
corresponding shifts in composition can be clearly seen in the lower
panel. These results demonstrate the importance of densely-sampled
weak rates to the evolution of the core.

\subsubsection{Accretion-Induced Collapse}
\label{s.applications-aic}

When the ground state to ground state transition is forbidden (even
nuclei), the first transition to become significant is typically an
allowed transition into an excited state.  In these cases, the beta
decays from the daughter ground state are blocked and decays from
daughter excited states are strongly suppressed by the Boltzmann
factor.  Therefore significant cooling via the Urca process does not
occur.  Instead, since the captures are preferentially to an excited
state, significant heating occurs via gamma-ray emission as level
populations relax to a thermal distribution.

\begin{figure}
  \centering
  \includegraphics[width=\apjcolwidth]{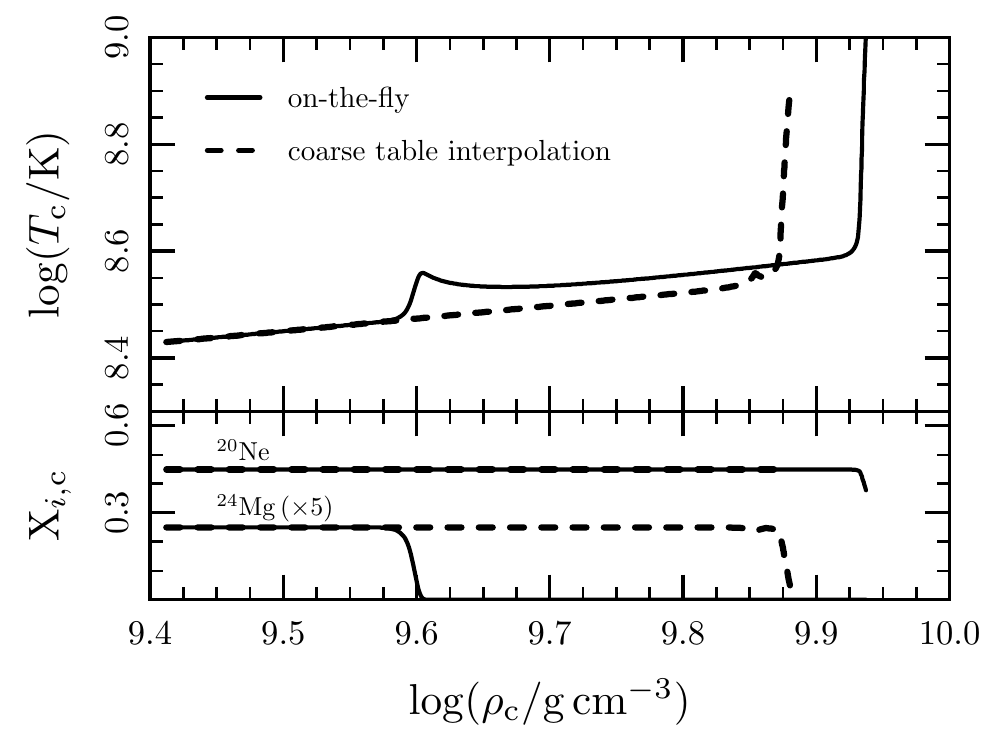}
  \caption{The evolution of a cold ONeMg WD towards AIC.  The top
    panel shows the evolution of $\rhoc$ and $\Tc$.  The bottom shows
    the central $\magnesium$ and $\neon$ mass fractions.  The solid
    lines show the evolution using the on-the-fly rates; the dashed
    lines show the results when interpolating in coarse tables.}
  \label{fig:aic}
\end{figure}

Two important capture chains occur in oxygen-neon-magnesium (ONeMg)
cores: $\magnesium[24] \to \sodium[24] \to \neon[24]$ and
$\neon[20] \to \fluorine[20] \to \oxygen[20]$.  For these sequences of
captures, the excess electron energy is thermalized. These are the key
reactions in electron-capture supernovae and the accretion-induced
collapse (AIC) of ONeMg white dwarfs \citep[e.g.,][]{Miyaji80}.  As the
degenerate core approaches the Chandrasekhar mass, the electron
captures remove the pressure support and heat the plasma.
Figure~\ref{fig:aic} shows the evolution of a cold ONeMg WD
($X_{\mathrm{O}} = 0.5$, $X_{\mathrm{Ne}} = 0.45$,
$X_{\mathrm{Mg}} = 0.05$) accreting at
$\dot{M} = 10^{-6}\,M_\odot\,\mathrm{yr}^{-1}$.  The solid lines show
the evolution using the on-the-fly rates described in this section;
the dashed lines show the results when interpolating in coarse tables.
When using the coarse tables, the electron captures on
$\magnesium[24]$ do not occur until approximately a factor of two
larger density.  At this greater density, the energy
deposition from each capture is higher and this leads to a large
temperature change due to the $A=24$ captures alone.  In contrast, the
on-the-fly rates show the behavior demonstrated in previous studies of
this evolution that did not use sparse tables
\citep[e.g.,][]{Miyaji87}: the $A=24$ captures heat the plasma and
accelerate the contraction; the $A=20$ captures, due to the higher
$\neon$ abundance and a higher energy release per capture, cause a
thermal runaway and the formation of an oxygen deflagration
\citep{Schwab15}.

\subsection{Guidelines}
\label{s.weak-limitations}

\mesa\ provides the nuclear data used in the calculation of the
reactions specifically discussed in this section
\citep{Tilley98,Firestone07a,Firestone07b,Firestone09,ShamsuzzohaBasunia11,MartinezPinedo14}.
To consider additional reactions, a list of nuclear levels and
$(ft)$-values must be specified.

The expressions in Section~\ref{s.weak-details} assume degenerate,
relativistic electrons. As $\mue$ increases, additional transitions to
higher energy states of the daughter nuclei and must be included.  At
higher temperatures, excited states of the parent nucleus will begin
to be thermally populated and captures or decays from those states and
must be included.  At temperatures and densities where the composition
approaches NSE, these methods are particularly inappropriate, as it is
necessary to consider large pools of isotopes \citep{Juodagalvis10}.



\section{Chemical  Diffusion}
\label{s.mass_diffusion}

\mesa's early implementation of microscopic element diffusion
incorporated the approach used by \citet{Thoul94} in their seminal
work on understanding the sedimentation of helium in the solar
interior.  The fundamental starting point for this treatment of
diffusion is the Boltzmann equation with the assumption of binary
collisions where the particle's mean free path is much larger than the
average particle spacing. This formalism, encoded in the Burgers
equations \citep{Burgers69}, assumes that ions interact with an
effective potential that governs isolated interactions between only
two particles at a time. For more strongly coupled plasmas, as
$\Gamma\approx e^2/(\lambda_{\rm ion} \kB T)$ exceeds unity (where
$\lambda_{\rm ion} = (3/4\pi n_{\rm ion})^{1/3}$ is the mean
inter-ion spacing, and $n_{\rm ion}$ is the total ion number density),
it is no longer clear that this assumption remains valid. Later
updates to \mesa\ incorporated the work of \citet{Hu11} on radiative
levitation and incorporated the resistance coefficients calculated by
\citet{Paquette86} for approaches to the denser plasma regime as
$\Gamma \rightarrow 1$.

Here we describe \MESA's current implementation of chemical diffusion
and then discuss the path forward for diffusion implementations in the
$\Gamma> 1$ regime, needed for accurate studies of diffusion in the
interiors of white dwarfs or surfaces of neutron stars.  Recent
theoretical work in this strongly coupled regime
\citep{Baalrud13,Baalrud14,Beznogov14} provides support for a future
update of \MESA.


\subsection{Current Methods in \mesastar}

We now describe the formalism and assumptions underlying the approach
to diffusion currently present in \MESA.  This is followed by a
discussion of the framework for numerical implementation of this
formalism provided by \citet{Thoul94} and key modifications present in
the current version of the \mesa\ diffusion routine.

\subsubsection{Burgers Equations and the Low Density Limit}
\label{s:Burgers}

The Burgers equations for diffusion in an ionized plasma are derived
using the Boltzmann equation for the distribution function
$F_s(\mathbf x,\boldsymbol\xi,t)$ for particles of type $s$
\begin{equation}
\frac{\partial F_s}{\partial t} + \sum_i \xi_{i} \frac{\partial F_s}{\partial x_i}
+ \sum_i\frac{f_{si}}{m_s} \frac{\partial F_s}{\partial \xi_{i}} = \left( \frac{dF_s}{dt} \right)_{\rm{collision}},
\end{equation}
where $x_i$ are the components of the position vector, $\xi_{i}$ are
the components of the velocity vector, $f_{si}$ are components of the
forces on particles of type $s$, and $m_s$ is the mass for those
particles.  Throughout this section, the indices $s$ and $t$ refer to
particle species, while $i$ and $j$ are used to index other quantities
such as spatial components of vectors.

Burgers adopts the 13-moment approximation due to~\citet{Grad49} as a
closure scheme for taking moments of the Boltzmann equation. Burgers
also assumes an approximately Maxwellian distribution function
\begin{equation}
F_s = \frac{n_s}{\pi^{3/2}a_s^3} \exp \left( \frac{-c_s^2}{a_s^2} \right) (1+\phi_s),
\end{equation}
where $a_s = (2 \kB T/m_s)^{1/2}$, $c_{si} = \xi_{i} - u_{si}$
represents the components of the deviation of the velocity
from the mean flow velocity $\mathbf{u}_s$ of the species, and
\begin{equation}
\phi_s = \sum_{i,j} B_{sij} c_{si}c_{sj} + \sum_i C_{si}\left(c_s^2 - \frac 5 2 a_s^2\right) c_{si}
\end{equation}
is the small deviation (${\phi_s \ll 1}$) from the Maxwellian
distribution. The coefficients $B_{sij}$ and $C_{si}$ are defined such
that the distribution function has a total of 13 free parameters
corresponding to the 13 moments of the closure scheme
\citep[see][]{Burgers69}.

Burgers derives the collision integrals ($S_{st}^{(l)}$) and
cross-sections ($\Sigma_{st}^{(lj)}$) that result from taking moments
of the right hand side of the Boltzmann equation
\begin{equation}
\label{eq:collision_integrals}
S_{st}^{(l)} = 2 \pi \int_0^\infty ( 1 - \cos^l \chi_{st}) b \, db,
\end{equation}
\begin{equation}
\label{eq:sigmas}
\Sigma_{st}^{(lj)} = \frac{4 \pi}{\pi^{3/2}} \int_0^\infty dv \, \exp \left(\frac{-v^2}{\alpha_{st}^2} \right) \frac{v^{2j+3}}{\alpha_{st}^{2j+4}} S_{st}^{(l)},
\end{equation}
where $\alpha_{st}^2 = 2 \kB T/\mu_{st}$,
$\mu_{st} = m_s m_t/(m_s + m_t)$, $v$ represents the relative velocity
of colliding particles, and the angle of deviation $\chi_{st}$ is a
function of both $v$ and the impact parameter $b$ that depends on the
physics of the two-particle interaction between colliding particles in
the gas.  Burgers then defines the dimensionless coefficients
$z_{st}$, $z_{st}'$, $z_{st}''$, and $z_{st}'''$, along with
resistance coefficients ($K_{st}$) in terms of the collision
integrals:
\begin{align}
\begin{split}
K_{st} = K_{ts} &= \frac 2 3 n_s n_t \mu_{st} \alpha_{st} \Sigma_{st}^{(11)}, \\
\Sigma_{st}^{(12)}/\Sigma_{st}^{(11)} &= \frac 5 2 ( 1 - z_{st}), \\
\Sigma_{st}^{(13)}/\Sigma_{st}^{(11)} &= \frac{25}{4} - \frac{25}{2} z_{st} + \frac 5 2 z_{st}', \\
\Sigma_{st}^{(22)}/\Sigma_{st}^{(11)} &= z_{st}'', \\
\Sigma_{st}^{(23)}/\Sigma_{st}^{(11)} &= z_{st}'''.
\end{split}
\end{align}

In the ``single-fluid picture'' the diffusion velocities are defined
with reference to the mean velocity of the gas as a whole
($\mathbf u$), rather than with respect to the mean species velocity
($\mathbf u_s$):
\begin{equation}
u_{si} = \frac 1 {n_s} \int d \boldsymbol \xi \, \xi_{i} F_s,
\quad
\mathbf u = \frac 1 \rho \sum_s \rho_s \mathbf u_s,
\quad
\mathbf w_s = \mathbf u_s - \mathbf u.
\end{equation}
Burgers defines residual heat flow vectors
\begin{equation}
\label{eq:residual}
r_{si} = \left(\frac{m_s}{2 n_s \kB T} \int d^3 \boldsymbol \xi \, (\xi_{i} - u_i) |\boldsymbol\xi - \mathbf u|^2 F_s \right)
- \frac 5 2 w_{si}.
\end{equation}
As shown in section 18 of \citet{Burgers69} if we assume
\hbox{$|\mathbf w_s |\ll a_s$} and the absence of magnetic fields, the basic
equations of diffusion are
\begin{equation}
\label{eq:Burgers1}
\nabla p_s - \rho_s \mathbf g - \rho_{es} \mathbf E
= \sum_{t \neq s} K_{st} (\mathbf w_t - \mathbf w_s) + \sum_{t \neq s} K_{st} z_{st} \frac{m_t \mathbf r_s - m_s \mathbf r_t}{m_s + m_t},
\end{equation}
\begin{align}
\begin{split}
\label{eq:Burgers2}
\frac 5 2 n_s \kB \nabla T &= - \frac 2 5 K_{ss}z_{ss}''\mathbf r_s
- \frac 5 2 \sum_{t \neq s} K_{st} z_{st} \frac{m_t}{m_s+m_t}(\mathbf w_t - \mathbf w_s)   \\
 & \quad\; - \sum_{t \neq s} K_{st} \left[\frac{3 m_s^2 + m_t^2z_{st}'}{(m_s+m_t)^2} +
\frac 4 5 \frac{m_sm_t}{(m_s + m_t)^2}z_{st}'' \right]\mathbf r_s \\
&  \quad\; + \sum_{t \neq s} K_{st} \frac{m_sm_t}{(m_s + m_t)^2} \left(3 + z_{st}' - \frac 4 5 z_{st}'' \right) \mathbf r_t,
\end{split}
\end{align}
where $\mathbf E$ is the quasi-static electric field and $\rho_{es}$
is the average charge density of species $s$. These equations are
still general, with the form of the resistance coefficients not yet
fully specified. The physics of the particular types of interactions
within ideal gases is fully contained in the coefficients $K_{st}$,
$z_{st}$, $z_{st}'$, $z_{st}''$, and $z_{st}'''$.

For ionized gases, the resistance coefficients require evaluation of
collision integrals that diverge for a pure Coulomb
potential. However, since the two-particle interaction potential is
only truly applicable on short length scales, an integration cutoff or
screened potential is commonly adopted. Burgers chooses to calculate
resistance coefficients using a pure Coulomb potential truncated at
the Debye radius
\begin{equation}
R_\mathrm{D} = \left(4 \pi \sum_s \frac{n_s Z_s^2 e^2}{\kB T} \right)^{-1/2},
\end{equation}
which is assumed to be much larger than the inter-ion spacing.
Indeed, for a plasma of one species,
${R_\mathrm{D}/\lambda_{\rm ion}=(3\Gamma)^{-1/2}}$.  Applying this
form of interaction to the collision integrals, the $l=1$ integrals
defined in Equation~\eqref{eq:collision_integrals} can be
evaluated~\citep{Baalrud14}
\begin{equation}
\label{eq:exactCollision}
S_{st}^{(1)} = \frac{2 \pi R_\mathrm{D}^2 \alpha_{st}^4}{\Lambda_{st}^2 v^4} \ln\left[1 + \Lambda_{st}^2
\left(\frac{v}{\alpha_{st}} \right)^4\right],
\end{equation}
where
$\Lambda_{st} = \mu_{st} \alpha_{st}^2 R_\mathrm{D} /(Z_s Z_t e^2)$.
In order to perform the integral in Equation~\eqref{eq:sigmas},
Burgers notes that the dependence of $S_{st}^{(l)}$ on $v$ inside the
logarithmic term is weak, so that we can replace $v^2$ there with its
average value $\langle v^2 \rangle = 3 \kB T/\mu_{st}$. Assuming a
very dilute plasma, so that
$\Lambda_{st}^2 \langle v^2 \rangle^2/\alpha_{st}^4 \gg 1$, Burgers
then writes
\begin{equation}
S_{st}^{(1)} \approx \frac{4 \pi R_\mathrm{D}^2 \alpha_{st}^4}{\Lambda_{st}^2 v^4}
\ln\left(\frac{3 \kB T R_\mathrm{D}}{Z_s Z_t e^2}\right),
\end{equation}
and the final result for the resistance coefficients follows as
\begin{equation} \label{eq:BurgersResist1}
K_{st} \approx \frac{16 \sqrt \pi}{3} \frac{n_s n_t Z_s^2 Z_t^2 e^4}{\mu_{st} \alpha_{st}^3}
\ln \left( \frac{3 \kB T R_\mathrm{D}}{Z_s Z_t e^2} \right),
\end{equation}
\begin{equation} \label{eq:BurgersResist2}
z_{st} = \frac 3 5, \qquad z_{st}' = \frac{13}{10}, \qquad z_{st}'' = 2, \qquad z_{st}''' = 4.
\end{equation}
With these coefficients now fully specified, Burgers diffusion
equations along with constraints such as charge neutrality and current
neutrality form a closed set of equations, which can be solved for
$\mathbf w_s$, $\mathbf r_s$, $\mathbf E$, and $\mathbf g$ from the
input of a stellar profile.


\subsubsection{\MESA's Implementation of Thoul et al.'s Approach}
\label{s:Thoul}

The diffusion routine originally implemented in \MESA\ was based on
the work of \citet{Thoul94}. They start with the Burgers equations,
written in a compact notation following
\citet{Noerdlinger77,Noerdlinger78} that is equivalent to
Equations~\eqref{eq:Burgers1} and \eqref{eq:Burgers2} in one
dimension.
However, the approach of \citet{Thoul94} differs from Burgers'
original treatment in one important respect: the resistance
coefficients are based on a modified result for the collision
integrals. They follow Equation~\eqref{eq:BurgersResist2} for the
various $z_{st}$ coefficients, which uses a pure Coulomb potential
with a cutoff at the Debye length, but the $K_{st}$ coefficients were
derived from an alternative fitting of the Coulomb logarithms
introduced by \citet{Iben85}. For these coefficients, they define
$\lambda =\max(R_\mathrm{D},\lambda_{\rm ion})$, and use
\begin{align} \label{eq:ThoulResist}
\begin{split}
K_{st} = & \frac{16 \sqrt \pi}{3} \frac{n_s n_t Z_s^2 Z_t^2 e^2}{\mu_{st} \alpha_{st}^3} \\
& \times \frac{1.6249}{2} \ln\left[1+ 0.18769\left( \frac{4 \kB T \lambda}{Z_s Z_t e^2} \right)^{1.2} \right].
\end{split}
\end{align}
This expression is a fit to the numerical results of
\cite{Fontaine79}, motivated by white dwarf conditions where Burgers'
approximations for dealing with Equation~\eqref{eq:exactCollision} are
not valid (${\Gamma>1}$). Since this fit focuses on the strong
coupling regime, and differs from Equation~\eqref{eq:BurgersResist1},
these results can be incorrect in the limit of a dilute plasma as we
discuss later. Nevertheless, \cite{Thoul94} elected to use
Equation~\eqref{eq:ThoulResist} under all conditions, since it
provides an approximately correct solution in a convenient closed
form.

Using Equations \eqref{eq:Burgers1} and \eqref{eq:Burgers2} along with
the constraints of current neutrality ($\sum_s \rho_{es} w_s = 0$) and
local mass conservation ($\sum_s \rho_s w_s = 0$), \citet{Thoul94}
express an entire closed system of equations in a dimensionless matrix
form suitable for numerical evaluation:
\begin{equation}
\label{eq:Thoul3}
\frac p{K_0} \Bigg( \alpha_i \frac{d \ln p}{dr} + \nu_i \frac{d \ln T}{dr} +
\sum_{\substack{j=1 \\ j \neq \rm e}}^S \gamma_{ij} \frac{d \ln C_j}{dr} \Bigg) = \sum_{j = 1}^{2S+2} \Delta_{ij} W_j,
\end{equation}
where $S$ is the total number of species in the gas (including
electrons) and $C_j = n_j/n_{\rm e}$ is the concentration of the $j$th
species. Consult \citet{Thoul94} for definitions of $K_0$, $\alpha_i$,
$\nu_i$, $\gamma_{ij}$, and $\Delta_{ij}$. The definition of $W_{j}$
is
\begin{equation} \label{eq:W}
W_j = \begin{cases}
w_j & \text{for } j = 1 \ldots S,\\
r_j & \text{for } j = S+1 \ldots 2S,\\
K_0^{-1} n_{\rm e} e E & \text{for } j = 2S+1,\\
K_0^{-1} n_{\rm e} m_{\rm p} g & \text{for } j = 2S+2.
\end{cases}
\end{equation}
This is the vector containing the unknown quantities solved for after
specifying $K_0$, $\alpha_i$, $\nu_i$, $\gamma_{ij}$,
and~$\Delta_{ij}$. The routine provided by \citet{Thoul94} inverts
Equation~\eqref{eq:Thoul3} for one term in the left hand side at a
time so as to find the ``generalized diffusion coefficients,'' which
can be used to construct diffusion velocities or contributions from
pressure, temperature, or concentrations individually.


\subsubsection{Modified Coefficients and Radiative Levitation as Implemented by Hu et al.}
\label{s:Hu}

\citet{Hu11} extend the methods of \citet{Thoul94} by introducing some
key modifications. First, they include an extra force term due to
radiative levitation, so that Equation~\eqref{eq:Burgers1} becomes
\begin{align}
\begin{split}
&\frac{d p_s}{dr} + \rho_s(g - g_{\mathrm{rad},s}) - n_s \bar Z_s e E \\
&= \sum_{t \neq s} K_{st}(w_t - w_s) + \sum_{t \neq s} K_{st} z_{st} \frac{m_tr_s - m_s r_t}{m_s + m_t},
\end{split}
\end{align}
where $g_{\mathrm{rad},s}$ refers to the radiative acceleration on
species $s$.  $\bar{Z}_s$ is the average charge of species $s$,
allowing an account of partial ionization so that
$n_s \bar Z_s e = \rho_{es}$.  They do not modify
Equation~\eqref{eq:Burgers2}.\footnote{As written in equation (3) of
  \cite{Hu11}, their expression has two errors in the first term on
  the right hand side of the first line: the sign is wrong, and it is
  missing resistance coefficients $K_{ij}$. Since neither of these
  errors propagates into later sections of the paper, it appears that
  both are simply typos, and otherwise their expression matches
  Equation~\eqref{eq:Burgers2} exactly.}

In contrast to Thoul's original routine, \citet{Hu11} use the
resistance coefficients from \citet{Paquette86}, which were generated
based on substantial improvements to \citet{Fontaine79}.  In
evaluating the collision integrals, \citet{Paquette86} use a screened
Coulomb potential of the form
\begin{equation}
V_{st}(r) = \bar Z_s \bar Z_t e^2 \frac{\exp(-r/\lambda)}{r},
\end{equation}
where, once again, $\lambda =\max(R_\mathrm{D},\lambda_{\rm ion})$.
As we note below, this choice of $\lambda$ makes a substantial
difference in strongly coupled plasmas, where the Debye radius no
longer corresponds to a distance at which other nearby charged
particles can significantly screen the Coulomb field.  After setting
up the algebra for a matrix solution very similar to that of
\citet{Thoul94}, \citet{Hu11} solve for the vector $W_j$ (as defined
in Equation \ref{eq:W}) appearing in the equation
\begin{align}
\begin{split}
\frac{p}{K_0} \Bigg( &- \frac{\alpha_i m_i g_{\mathrm{rad},i}}{\kB T} + \alpha_i \frac{d \ln p}{dr} \\
&+ \nu_i \frac{d \ln T}{dr} + \sum_{\substack{j=1 \\ j \neq \rm e}}^S \gamma_{ij} \frac{d \ln C_j}{dr} \Bigg) 
= \sum_{j=1}^{2S + 2} \Delta_{ij} W_j.
\end{split}
\end{align}
Many of the quantities appearing in this equation are defined
differently than in \citet{Thoul94}; see \citet{Hu11} for details.  We
can also solve this equation directly for the vector $W_j$ to obtain
\begin{equation} \label{eq:diffusion_efield}
\frac{W_{2S+1}}{W_{2S+2}} = \frac{K_0^{-1} n_{\rm e} e E}{K_0^{-1} n_{\rm e} m_{\rm p} g} = \frac{eE}{m_{\rm p} g},
\end{equation}
the strength of the electric field relative to gravity.


\subsection{Analytic Expression for the Electric Field}
\label{sec:Analytic}

In some simple cases, Burgers equations can be solved to yield an
analytic expression for the electric field, providing a useful test
for \MESA. Starting directly with his diffusion equations,
\citet{Burgers69} arrives at the following expressions for a pure
plasma of electrons along with one species of ions (charge $Ze$):
\begin{align}
\frac{\nabla p_{\rm e} + n_{\rm e} e \mathbf E}{Z K_0} &= \mathbf w
+ \frac 3 5 \mathbf r_{\rm e},  \\
\frac 5 2 \frac{n_{\rm e} \kB \nabla T}{Z K_0} &= - \frac 3 2 \mathbf w
-\left(\frac 2 5 \frac{K_{\rm ee}z_{\rm ee}''}{K_{\rm ie}} + z_{\rm ie}' \right) \mathbf r_{\rm e},
\end{align}
where $\mathbf w = \mathbf w_{\rm i} - \mathbf w_{\rm e}$.
For a plasma with only one ion species in diffusion equilibrium, the
constraints of current neutrality and local mass conservation
give $\mathbf w = 0$.
In the case of a pure hydrogen plasma, $p = 2 p_{\rm e}$, and in hydrostatic
equilibrium $\nabla p_{\rm e} = \nabla p / 2  = \rho \mathbf g /2$.
Hence, we can solve the above set of equations to find
\begin{equation}  \label{eq:BurgersEfield}
e \mathbf E = - \frac 1 2 m_{\rm p} \mathbf g - \frac{3}{2}
\left( \frac 2 5 \frac{K_{\rm ee}z_{\rm ee}''}{K_{\rm ie}}
+ z_{\rm ie}'\right)^{-1} \kB \nabla T.
\end{equation}
The coefficient for the temperature gradient term depends directly on
the nature of the resistance coefficients in the Burgers formalism, so
different models of Coulomb collisions in ionized plasma will lead to
different results for the electric field.

As a slight generalization of Equation~\eqref{eq:BurgersEfield} in one dimension, we write
\begin{equation} \label{eq:generic_hydrogen_efield}
\frac{eE}{m_{\rm p}g} = \frac 1 2 - \alpha_e \frac{\kB}{m_{\rm p} g} \frac{dT}{dr}.
\end{equation}
If we calculate the coefficient $\alpha_e$ using the Burgers' formalism with
Equations~\eqref{eq:BurgersResist2} and \eqref{eq:ThoulResist},
we find
\begin{equation} \label{eq:BurgersChapman}
\alpha_e = \frac{3}{2} \left( \frac 2 5 \frac{K_{\rm ee}z_{\rm ee}''}{K_{\rm ie}}
+ z_{\rm ie}'\right)^{-1} = 0.804
\end{equation}
A comparable analytic expression for the electric field is provided by
\citet{Dupre81}, who applies a Boltzmann-Fokker-Planck approach to
finding diffusion coefficients for trace elements in hydrogen
plasma. His treatment of diffusion is more precise than the Burgers'
formalism, but has the limitation of only being applicable in the case
of nearly pure hydrogen with a diffusing trace element. His result for
the electric field matches the form of
Equation~\eqref{eq:generic_hydrogen_efield} with the coefficient
$\alpha_e = 0.703$.  This provides another useful point of comparison
in the specific case of nearly pure hydrogen plasmas. Below we use
this analytic expression as a test of the updated resistance
coefficients employed by \citet{Hu11}.


\subsection{Results and Comparisons}

We have constructed several simple \MESA\ test cases in order to
illustrate the effects of radiative levitation and different
resistance coefficients. Where possible, we compare \mesa\ output to
corresponding analytic expressions.

\subsubsection{Electric Fields}

By default, \MESA\ uses the resistance coefficients provided by
\citet{Paquette86}, but it can also use the resistance coefficients
defined by \citet{Iben85}, given here in
Equation~\eqref{eq:ThoulResist}.  In the case of a pure hydrogen star,
the coefficients given in Equation~\eqref{eq:ThoulResist} lead
directly to Equation~\eqref{eq:BurgersChapman}, so these coefficients
are especially useful in performing simple comparisons of \MESA\
output to a corresponding analytic expression.  Due to the complicated
numerical methods used to obtain the resistance coefficients of
\citet{Paquette86}, it is not possible to write down a directly
corresponding closed form analytic expression for the electric field,
but results based on these more precise calculations compare favorably
to those of \citet{Dupre81} in the case of a pure hydrogen plasma.
Starting with the \MESA\ test suite, we constructed a solar mass pure
hydrogen star, and we ran just long enough to turn on the diffusion
routine and gather output for electric and gravitational fields.  For
such a star, we can compare \MESA\ results for the electric field
directly to the analytic expression given in
Equation~\eqref{eq:generic_hydrogen_efield}, with $\alpha_e = 0.804$
in the solution of \citet{Burgers69} and $\alpha_e = 0.703$ for
\citet{Dupre81}.

\begin{figure}[hbtp]
\begin{center}
\includegraphics[width=\apjcolwidth]{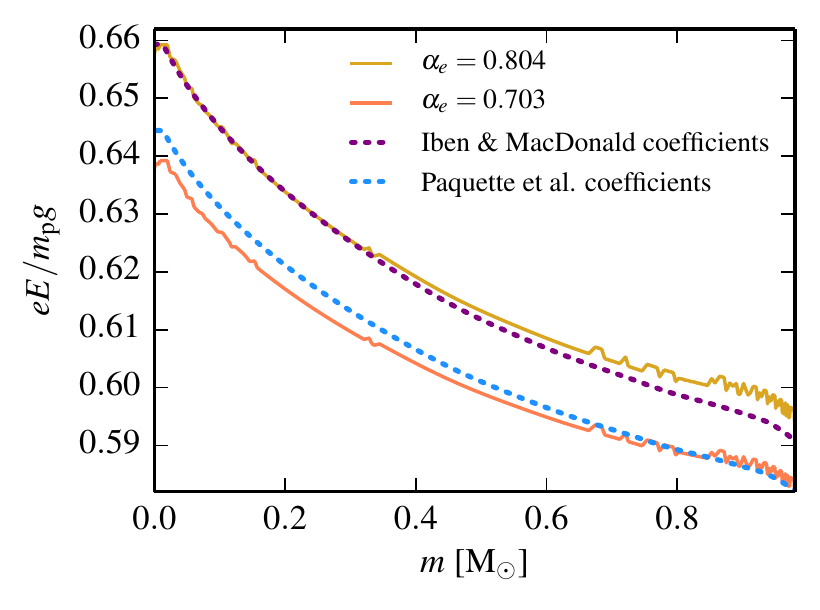}
\caption[Electric fields in a pure hydrogen star]{Comparison of
  electric field strengths relative to gravity in a pure hydrogen star
  (${M = 1.0\,\mathrm{M}_\odot}$,
  ${T_\mathrm{eff} = 5.74 \times 10^3 \, \mathrm{K}}$,
  $L = 0.576 \, \rm{L}_\odot$) with nuclear burning artificially
  suppressed in the \MESA\ routine to avoid any helium contamination.
  Solid lines represent the analytic expression given by
  Equation~\eqref{eq:generic_hydrogen_efield} for two different values
  of the coefficient $\alpha_e$. Dashed lines represent output from
  the \MESA\ diffusion routine as described in
  Equation~\eqref{eq:diffusion_efield}, with the only difference being
  the resistance coefficients used to solve the Burgers equations.}
\label{fig:PureHydrogen}
\end{center}
\end{figure}

Figure~\ref{fig:PureHydrogen} plots the result of
Equation~\eqref{eq:generic_hydrogen_efield} for both values of
$\alpha_e$, along with the results from the diffusion routine
(Equation~\ref{eq:diffusion_efield}) for each type of resistance
coefficients available in \MESA. As expected, the curve calculated
from the \MESA\ diffusion routine output using the resistance
coefficients of \cite{Iben85} closely matches the analytic expression
with $\alpha_e = 0.804$ as calculated by \cite{Burgers69} using his
similar coefficients.  When using the more detailed numerical
calculations for the resistance coefficients provided by
\cite{Paquette86}, the diffusion routine output closely resembles the
more precise analytic calculation given by \cite{Dupre81}.

\begin{figure}[hbtp]
\begin{center}
\includegraphics[width=\apjcolwidth]{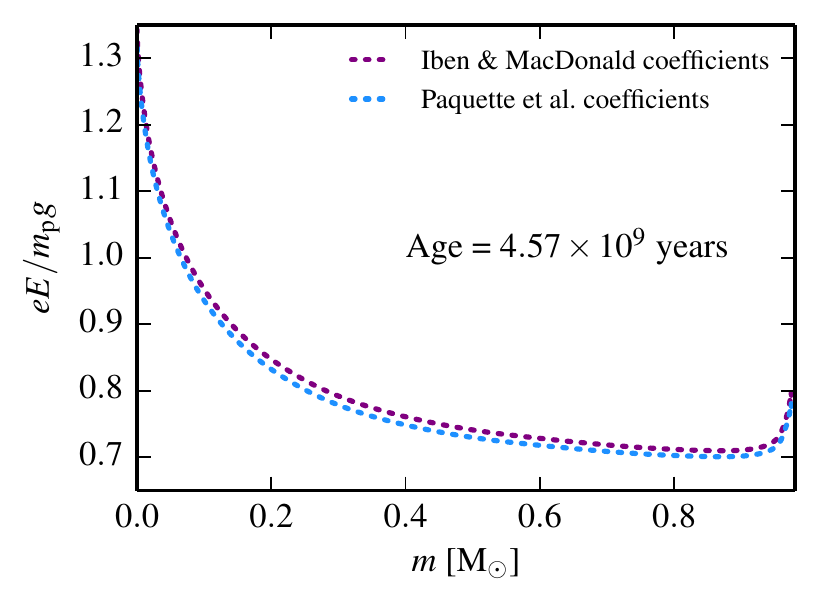}
\caption[Comparison of electric fields in the sun]{Comparison of electric field strengths
relative to gravity using different resistance coefficients in a solar model.}
\label{fig:SunComparison}
\end{center}
\end{figure}

The Sun provides another interesting test case for comparing the
effects of using different resistance coefficients.  An example solar
model from the \MESA\ test suite was run with different choices of the
resistance coefficients.  Figure~\ref{fig:SunComparison} shows a
slight difference between the electric field strengths relative to
gravity given by the \citet{Paquette86} coefficients and those by
\citet{Iben85}.

\subsubsection{Gravitational Fields}

\begin{figure}[hbtp]
\begin{center}
\includegraphics[width=\apjcolwidth]{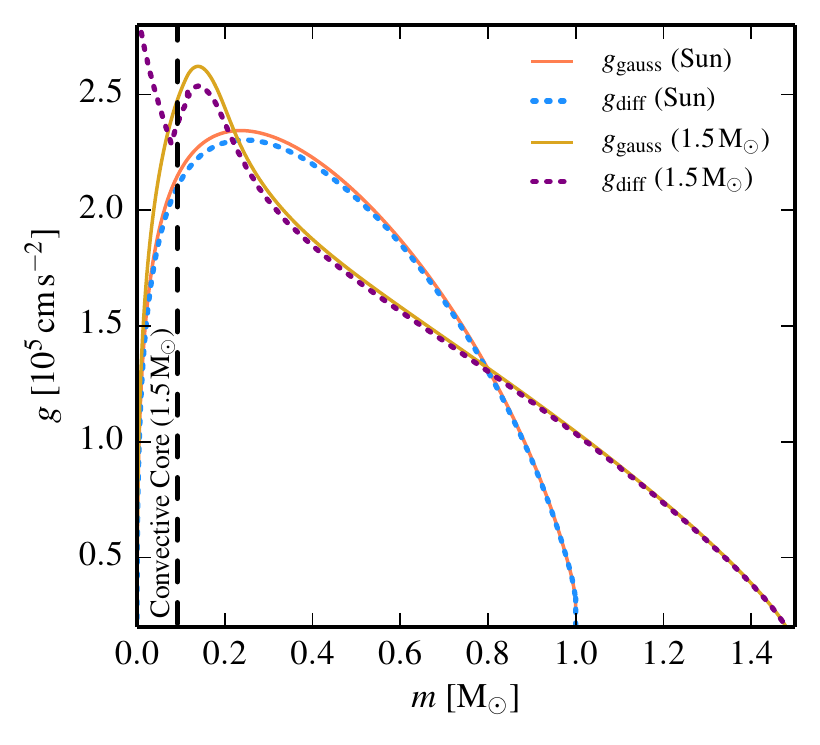}
\caption[Comparison of gravitational fields in the sun]{Comparison of
  gravitational fields obtained from $g_{\rm diff}$ and
  $g_{\rm gauss}$ in two \MESA\ test suite cases.  The two lines
  representing the Sun (${\rm age = 4.57 \, Gyr}$) show good
  agreement, while the two lines representing a $1.5 \, \rm{M}_\odot$
  star disagree in regions with large convective flux where diffusion is
  inconsequential.}
\label{fig:gfields}
\end{center}
\end{figure}

The \MESA\ diffusion routine treats both the electric field and local
gravitational acceleration as unknown quantities. \MESA\ records the
quantity $W_{2S+2}$ (Equation~\ref{eq:W}), used to calculate the
gravitational acceleration from the diffusion routine:
\begin{equation}
\label{eq:gDiff}
g_\mathrm{diff} = \frac{K_0W_{2S+2}}{n_{\rm e} m_{\rm p}}.
\end{equation}
This expression for $g_{\rm diff}$ is independent of the simpler
expression for local gravitational acceleration
${g_{\rm gauss} = Gm/r^2}$.  Figure~\ref{fig:gfields} compares
$g_\mathrm{gauss}$ and $g_\mathrm{diff}$ for a typical profile found
using the example solar model from the \MESA\ test suite.  In
Figure~\ref{fig:gfields} a profile from a star of larger mass
($M = 1.5 \, \mathrm{M}_\odot$) shows disagreement between the gravity
outputs in the convective core.
The Burgers formalism assumes heat transfer that is correlated with
temperature gradients through the residual heat flow vectors defined in 
Equation~\eqref{eq:residual}.
This assumption breaks down when most of energy is transported by
convection;
however, the effects of diffusion in this region are completely overwhelmed by
convective mixing and are therefore inconsequential.

\subsubsection{Radiative Levitation}

\begin{figure}[hbtb]
\begin{center}
\includegraphics[width=\apjcolwidth]{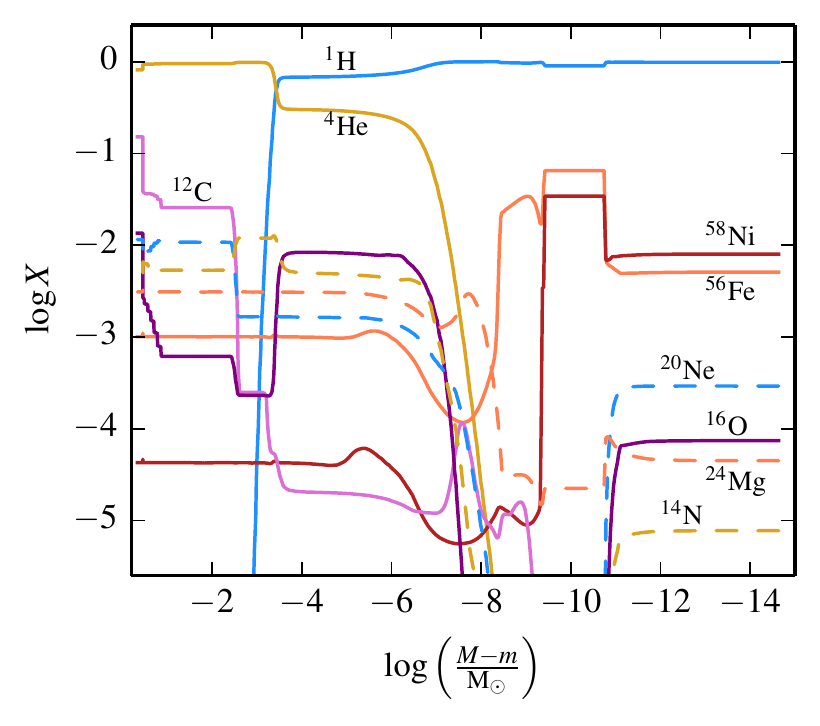}
\caption[Abundance profile of an sdB star]{Abundance profile of a
  subdwarf B star model (${M = 0.462 \, \rm{M}_\odot}$,
  ${T_{\mathrm{eff}} = 2.67 \times 10^{4} \, \rm{K}}$,
  ${L = 1.12 \, \rm{L}_\odot}$, $\mathrm{age}=5 \, \mathrm{Myr}$)
  showing the effects of radiative levitation with a layer of
  $^{56}\rm{Fe}/^{58}\rm{Ni}$ at the surface.  }
\label{fig:sdb}
\end{center}
\end{figure}

\MESA's implementation of radiative levitation is based on
\citet{Hu11}.  Figure~\ref{fig:sdb} shows an abundance profile of a
subdwarf B star model produced by \MESA, where radiative levitation
is responsible for the presence of $^{56}\rm{Fe}$, $^{58}\rm{Ni}$, and
other metals near the surface 
\citep[as also seen in figure~3 of][]{Hu11}.

\subsubsection{White Dwarf Sedimentation}

In a cooling WD, diffusion governs sedimentation over long
timescales. The assumptions behind the formalism of the Burgers
equations do not hold under white dwarf conditions:
\begin{itemize}
\item The Burgers equations assume all particle species satisfy an ideal gas
equation of state. In the context of a degenerate WD both  electrons and ions
violate this assumption.
\item The very dense, strongly coupled ($\Gamma > 1$) conditions of a WD
call into question the validity of the two-particle scattering picture used
to calculate the ion resistance
coefficients.
\end{itemize}
Nevertheless, for lack of a better option, previous studies have
relied on the Burgers equations with the coefficients of
\citet{Paquette86}. For example, see \citet{Argentina02}.

\begin{figure}[hbtb]
\begin{center}
\includegraphics[width=\apjcolwidth]{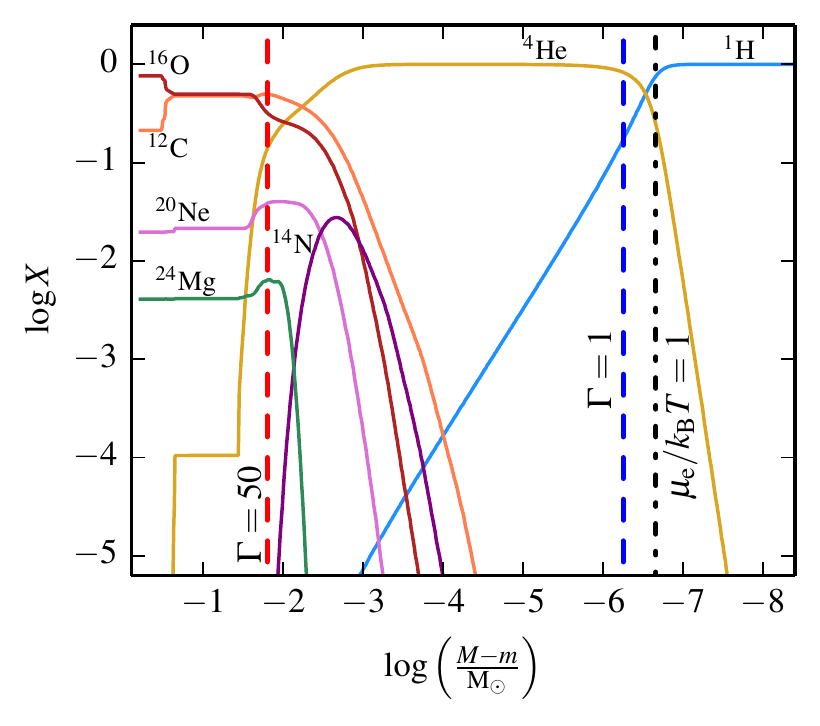}
\caption[Abundance profile of a CO white dwarf]{Abundance profile of a CO WD
($M$ = 0.611 $\rm{M}_\odot$, $T_{\mathrm{eff}}$ = 5.16 $\times$ 10$^3 \, \rm{K}$,
$L = 9.29 \times 10^{-5} \, \rm{L}_\odot$)
after 4 Gyr of WD evolution. The region left of the blue, dashed line is the interior
of the WD, where $\Gamma \geq 1$.
Left of the red, dashed line $\Gamma \geq 50$,
and diffusion has been turned off for this region. 
The electrons are an ideal gas to the right of the black dot-dash line.}
\label{fig:COwd}
\end{center}
\end{figure}

Figure~\ref{fig:COwd} shows an abundance profile produced by
\MESA\ for a CO WD after 4 Gyr of evolution, where diffusion
governs sedimentation in the outer layers.  The vertical lines in
Figure~\ref{fig:COwd} mark the outer boundaries of regions where the
two concerns listed above become significant. Nearly all of the WD
resides inside at least one of these regimes, and much of the
interesting diffusion sedimentation occurs inside regions that are
both significantly coupled and highly degenerate.  Thus, improvements
to the treatment of diffusion are clearly necessary before we are 
able to describe diffusion in WDs adequately.  This
\MESA\ run turns off diffusion for $\Gamma \geq 50$, where we expect
strong coupling to substantially modify the underlying equations.

\subsection{Expanding the Domain of Validity and Next Steps}
\label{sec:validity}

The validity of the Boltzmann approach becomes questionable as
$\Gamma>1$ and the ions become a liquid. \cite{BildstenHall}
estimated the diffusion coefficient in this liquid regime by using the
Stokes-Einstein relation. However, for a broad-based code such as
\MESA, we need to implement diffusion into the $\Gamma>1$ regime in a
manner that allows for a smooth transition between coupling regimes.

\cite{Paquette86} successfully described diffusion in a regime of
intermediate coupling through the use of screened potentials, 
which are a way to account for the collective nature of interactions
in a dense plasma.  Though there is no rigorous reason to expect that
a formalism based on the two-particle scattering picture should work
well as $\Gamma \rightarrow 1$, their comparison to simulations
verified that this description of diffusion is very accurate for
$\Gamma \lesssim 1$. 

\begin{figure}[hbtp]
\begin{center}
\includegraphics[width=\apjcolwidth]{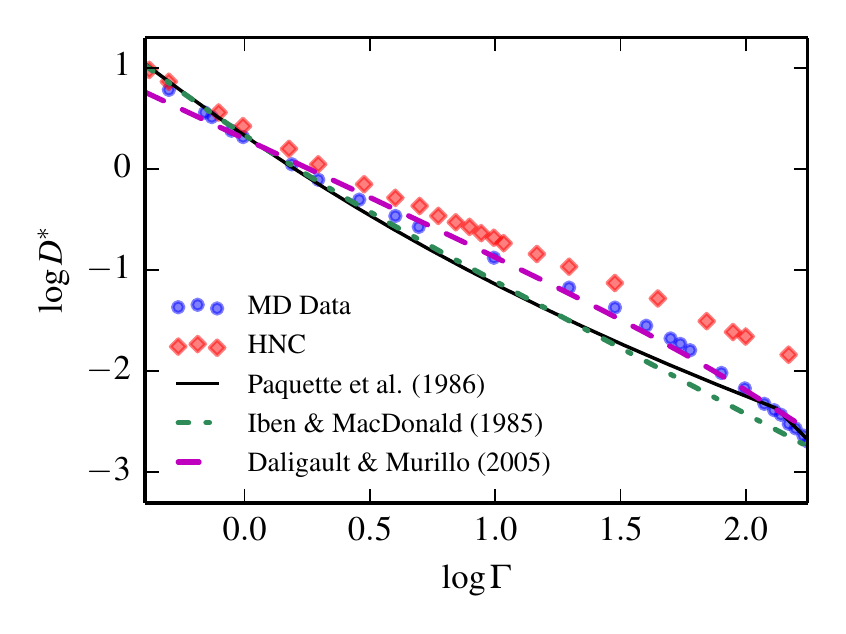}
\caption[Self diffusion coefficient from several methods]{Compilation of the self-diffusion
coefficients obtained from different methods. ``MD Data'' and ``HNC'' points
are taken from \cite{Baalrud13}. The solid black line is the result of the \MESA\
calculation using the coefficients of \cite{Paquette86}. The dashed green line is the
result of the calculation using the resistance coefficients from the original routine of
\cite{Thoul94} based on the fit to the Coulomb logarithm found in \cite{Iben85}, given
here in Equation~\eqref{eq:ThoulResist}. The dashed purple line represents the fit
to MD data given here in Equation~\eqref{eq:MDfit}.
}
\label{fig:diffusion_coefficient}
\end{center}
\end{figure}

Can these approximations be extrapolated to the strongly coupled
regime of $\Gamma>1$?  \cite{Baalrud13} provide a method for
numerically calculating resistance coefficients using a hypernetted
chain (HNC) approximation from effective potentials. 
Figure~\ref{fig:diffusion_coefficient} compares their HNC results 
(diamonds) to their Molecular Dynamics (MD) simulations of a
one-component plasma (OCP, circles) for the self-diffusion
coefficient $D^*$, defined by
\begin{equation}
D^* = \frac{D}{\lambda_{\rm ion}^2 \omega_p},
\end{equation}
where $\omega_p$ is the plasma frequency and $D = 2 D^{(2)}_{ss}$ (the
factor of $2$ in this definition ensures that if we redefine species
$s$ in terms of two subspecies $s_1$ and $s_2$, then $D =
D^{(2)}_{s_1s_2}$). The general expression for the interdiffusion coefficient
is 
\begin{equation}
D_{st}^{(2)} = \frac{n_s n_t}{n_s + n_t} \frac{\kB T}{K_{st}(1-\Delta)},
\end{equation}
where the $1- \Delta$ term in the denominator accounts for a second
order correction that can be defined using
\begin{equation}
\Delta = \frac{(2 \Sigma_{st}^{(12)}- 5 \Sigma_{st}^{(11)})^2/\Sigma_{st}^{(11)}}{55\Sigma_{st}^{(11)} - 20 \Sigma_{st}^{(12)} + 4 \Sigma_{st}^{(13)} + 8 \Sigma_{st}^{(22)}}.
\end{equation}
For reference, we also include a direct fit of \cite{Daligault05} to
the MD data of \cite{Ranganathan03}, given by
\begin{equation}
\label{eq:MDfit}
D^* = 0.0028 + 0.00525 \left( \frac{173}{\Gamma} - 1\right)^{1.154}.
\end{equation}
The agreement between the HNC and MD simulations shows that the HNC does
a better job of accounting for correlation physics in strongly coupled
plasmas than a simple screened Coulomb potential and allows for a
surprising (and still physically unexplained) extension of the Burgers
formalism into the strongly coupled regime.  This recent work allows
us to go into the large $\Gamma$ limit with the
Burgers formalism, but the question remains as to how we obtain
diffusion coefficients in a reliable manner.

The self-diffusion coefficients from the two options in \MESA\ are
shown in Figure~\ref{fig:diffusion_coefficient} and correlate with the
MD data better than expected for the high $\Gamma$ regime.  In
particular, the agreement is much better than that shown in figure~2
of \cite{Baalrud13} for either ``cutoff'' or ``screened'' Coulomb
methods.  The reason for this agreement is that both
\MESA\ implementations use the inter-ion spacing rather than the Debye
length once $\Gamma>1/3$, which yields favorable scalings in the high
$\Gamma$ limit.  \citet{Iben85} constructed their fitting formula
based on a few numerical results for $\Gamma >$1.  \cite{Paquette86}
also showed that their formalism can be extended to $\Gamma>1$ as long
as the inter-ion spacing is used rather than the Debye radius for the
screening length.

Though \MESA\ does not yet provide the capability of implementing
resistance coefficients based on the HNC method, we hope to accomplish this 
in the near future by means of a table similar to that
provided for the coefficients of~\cite{Paquette86}. For a more
thorough discussion of these methods and the likely path of
application to mixtures, consult \cite{Beznogov14}.  We will also need
to correctly account for the electron degeneracy and the non-ideal equation of state 
for the ions, both of which modify the electrostatic field needed to
correctly determine the forces that drive diffusion.



\section{Software Infrastructure}\label{s.infrastructure}

Here we describe a number of changes to \mesa\ that have
occurred since \mesatwo\ and are of potential interest to users of
\mesa\ or developers of similar software.

\mesa\ can be compiled with either the GNU or Intel Fortran compilers,
runs on multiple operating systems (Windows, OS X, and Linux), and can
use different numbers of OpenMP threads.  It is necessary to regularly
test that the code is performing correctly across the different
combinations of compiler, OS, thread count.  To this end, developers
and engaged users run the \mesa\ test suite on a wide range of systems
before each release.

Previously, test cases in the \mesa\ test suite accepted different
results so long as they were within a certain tolerance, an appropriate
choice for testing scientific results where the physical uncertainties
are much greater than the numerical ones.  However, we found that this
made detecting and tracking bugs across platforms difficult.  For the
purposes of code testing, it is much better to insist that any
inconsistency is a problem, no matter how small.


Motivated by this challenge, \mesa\ now provides bit-for-bit
consistency for all results across all the supported platforms.  It is
essential to emphasize that the goal of this achievement is to enable
better testing.  It allows users to exactly reproduce the results of
others, independent of platform differences, which is especially
useful to developers attempting to reproduce bugs.  The achievement of
bit-for-bit consistency is not a claim that the results of \mesa\
calculations are physically accurate or numerically converged to any
specific degree.

This bit-for-bit consistency was achieved via the following choices:
\begin{itemize}

\item Using parallel algorithms that give identical results
  independent of number of threads or order of thread execution.
  \MESA's linear algebra solver is based on BCYCLIC
  \citep{hirshman_2010_aa}. It sub-divides the work between threads
  based on the the size of the matrix rather than on the number of
  threads available.  It is also necessary to avoid OpenMP reduction
  clauses, which provide no guarantees on ordering of operations.
 
\item Specifying compiler flags that forbid the compiler from making
  any optimization that can affect floating point precision (e.g.,
  forbid re-association and fast math operations). Most optimizations
  are still allowed.

\item Using an I/O library that does precise conversion from binary to
  ASCII for double precision numbers.  

\item Using a math library that gives consistent results for
  operations such as \texttt{log}, \texttt{exp}, \texttt{sin},
  \texttt{cos}, \texttt{pow}. \mesa\ uses
  CRLIBM\footnote{http://lipforge.ens-lyon.fr/www/crlibm/index.html}
  in round towards zero mode. The choice to use a math library that
  gives exact results is not because 16 digit accuracy from the math
  routines in necessary.  Rather, we want consistent results across
  supported platforms and this is the best way to achieve this
  consistency.

\item Replacing integer power expressions (i.e., \texttt{x**3}) by
  repeated multiplications (i.e., \texttt{x*x*x}).  Different
  compilers implement integer powers differently, giving different
  results.

\end{itemize}

Having achieved bit-for-bit identical results, we can test files for
exact equality.  This applies both to the module-by-module tests that
run at installation time and the case by case tests in the star and
binary test suites.  These test cases compare the final model from the
test run to a saved result from a previous \mesa\ version.  If they
are not exactly the same, the test fails.  The test is also restarted
from an intermediate state to confirm that runs which are stopped and
restarted yield exactly the same results as those that are not.

While \MESAstar\ is parallelized via OpenMP, the install process has
historically been serial.  \mesa\ contains approximately 1000 Fortran
files and so the ability to compile more than one file simultaneously
has the potential to provide significant reductions in the time needed
to install \mesa.  Recently the compilation step has been
parallelized, enabled by the automated dependency generation tool
\texttt{makedepf90},\footnote{http://personal.inet.fi/private/erikedelmann/makedepf90/}
allowing multiple instances of the Fortran compiler to be invoked
simultaneously.  This is of particular utility for developers who may
recompile \mesa\ frequently.

Since \mesatwo\ the main \mesa\
website\footnote{http://mesa.sourceforge.net} has undergone
significant revision, making it easier for new users to get started
with \mesa.  This restructuring has also made it easier for the
developers to keep material up-to-date as \mesa\ evolves.  One of the
most important improvements is that the files that document the
default value of each \mesa\ option use the
Markdown\footnote{http://daringfireball.net/projects/markdown/} markup
language.  This allows documentation web pages to be generated
automatically for each \mesa\ release.

Improvements have also been made to the distribution of \mesa.
Previously, \mesa\ was available only by checking out the source code
using the Subversion\footnote{https://subversion.apache.org/} version
control system.  Now, every release version of \mesa\ (including past
releases) is available for download as a ZIP archive.  This is simpler
and saves bandwidth and disk space.  It has quickly become the
preferred way to install \mesa\ with the ZIP file of the current
release being downloaded tens of times per week.


\section{Summary and Conclusions}\label{s.conclusions}

We have explained and, where possible, verified or validated, major
new capabilities and improvements implemented in \MESA \ since the
publication of \mesaone\ and \mesatwo.  These advancements include
interacting binary systems (Section~\ref{s.binaries}), implicit
hydrodynamics and shocks (Section~\ref{s.hydro}), in-situ usage of
large reaction networks, especially for X-ray bursts and core-collapse
supernova progenitors (Section \ref{s.advburn}), and the explosion of
massive stars (Section \ref{s.ccsn}). These new capabilities will
allow for extended exploration of core collapse progenitors and the
sensitivity of shock nucleosynthesis to their explosion mechanism.
The full coupling of \mesa\ to the \GYRE\ non-adiabatic pulsation
instrument (Section~\ref{s.pulse}) has already revealed the richness
of the instability strips for massive stars and enables the continued
growth of astero-seismology across the HR diagram.  Progress in the
treatment of mass accretion (Section~\ref{s.accretion}) and weak
reaction rates (Section~\ref{s.weak}) will improve studies of their
impact on stellar evolution.  We also discuss the domain of validity
for particle diffusion within \mesa\ and describe a path forward for
extending diffusion into the regime relevant to WD sedimentation
(Section~\ref{s.mass_diffusion}).  We also describe significant
improvements to the infrastructure of
\mesa\ (Section~\ref{s.infrastructure}).  \MESAstar\ input files and
related materials for all the figures are available at
\href{http://mesastar.org}{http://mesastar.org}.

These hitherto unpublished advancements have already enabled a number
of studies in interacting binary systems
\citep{wolf_2013_aa,pavlovskii_2015_aa,b:vos2015}
 and stellar pulsations
\citep{Papics:2014aa, stello_2014_aa, quinn_2015_aa, cunha_2015_aa}, and led to the
discovery of new features in the thermal runaway during the evolution
of ONeMg cores towards AIC \citep{Schwab15}. It
also enabled the first three dimensional simulations of the final
minutes of iron core growth in a massive star up to and including the
point of core gravitational instability and collapse
\citep{couch_2015_aa}.  In addition, these enhanced capabilities have
allowed for applications of \MESAstar \ that were not initially
envisioned, such as the treatment of Magneto-Rotational Instability in stars
\citep{wheeler_2015_aa}, effects of axions on nucleosynthesis in
massive stars \citep{aoyama_2015_aa}, and particle physics beyond the
Standard Model \citep{curtin_2014_aa}.

As a community software instrument for stellar astrophysics new
directions for \mesa\ will be driven by: features useful to the
\MESA\ user community, advances in the physics modules, algorithmic
developments, and architectural evolution.  Potential examples include
a treatment of ionization in the equation of state for an arbitrary composition
across an expanded region in the \hbox{$\rho$-$T$} plane, non-linear
pulsations, Monte Carlo based thermonuclear reaction rates, modules
for subsonic flame propagation, ports to additional architectures, and a
web-interface to \MESA \ for education.


\acknowledgements


It is a pleasure to thank
Nilou Afsari,
Dave Arnett,
Warrick Ball,
Ed Brown,
Jieun Choi,
Joergen Christensen-Dalsgaard,
Andrew Cumming,
Sebastien Deheuvels,
Aaron Dotter,
Chris Fryer,
Duncan Galloway,
Pascale Garaud,
Alfred Gautschy,
Samuel Jones,
Max Katz,
Eli Livne,
Marcin Mackiewicz,
Chris Mankovich,
Casey Meakin, 
Broxton Miles,
Ehsan Moravveji,
Kevin Moore,
Eliot Quataert,
Jeremy Sakstein,
Richard Stancliffe,
Willie Strickland,
Anne Thoul,
and Joris Vos.
We also thank the participants of the 2013 and 2014 MESA Summer Schools
for their willingness to experiment with new capabilities.


This project was supported by NSF under the SI$^2$ program grants
(ACI-1339581, ACI-1339600, ACI-1339606) and NASA under the 
TCAN program grants (NNX14AB53G, NNX14AB55G, NNX12AC72G).
The work at UC Santa Barbara was also supported by the NSF
under grants PHY 11-25915, AST 11-09174, AST 12-05574.
J.S. is supported by the NSF Graduate Research
Fellowship Program under grant DGE 11-06400.
L.D. acknowledges financial support by the ``Agence Nationale de la Recherche''
under grant ANR-2011-Blanc-SIMI-BS56-0007.
D.T. acknowledges support under HST-GO-12870.14-A from the Space
Telescope Science Institute, which is operated by the Association of
Universities for Research in Astronomy, Inc., for NASA, under contract
NAS 5-26555.
R.H.D.T. acknowledges resources provided by of the
University of Wisconsin-Madison Advanced Computing Initiative.
F.X.T. acknowledges support from the Simons Foundation.



\bibliographystyle{apj}
\bibliography{bibs/mesa,bibs/luc,bibs/josiah,bibs/evan,bibs/pulsations,bibs/mass_accretion,bibs/rob,bibs/binaries}

\end{document}